\documentclass[a4paper,11pt]{article}
\pdfoutput=1 

\usepackage{jheppub} 
\usepackage{xspace}
\usepackage[T1]{fontenc} 
\usepackage{amsfonts}
\usepackage{amsmath,amssymb,amsthm,bbm}
\usepackage{color}
\usepackage[normalem]{ulem}
\usepackage{verbatim}
\usepackage{braket}
\usepackage{multirow}
\usepackage{inputenc} 
\usepackage{lmodern}
\usepackage{anyfontsize}
\usepackage{todonotes}
\usepackage{xspace}
\presetkeys{todonotes}{inline}{}
\usepackage{colortbl}
\usepackage{xcolor}
\usepackage{multirow}
\usepackage{hhline}
\usepackage[british]{babel}
\usepackage{cleveref}
\usepackage{hyperref}
\usepackage[acronym]{glossaries}
\usepackage{graphicx,booktabs,latexsym,bm}
\usepackage{epsfig}
\usepackage{euscript}
\usepackage{pslatex}
\usepackage{fancybox}
\usepackage{enumerate}
\usepackage{widetext}
\usepackage{wasysym}
\usepackage[figuresright]{rotating}
\usepackage{bbold}
\usepackage{mathtools}
\usepackage[caption=false]{subfig}
\usepackage{slashed}

\newcommand{\Pe}{\ensuremath{\text{e}}\xspace}

\newcommand{\PW}{\ensuremath{\text{W}}\xspace}

\newcommand{\ord}{\mathcal{O}}

\newcommand{\eff}{\mathrm{eff}}
\newcommand{\MW}{M_\mathrm{W}}

\newcommand{\GamW}{\Gamma_\mathrm{W}}

\newacronym{mc}{MC}{Monte Carlo}

\newcommand{\done}{{\mathrm d}}

 \newcommand{\dd}[1]{\mathrm{d}{#1}}

\newcommand{\oforder}[1]{\ensuremath{\mathcal{O}\left(#1 \right)}\xspace}

\newcommand{\bea}{\begin{eqnarray}}
\newcommand{\eea}{\end{eqnarray}}
\newcommand{\beq}{\begin{equation}}
\newcommand{\eeq}{\end{equation}}

\newcommand{\bs}{\begin{split}}
\newcommand{\es}{\end{split}}

\newcommand{\bi}{\begin{itemize}}
\newcommand{\ei}{\end{itemize}}
\newcommand{\bc}{\begin{center}}
\newcommand{\ec}{\end{center}}
\newcommand{\bac}{\begin{array}{c}}
\newcommand{\bacc}{\begin{array}{cc}}
\newcommand{\ea}{\end{array}}

\def\spa#1.#2{\langle#1\,#2\rangle}
\def\spb#1.#2{[#1\,#2]}

\newcommand{\ww}{\ensuremath{W^+W^-}\xspace}

\newcommand{\epluseminus}{\ensuremath{e^+ e^-}\xspace}
\newcommand{\ee}{\epluseminus}

\DeclareRobustCommand{\PW}{{\ensuremath{\mathrm{W}}}}

\newcommand{\NNLOQED}{\ensuremath{\mathrm{NNLO_{QED}}}\xspace}
\newcommand{\NLOEW}{\ensuremath{\mathrm{NLO_{EW}}}\xspace}

\newcommand{\sla}[1]{\ensuremath{{#1\kern-0.45em/}}}

\newcommand{\eik}[1]{\ensuremath{\tilde{S}\left(k_{#1}\right)}}
\newcommand{\eikK}{\ensuremath{\tilde{S}\left(k\right)}}

\newcommand{\OpenLoops}{O\protect\scalebox{0.8}{PEN}L\protect\scalebox{0.8}{OOPS}\xspace}
\newcommand{\Recola}{R\protect\scalebox{0.8}{ECOLA}}

\newcommand{\Sherpa}{S\protect\scalebox{0.8}{HERPA}\xspace}
\newcommand{\Comix}{C\protect\scalebox{0.8}{OMIX}\xspace}

\newcommand{\Amegic}{A\protect\scalebox{0.8}{MEGIC}\xspace}
\newcommand{\CSS}{CSS\protect\scalebox{0.8}{HOWER}\xspace}
\newcommand{\Dire}{D\protect\scalebox{0.8}{IRE}\xspace}

\newcommand{\Photons}{P\protect\scalebox{0.8}{HOTONS}\xspace}
\newcommand{\LHAPDF}{L\protect\scalebox{0.8}{HAPDF}\xspace}

\newcommand\aem{\alpha}

\newcommand\aNLO{{\sc\small MadGraph5\_aMC@NLO}}

\newcommand\MSb{\overline{\rm MS}}

\newcommand\ePDF{{\sc\small ePDF}}

\newcommand{\babayagaNLO}{\texttt{\textbf{BabaYaga@NLO}}}

\newcommand{\babayaga}{\texttt{\textbf{BabaYaga}}}

\newcommand{\Meu}{\EuScript{M}}

\newcommand{\KK}{${\cal KK}$}

\def\st{\hbox{}} 

\def\myat{@}

\def\D{\mathrm{d}}

\newcommand{\M}[2]{\mathcal{M}_{#1}^{(#2)}}
\newcommand{\fM}[2]{\mathcal{M}_{#1}^{(#2)f}}
\newcommand\ieik{\hat{\mathcal{E}}}
\def\xc{\xi_{c}}
\newcommand{\cdis}[2][c]{\left(\frac{1}{#2}\right)_{\hspace*{-3pt}#1}}

\newcommand\Eqn[1]     {Eq.\,(\ref{#1})}

\newcommand\eqn[1]     {eq.\,(\ref{#1})}

\newcommand{\NLP}{\mathrm{NLP}}

\newcommand{\smallz}{{\scriptscriptstyle Z}}
\newcommand{\mz}{m_\smallz}
\newcommand{\mzsq}{m_\smallz^2}

\title{\boldmath Initial state QED radiation aspects for future
  $e^+e^-$ colliders}

\author[1]{ \textmd{Conveners:\;} S. Frixione,} %
\author[2,3]{E. Laenen} %
\author{{}\\}

\author[4]{C.M. Carloni Calame,}%
\author[5]{A. Denner,}%
\author[6]{S. Dittmaier,}%
\author[7,8]{T. Engel,}%

\author[9]{L. Flower,}%

\author[10]{L. Gellersen,}%
\author[11]{S. Hoeche,}%
\author[12]{S. Jadach,}%
\author[13]{M.R. Masouminia,}%
\author[4,14]{G. Montagna,}%
\author[4]{O. Nicrosini,}%
\author[4]{F. Piccinini,}%
\author[15,16]{S. Pl\"atzer,}%
\author[17]{A. Price,}%
\author[18]{J. Reuter,}%
\author[7]{M. Rocco,}%
\author[9]{M. Sch\"onherr,}%
\author[7,8]{A. Signer,}%
\author[10]{T. Sj\"ostrand,}%
\author[8]{G. Stagnitto,}%
\author[9]{Y. Ulrich,}%
\author[20]{R. Verheyen,}%
\author[21,2]{L. Vernazza,}
\author[22]{A. Vicini,}
\author[23]{B.F.L. Ward,}
\author[24]{M. Zaro}

\affiliation[1]{INFN, Sezione di Genova, Via Dodecaneso 33, I-16146, Genoa, Italy}
\affiliation[2]{Nikhef, Science Park 105, 1098 XG Amsterdam,
  Netherlands}
\affiliation[3]{Institute of Physics, University of Amsterdam, Science Park 904, 1098 XH Amsterdam, Netherlands}
\affiliation[4]{INFN, Sezione di Pavia,\\ via A. Bassi 6, 27100 Pavia, Italy}
\affiliation[5]{Universit\"at W\"urzburg, %
        Institut f\"ur Theoretische Physik und Astrophysik,  %
         Emil-Hilb-Weg 22,
        97074 W\"urzburg, %
        Germany}%
\affiliation[6]{Albert-Ludwigs-Universit\"at Freiburg, %
        Physikalisches Institut,  %
        Hermann-Herder-Str. 3, %
        79104 Freiburg, %
        Germany}
\affiliation[7]{Paul Scherrer Institut,
CH-5232 Villigen PSI, Switzerland}
\affiliation[8]{Physik-Institut, Universit\"at Z\"urich, 
  Winterthurerstrasse 190, CH-8057 Z\"urich, Switzerland}
\affiliation[9]{Institute for Particle Physics Phenomenology, Durham University, Durham DH1 3LE, United Kingdom}

\affiliation[10]{Dept. of Astronomy and Theoretical Physics, Lund University,
  S\"olvegatan 14A, SE-223 62 Lund, Sweden}
\affiliation[11]{Fermi National Accelerator Laboratory, Batavia, IL, 60510, USA}
\affiliation[12]{Institute of Nuclear Physics Polish Academy of
  Sciences, Cracow, Poland}

\affiliation[14]{Dipartimento di Fisica, Universit\`a di Pavia,\\ via
  A. Bassi 6, 27100 Pavia, Italy}
\affiliation[15]{Institute of Physics, NAWI Graz, University of Graz, Universit\"atsplatz 5, A-8010 Graz, Austria}
\affiliation[16]{Particle Physics, Faculty of Physics, University of
  Vienna, Boltzmanngasse 5, A-1090 Wien, Austria}
\affiliation[17]{Theoretische Physik 1,
  Naturwissenschaftlich-Technische Fakult\"{a}t, Universit\"{a}t
  Siegen, Walter-Flex-Strasse 3, D-57068 Siegen, Germany}
\affiliation[18]{Deutsches Elektronen-Synchrotron DESY, Notkestr. 85,
  22607 Hamburg, Germany}
\affiliation[20]{Dept. of Physics and Astronomy, UCL, Gower St, Bloomsbury,
London WC1E 6BT, United Kingdom}
\affiliation[21]{INFN, Sezione di Torino, Via P. Giuria 1, I-10125 Torino, Italy}
\affiliation[22]{Dipartimento di Fisica ``Aldo Pontremoli'',
  University of Milano and INFN, Sezione di Milano, I-20133 Milano, Italy}
\affiliation[23]{Baylor University, Waco, TX, USA}
\affiliation[24]{TIFLab, Dipartimento di Fisica, Universit\`a degli
Studi di Milano and INFN, Sezione di Milano,\\ Via Celoria 16, 
20133 Milano, Italy}

\abstract{This white paper concerns theoretical and phenomenological aspects relevant 
to the physics of future $e^+e^-$ colliders, in particular regarding
initial-state QED radiation. The contributions each contain key technical 
aspects, and are formulated in a pedagogical manner so as to render them 
accessible also to those who are not directly working on these and 
immediately-related topics. This should help both experts and non-experts 
understand the theoretical challenges that we shall face at future $e^+e^-$ 
colliders. Specifically, this paper contains descriptions of the treatment 
of initial state radiation from several Monte Carlo collaborations, as well 
as contributions that explain a number of more theoretical developments with
promise of future phenomenological impact.}

\begin{document}
\maketitle

\section{Introduction\label{sec:introee}}
While the recent, current, and near-future activity of the high-energy
particle physics community has been and will be focused on the experimental 
and theoretical studies of hadronic collisions, owing to the operation
of the LHC at CERN, in the longer term there is some consensus towards
building an $e^+e^-$ machine. In view of the fact that its basic
parameters (such as whether it will be linear or circular, and its
running energy/energies) are far from being agreed upon, it is important
at this stage to start reviewing the theoretical tools available for
$e^+e^-$-collision simulations. More often than not, these have been
developed during the LEP era, and have not undergone those major upgrades
which will almost certainly be necessary at a future collider. On the
other hand, enormous technical progress has been made in the past twenty
years, thanks to the challenges posed by the LHC; we expect that a 
non-negligible part of that can be ported, with a relatively minor
effort, to an $e^+e^-$ environment.

This white paper, which is part of the Snowmass 2021 proceedings,
presents a review of the computer programs that have been used in the past
at LEP, and/or are being used presently at the LHC, as well as some
more theoretical contributions which may inform future phenomenological
developments. The status of the codes discussed below is heterogeneous,
in that their readiness to tackle immediately complex $e^+e^-$ simulations
varies greatly. We thus encourage the reader to see this paper as giving
the descriptions of the various starting points towards the possible 
achievement of the tools necessary for future sophisticated analyses.
At the same times, the reader is urged to not take the successes of the
LEP era for granted: it will be essential to thoroughly vet the capabilities, 
physics scope, and accuracy of each code in an unbiased manner, regardless
of whether it has been used frequently, sparingly, or not at all at LEP
(or at even smaller energies).

This paper is organised as follows. Sects.~\ref{sec:PY}--\ref{sec:she}
review the status of general-purpose parton shower Monte Carlos (PSMCs), 
with an emphasis on their features relevant to $e^+e^-$ simulations.
Sects.~\ref{sec:whiz}--\ref{sec:mcm} present a variety of tools
whose scope is either generally narrower than that of the PSMCs 
discussed before, or that specialise on matrix-element computations
(that can possibly be interfaced to PSMCs). Finally, sects.~\ref{sec:coll}
and~\ref{sec:pw} discuss, from a more theoretical viewpoint, various
issues that are ubiquitous at $e^+e^-$ colliders, and are related to the
treatment of collinear/soft radiation, which is in turn essential
for a phenomenological-sensible description of the production process.

We finally point out that, in the context of the Snowmass 2021 proceedings,
simulation codes relevant to all possible types of high-energy interactions
are reviewed in ref.~\cite{Campbell:2022qmc}.

\section{The PYTHIA QED showers\label{sec:PY}}
\begin{center}
  \emph{ Leif Gellersen, Torbj\"orn Sj\"ostrand, Rob Verheyen}
\end{center}







\subsection{Introduction}

\textsc{Pythia} \cite{Sjostrand:2006za,Sjostrand:2014zea} has three
parton-shower options, the older simple one that still remains default,
and the newer Vincia and \textsc{Dire} ones. Each of these include QCD and QED
evolution, and also a varying selection of further branching kinds.
Here we only summarize the key QED aspects. 

The three parton shower options in \textsc{Pythia} span a wide variety of 
ideas and models, from a purely positive rate dipole model (``simple shower"), 
an indefinite-sign dipole model (\textsc{Dire}), to a multipole radiation 
approach (Vincia). The models further employ different evolution variables, 
splitting functions, and phase space parametrizations. As an example, 
$\alpha_{\mathrm{EM}}$ can be chosen either fixed at vanishing momentum
transfer or running at the shower evolution scale in the simple and Vincia
showers, with the latter default, while \textsc{Dire} only admits the former
option. These different choices make \textsc{Pythia} a valuable tool to
assess QED modelling uncertainties.

None of the three showers has an internal implementation of beamstrahlung,
and currently there is no way to input beamstrahlung consistently separated
from bremsstrahlung.

\subsection{The simple shower QED handling}

In $e^+ e^-$ annihilation events the simple shower handling of QED
initial-state radiation (ISR) and final-state radiation (FSR) is decoupled,
i.e.\ interference effects are neglected. For on-shell $Z^0$, $W^{\pm}$
and $H^0$ production this is a good approximation for photon energies
above the respective resonance width. Emissions are ordered in terms of
a decreasing $p_{\perp\mathrm{evol}}^2$ scale, i.e.\ by backwards evolution
for ISR \cite{Sjostrand:1985xi}. Here
\begin{equation}
p_{\perp\mathrm{evol}}^2 =
\begin{cases}
z (1-z) Q^2 & \text{for FSR}~, \\
(1-z) Q^2 & \text{for ISR}
\end{cases}
\end{equation}
\cite{Sjostrand:2004ef}, where $Q^2$ is the time- or spacelike virtuality
associated with a branching, and $z$ describes the energy sharing in it,
by standard DGLAP splitting kernels, like
\begin{equation}
\mathrm{d}\mathcal{P}_{f \to f\gamma} =
e_f^2 \, \frac{\alpha_{\mathrm{EM}}}{2\pi} \,
\frac{\mathrm{d}p_{\perp\mathrm{evol}}^2}{p_{\perp\mathrm{evol}}^2} \,
\frac{1 + z^2}{1-z} \mathrm{d}z ~,
\end{equation}
where $e_f$ is the electrical charge of the fermion $f$.

FSR is handled in a dipole picture. In it, the radiation of a
$f\overline{f}'$ pair is split as a sum of radiation from the two ends.
Specifically, consider the branching $a \to b c$ with a recoiler $r$
at the other end of the $(a,r)$ dipole, and $c$ here representing the radiated
photon. The $p_{\perp\mathrm{evol}}^2$ and $z$ values of the branching
defines the virtuality $m_a^2 = Q^2 = p_{\perp\mathrm{evol}}^2 / z(1-z)$,
which leads to a redistribution of $E_a$ and $E_r$. Thereafter
$E_b = z E_a$ and $E_c = (1-z) E_a$. This procedure reproduces the
singularity structure of relevant matrix elements. The
sum of radiation from the two dipole ends gives
a numerator that exceeds the matrix-elements one, so the veto algorithm
\cite{Sjostrand:2006za} can be used to correct the emission rate to
agree with matrix elements for the first emission \cite{Norrbin:2000uu},
modulo Sudakov factors. Fermion mass effects are included in the shower
to still match singularities and matrix elements for $Z^0$ and $H^0$
decays, whereas the $W^{\pm} \to f \overline{f}' \gamma$ matrix element
is only coded in the massless limit.

A dipole picture is also used to allow $\gamma \to f \overline{f}$
branchings, where the mother parton ($b$ in $a \to b c$) acts as a
recoiler. This opens up for the creation of several charge pairs in a
resonance decay. There is no multipole handling, but instead each new
fermion pair is handled as a dipole fully separated from other dipoles
in the event.

For the backwards evolution of ISR, ratios of an already existing PDF 
need to be used. The starting point is an expression resummed to LL
and matched to an exact ${\cal O}(\alpha^2)$ result
\cite{Kleiss:1989de, Nicrosini:1986sm}, too lengthy to be reproduced here, 
whose leading $x\rightarrow 1$ behaviour reads
\begin{equation}
f_e^e(x,Q^2) \approx \frac{\beta}{2} (1-x)^{\beta/2-1} ~; ~~~~
\beta = \frac{2 \alpha_{\mathrm{EM}}}{\pi} \left( \ln \frac{Q^2}{m_e^2} - 1
\right) ~.
\end{equation}
The form is divergent but integrable for $x \to 1$, which gives numerical 
problems in this limit. To this end the PDF is set to zero for
$x > 1-10^{-10}$, and is rescaled upwards in the range
$1-10^{-7} < x < 1-10^{-10}$, in such a way that the total area under the
parton distribution is preserved:
\begin{equation}
f_{e,\mathrm{mod}}^e(x,Q^2) =
\begin{cases}
 f_e^e(x,Q^2) & 0 \leq x \leq 1-10^{-7} \\
 \frac{\displaystyle 1000^{\beta/2}}{\displaystyle 1000^{\beta/2}-1}
\, f_e^e(x,Q^2) &  1-10^{-7} < x < 1-10^{-10}  \\
0 & x > 1-10^{-10} \, ~.
\end{cases} 
\end{equation}
The fact that the backwards evolution does not end up with an electron
at $x \equiv 1$ is compensated by inserting an extra photon along the
beam axis with the remaining energy.

The $z$ variable of the splitting kernel in ISR is given by the mass-square
ratio of the radiator+recoiler system before to after the backwards step.
This again guarantees the same soft- and collinear-photon singularities
as in matrix elements. For the $e^+e^- \to \gamma^*/Z^0$ process this
can be used to perform a matrix-element correction for the first/hardest
emission \cite{Miu:1998ju}.  

\subsection{Vincia's QED shower}

Vincia is a $p_{\perp}$-ordered parton shower based on the antenna formalism. 
The fundamental branching step is $2 \to 3$, where the two pre-branching momenta are treated on equal footing.
The soft eikonal contribution to the radiation pattern remains unpartitioned, and is fully contained in Vincia's branching kernels, which are referred to as antenna functions. 
By contrast, the more common dipole showers partition the antenna into two separate dipole functions, in which one of the pre-branching momenta is assigned to be the splitter and acquires all transverse recoil.

For a generic branching $IK \to ijk$, where $j$ is the radiated momentum, the ordering scale is given by 
\begin{equation} \label{eq:vincia-ordering-scale}
    p_{\perp}^2 = \frac{\bar{q}_{ij}^2 \bar{q}_{jk}^2}{s_{\text{max}}}, \text{ with } \bar{q}_{ij} = 
    \begin{cases}
        s_{ij} + m_i^2 + m_j^2 - m_I^2 & i \text{ is final-state}\\
        s_{ij} - m_i^2 - m_j^2 + m_I^2 & i \text{ is initial-state}
    \end{cases},
\end{equation}
where $s_{ij} = 2 p_i{\cdot}p_j$ and $s_{\text{max}}$ is the largest antenna invariant mass, which depends on which momenta are initial-state or final-state. 
The other major components of the shower algorithm are the phase space factorization, as well as the associated kinematic mapping, and the exact form of the antenna functions. 
These, and many other details may be found in \cite{Brooks:2020upa}.

Vincia's QED shower has been developed in \cite{Kleiss:2017iir,Skands:2020lkd}, and is described there in detail.
An important component is given by the photon emission antenna function, which, for final-state radiators, is given by
\begin{equation} \label{vincia-qed-antenna}
    A_{\gamma/IK}(s_{ij}, s_{jk}, m_i^2, m_j^2, s_{IK}) = g^2 Q_I Q_K \bar{A}_{\gamma/IK}(s_{ij}, s_{jk}, m_i^2, m_j^2, s_{IK}),
\end{equation}
where 
\begin{align}
    \bar{A}&_{\gamma/IK}(s_{ij}, s_{jk}, m_i^2, m_j^2, s_{IK}) = -2 g^2 Q_I Q_K  \nonumber \\
    \Bigg[ & 2 \frac{s_{ik}}{s_{ij} s_{jk}} - 2 \frac{m_i^2}{s_{ij}^2} - 2 \frac{m_k^2}{s_{jk}^2} + \frac{1}{s_{IK}} \left( \delta_{If} \frac{s_{ij}}{s_{jk}} + \delta_{Kf} \frac{s_{jk}}{s_{ij}} \right) + \nonumber \\
    & \delta_{IW} \frac{4}{3} \frac{s_{ij}}{s_{IK}^2} \left( \frac{s_{jk}}{s_{IK} - s_{jk}} + \frac{s_{jk}(s_{IK} - s_{jk})}{s_{IK}^2} \right) + \delta_{KW} \frac{4}{3} \frac{s_{jk}}{s_{IK}^2} \left( \frac{s_{ij}}{s_{IK} - s_{ij}} + \frac{s_{ij}(s_{IK} - s_{ij})}{s_{IK}^2} \right)\Bigg]. \nonumber \\
\end{align}
Here, $Q_I$ and $Q_K$ are the electromagnetic charges of $I$ and $K$, and the Kronecker deltas ensure the correct collinear terms are incorporated in the case where $I$ and $K$ are either fermions or $W$-bosons.
The corresponding initial-state antenna may be found by employing crossing symmetry. 

A straightforward shower implementation in the same spirit of the QCD shower using eq.~\eqref{vincia-qed-antenna} directly suffers from the issue that the product $-Q_I Q_K$ is negative for same-sign antennae, making a probabilistic implementation troublesome. 
Rather than resorting to negatively weighted events, Vincia's most sophisticated photon radiation algorithm solves this issue by defining a single branching kernel 
\begin{equation} \label{eq:vincia-coherent-antenna}
    A_{\gamma/\text{coh}} = \sum_{\{IK\}} \sigma_I Q_I \sigma_K Q_K \bar{A}_{\gamma/IK}(s_{ij}, s_{jk}, m_i^2, m_j^2, s_{IK}),
\end{equation}
where $\{IK\}$ runs over all pairs of charged particles, and $\sigma_I$ and $\sigma_K$ are $1\,(-1)$ for final-state (initial-state) momenta.
This branching kernel is positive, and contains the complete set of soft multipole eikonal factors, as well as all collinear limits \cite{Kleiss:2017iir}.
Its singular structure is however not straightforwardly regularized by the ordering scale of eq.~\eqref{eq:vincia-ordering-scale}.
The QED shower implements eq.~\eqref{eq:vincia-coherent-antenna} by sectorizing the phase space according to 
\begin{equation}
    A_{\gamma/\text{coh}} \to \sum_{\{ik\}} \Theta^{\text{sct}}_{ik} \, \bar{A}_{\gamma/\text{coh}}, \text{ where } \Theta^{\text{sct}}_{ik} = \theta\left( \min_{\{xy\}} p^2_{\perp, xy} - p^2_{\perp, ik} \right).
\end{equation}
That is, the radiative phase space is divided into regions such that the photon is always radiated by the two charged particles that have the smallest $p_{\perp}$ with respect to the photon.
This procedure is equivalent to ordering the shower evolution in terms of $\min_{\{ik\}} p^2_{\perp,ik}$, which does indeed regulate all singularities of eq.~\eqref{eq:vincia-coherent-antenna}.
This implementation thus accomplishes a modelling of fully coherent photon radiation without resorting to negative weights, and it is thus the default choice.

However, the above algorithm requires sampling emissions in all sectors, the number of which scales like $\mathcal{O}(n^2_{\text{chg}})$. 
An alternative, faster and unsectorized approach is also available which instead approximates the radiation pattern as 
\begin{equation} \label{eq:vincia-pairing-antenna}
    A_{\gamma/\text{pair}} = \sum_{[IK]} Q_{[IK]}^2 A_{\gamma/IK}(s_{ij}, s_{jk}, m_i^2, m_j^2, s_{IK}),
\end{equation}
where $[IK]$ runs over all pairings of charged particles with opposite charges, including the sign factors $\sigma_I$ and $\sigma_K$. 
All contributions in eq.~\eqref{eq:vincia-pairing-antenna} are positive and can thus be sampled individually, similar to leading-$N_C$ QCD.
Charges are paired up by minimizing the sum of antenna invariant masses following the principle of maximum screening \cite{Skands:2020lkd}, 
which is technically accomplished through the Hungarian algorithm \cite{Kuhn55thehungarian,Munkres1957Assignment,10.1007/BF02278710}. 
While this algorithm is generally faster, it only approximates the soft multipole photon emission structure. 

The QED shower also includes photon splittings to charged fermions, both in the initial state and the final state. 
Photon radiation off quarks and photon splitting to quarks is cut off at a scale close to $\Lambda_{\text{QCD}}$, but radiation off and splitting to leptons is continued to a much lower scale. 
Radiation in resonance systems is also included following the procedures set out in \cite{Brooks:2019xso}.
However, QED radiation off charged hadrons and in hadron decays is not currently implemented.

\subsection{The \textsc{Dire} QED shower}

The \textsc{Dire} parton shower for \textsc{Pythia} was introduced in \cite{Hoche:2015sya} and combines the soft-emission treatment of $2\rightarrow 3$ dipole/antenna showers with the identification of collinear emissions familiar from traditional $1\rightarrow 2$ showers. The shower evolution variable $t$ corresponds to a scaled soft transverse momentum and is chosen to be symmetric between the identified radiating parton $i$ and the parton taking the recoil $k$ in a branching $(i1),k\rightarrow i,1,k$,
\begin{equation}
  t \propto \frac{(p_i p_1)(p_1 p_k)}{Q^2}\, .
\end{equation}
Here, $Q^2$ denotes a dipole-type-dependent maximum scale, and $p_i$, $p_1$ and $p_k$ represent the momenta of the involved partons after the branching.

The radiating dipoles in \textsc{Dire} allow for radiation from any combination of initial and final state partons, i.e. final state radiators with final state recoilers (FF), final state radiators with initial state recoilers (FI), initial state radiators with final state recoilers (IF) and initial state radiators with initial state recoilers (II). The branching rates are constructed from a partial fractioning of the soft radiation pattern augmented with collinear terms in both the massless and massive case following the ideas presented in \cite{Catani:1996vz} and \cite{Dittmaier:1999mb,Catani:2002hc}.

Details about the ordering, kinematic mappings for the respective dipole types and the branching rates may be found in \cite{Hoche:2015sya}. \textsc{Dire} allows for the use of negative splitting functions as described in \cite{Hoeche:2011fd,Hoche:2015sya}, and is thus not restricted to the formulation of branchings in terms of probabilities bounded by zero and one.

The treatment of QED radiation in \textsc{Dire} is described in \cite{Prestel:2019neg,Gellersen:2021caw} and is in parts inspired by \cite{Dittmaier:1999mb,Schonherr:2017qcj}. The QED branchings are similar to the QCD treatment, with the coupling factor replaced by the weak coupling, and colour charge factors replaced by QED charge correlators
\begin{align}
  q^2 = \begin{cases}-\frac{\eta_{i1}\eta_{k}q_{i1}q_{k}}{q^2_{i1}} & \text{for} \; f\rightarrow f \gamma \\
  \frac{1}{\mathrm{\#recoilers}} & \text{for} \; \gamma \rightarrow f \bar f\end{cases}\, ,
\end{align}
where $q_{i1}$ and $q_k$ denote the electric charges of the corresponding partons and $\eta$ is $+1$ for final state and $-1$ for initial state partons. As opposed to the leading color QCD case, any combination of charges might radiate, leading to a larger number of allowed dipoles. Depending on the sign of the correlator, these might enter with a negative sign, mandating the utilization of the weighted shower algorithm to deal with negative contributions. Following the arguments of \cite{Kniehl:1992ra}, the QED coupling is kept fixed at its value in the Thompson limit. This should be appropriate for on-shell final-state photons, but might lead to a suboptimal description of off-shell intermediate photons produced by $\gamma\rightarrow f \bar f$ branchings.

In \textsc{Dire}, the description of one or more QED emissions is fully integrated into the shower's merging machinery. In particular, this also means that QED radiation patterns can be improved by employing iterative matrix element corrections. Writing the shower splitting kernels as $P \equiv P(\Phi_{n+1}/\Phi_{n})$, where $\Phi$ denotes the phase space point for $n$ or $n+1$ partons before and after a branching, the shower generates a $\Phi_{n+1}$ configuration with the rate
\begin{equation}
  \sum_{\Phi_n,P}\frac{8\pi\alpha P(\Phi_{n+1}/\Phi_n)}{Q^2(\Phi_n)}\left|\mathcal M(\Phi_n)\right|^2\, .
\end{equation}
This expression may contain both QCD- and QED-like emissions encoded in the splitting kernels, and the coupling $\alpha$ may represent the strong or electroweak coupling depending on the corresponding kernel. For a given state, the reconstruction of all possible shower emission histories leading to this state allows for calculation of a corrective factor
\begin{equation}
  \mathcal{R}(\Phi_{n+1}) = \frac{\left|\mathcal{M}(\Phi_{n+1})\right|^2}{\sum_{\Phi'_n}\frac{8\pi\alpha P(\Phi_{n+1}/\Phi'_n)}{Q^2(\Phi'_n)}\left| \mathcal M(\Phi'_n)\right|^2}\, ,
\end{equation}
which is applied again using the weighted shower algorithm. For iterated branchings, the corresponding denominator is constructed by taking multiple branchings and their respective corrections into account. In this way, multiple emissions can be matrix-element corrected. This approach even allows for mixed QCD and QED shower histories, and by employing suitable matrix elements, also interference effects between QCD and QED. For further details, see \cite{Gellersen:2021caw}. This approach in principle also allows for the incorporation of QED interference effects between initial and final state radiation, which we have not yet studied in detail.

The treatment of initial-state lepton PDFs is similar to the simple shower model above, i.e.\ the lepton PDF is split into the same three regions, to avoid evaluating splittings close to integrable singularities. However, in the case of a leptonic initial state recoiler, even a final state emission might change the ``parton" flux of the collision event. Due to the steep rise of the lepton PDF $f_e^e(x,Q^2)$ for $x\rightarrow 1$, the corresponding PDF ratios might become (much) larger than unity. This effect is not of immediate concern for weighted shower algorithm, which would still yield correct emission rates without the need to limit the value of the PDF ratio. However, if the overestimate does not attempt to take this enhancement into account, the resulting sample might contain a few events with very high weights. Thus, the shower overestimates for splittings with initial-state lepton recoilers have been judiciously extended with additional terms that, for small $1-x$, allow to improve the statistical stability of the simulation by employing multi-channel sampling.

Finally, it is worth noting that QED emissions can be enhanced with a method that combines the ``boost" of \cite{Lonnblad:2012hz} with the weighted shower of \cite{Hoeche:2011fd}. Given that negative contributions to the emission rate -- which lead to event weights -- are not coupling-suppressed relative to positive contributions, very large ``boosts" $C$ with $C\alpha_\mathrm{EM}\gg \alpha_s$ can however lead to a poor statistical convergence.

\acknowledgments

L. Gellersen and T. Sjöstrand are supported by the
Swedish Research Council, contract number 2016-05996.
 Rob Verheyen is supported by the European Research Council (ERC) under the European Union’s Horizon 2020 research and innovation programme (grant agreement No. 788223, PanScales).



\section{QED radiation in Herwig\label{sec:HW}}
\begin{center}
\emph{Mohammad Masouminia, Simon Pl\"atzer, Peter Richardson}
\end{center}







QED radiation in Herwig is available in the angular ordered shower,
and we are also developing it for the dipole shower. In order to
understand why QED is not a simple 'downscaling' of the QCD evolution,
and how coherent branching algorithms ease the inclusion in existing
algorithms, we will briefly review generic features of QCD coherence
to introduce how QED radiation is actually handled.

QCD coherence has been a design criterion behind parton branching
algorithms since their early application to resummation of event shape
variables \cite{Catani:1992ua} and the first accurate parton shower
algorithms \cite{Marchesini:1987cf}. In particular, soft gluon
emission from collinear bundles of partons effectively appear as
emissions from the net colour charges carried by the collinear
bundle. Observables which globally measure the deviation from a
certain jet system, without any particular hierarchy in between the
structure of the jets, can thus be reliably described by ordering
emissions in decreasing opening angle. The modern version of this
evolution, improving on mass effects and a covariant description of
the angular ordering variable is at heart of the Herwig angular
ordered shower \cite{Gieseke:2003rz,Bahr:2008pv}. 

The non-abelian nature of QCD implies a change of the colour charge
distribution, which means that each emission restricts the opening
angle available to subsequent emissions. Angular ordering in this case
is also inherent in each dipole of colour charges, which, subject to
azimuthal averaging will then give rise to the respective destructive
interference outside of the opening cones of each dipole.  Dipole-type
parton showers, which locally account for this change in the radiation
pattern, are usually formulated in the limit of infinitely many colour
charges, in which a probabilistic sequence of emissions from a dipole
emerges in terms of a transition of one dipole into two colour charge
dipoles. Coherent showers, based on angular ordering, do not require
the large-$N$ limit and are able to predict full-colour accurate
contributions to event shapes deviating from the two-jet limit.

The conservation of colour charge in emission of spin one
particles,
\begin{equation}
  \sum_{i=1}^n {\mathbf T}_i | {\cal M}(p_1,...,p_n\rangle = 0 \ ,
\end{equation}
where ${\mathbf T}_i$ are colour-charge operators and ${\cal M}$ is a
gauge-invariant hard process amplitude, is crucial to simplify colour
correlations inherent to large-angle soft radiation and to translate
their contribution into the soft enhanced contributions of the
(colour-diagonal) QCD splitting functions. This employs the fact that
collinear radiation only refers to one leg and as such we can simplify
colour correlations using
\begin{equation}
  -\sum_{j\ne i} {\mathbf T}_i\cdot {\mathbf T}_j|{\cal M}\rangle = {\mathbf T}_i^2 |{\cal M}\rangle
\end{equation}
where the Casimir operator of leg $i$ is colour-diagonal.  This
relation enables one to define dipole shower algorithms including
hard-collinear contributions \cite{Forshaw:2019ver}, and is at the
heart of using coherence to properly describe the large-angle soft
radiation pattern which then subsequently is completed by collinear
radiation in an angular ordered shower. Spin correlations can also be
accounted for in such a framework \cite{Richardson:2018pvo}.

QED radiation is, due the spin one nature of the photon, and the
conservation of electric charge, in many facettes very similar to
QCD. Several simplifications and complications are present at the same
time: Due to the abelian nature of the underlying gauge group, the
charges $Q_i$ are simple numbers, not operators. To this extent, no
complicated tracking of flow of charge is needed. At the same point,
however, nothing like a large-$N$ limit exists in QED and while a
dipole-type shower (with no branching dipoles, but the same dipole
emitting multiple times) is possible, the corresponding charge
correlators $-Q_i\cdot Q_j$ can become negative for same-sign charges
(in QCD, $-{\mathbf T}_i\cdot {\mathbf T}_j$ is always positive
semi-definite in the large-$N$ limit). While we can account for these
negative contributions through newly developing weighted shower
algorithms \cite{Olsson:2019wvr}, with an implementation of QED
radiation in the Herwig dipole shower being in progress, coherence can
also assist in the QED case: A separation into several collinearly
enhanced directions $i$ allows to encode the electric charge
correlations to be translated into positive-definite contributions
according to the charge squared $Q_i^2$, and large-angle photon
radiation, which is resolved independently as deviating from a jetty
system with only small-angle QED radiation, can as well be encoded
through an initial condition which identifies angular ordered cones
from the electric charge partners in the hard process.

Generally speaking, finding of physically-allowed QED partners only
depends on the initial evolution scale of the shower progenitor and
the kinematical relationship between the pair, without dependence on
the dynamics.  The additional complication consists in the need for a
separate evolution variable describing angular ordering with respect
tot QED dipoles, which needs to be interleaved with the QCD angular
ordering variable. Extensions of this paradigm are currently explored
in the Herwig angular ordered shower for other interactions, while we
are also making progress in formulating an according dipole-type
picture.

Specifically for $e^+e^-$ collisions we still handle initial state
radiation with an approach that employs a Weizs\"acker-Williams 
function~\cite{vonWeizsacker:1934nji,Williams:1934ad,Budnev:1975poe}: 
we randomly sample it twice (once per incoming leg), in order to obtain 
the fraction of the incoming-beam energy lost to initial-state radiation, 
and thus the energy available to the hard scattering. However, given 
the structure of the angular ordered shower, backward evolution will 
be readily available once a full QED PDF is available through the LHAPDF 
library. The value of the QED coupling constant is set equal to the
Thomson value, unless the production or the branching mechanism(s)
involve heavy weakly-interacting bosons, in which case $\alpha(m_Z)$
is used instead. The effects of beam dynamics, and in particular of 
beamstrahlung, are ignored.



\section{Lepton collisions and QED FSR in \Sherpa\label{sec:she}}
\begin{center}
\emph{L. Flower, S. Hoeche, A. Price, M. Sch\"onherr}
\end{center}

\subsection{Introduction}

The \Sherpa event generator currently offers two different approaches
to treating QED ISR. The first approach uses the so called electron
structure function\footnote{In \Sherpa this is not a true PDF, as it contains no information on the photon content of the electron, and it is defined in the LL limit.
The current state of the art is NLL accurate as described in~\cite{Bertone:2019hks,Frixione:2019lga}}, which is a solution of the DGLAP evolution
equations~\cite{Altarelli:1977zs,Gribov:1972ri,Lipatov:1974qm,Dokshitzer:1977sg}
using leading order (LO) initial conditions~\cite{Skrzypek:1990qs}.
This analytic approach can be combined with a traditional parton shower,
extended to QED~\cite{Hoeche:2009xc}, to generate exclusive kinematic distributions 
for the collinear photons. In the second approach, the emission of
soft photons is resummed to all orders using the
Yennie-Frautschi-Suura~\cite{Yennie:1961ad}(YFS)
formalism. In this method, the photon
emissions are considered in a fully differential
form where the photons are explicitly created
and the treatment of their phasespace
is exact.

\subsection{Collinear Resummation}
The structure function approach~\cite{Kuraev:1985hb} resums the
leading logarithmic (LL) corrections, using the DGLAP evolution equations,
into universal factors. It allows to write the total cross section 
for the process $e^+e^-\to X$ in the factorized form
\begin{equation}\label{EQ::PDFEXS}
   \done\sigma(s) = \int \done x_1 \done x_2\; 
                    f_{e^+}(x_1,Q^2)\,f_{e^-}(x_2,Q^2)\; 
                    \done\hat{\sigma}(x_1x_2s)\;,
\end{equation}
where $\done\hat{\sigma}$ is the partonic cross section.
The structure functions $f_{e^\pm}(x,Q^2)$ that are commonly used today 
are obtained from solutions of the evolution equations using LO 
initial conditions~\cite{Skrzypek:1990qs}. For the electron this
initial condition is a Dirac delta function in $x$, $\delta(1-x)$.
At LL accuracy, $f_{e^+}(x)=f_{e^-}(x)$ is given by
\begin{equation}
\label{EQ::structure}
  f_{e^\pm}(x,Q^2) = \beta\, \frac{\exp\left(-\gamma_E\beta
          +\frac{3}{4}\beta_S\right)}{\Gamma\!\left(1+\beta\right)}
           (1-x)^{\beta-1}\,
          +\beta_H \sum_{n=0}^\infty \beta_H^n\,\mathcal{H}_n(x)\;,
\end{equation}
with $\beta=\tfrac{\alpha}{\pi}\left(\ln(s/m_e^2)-1\right)$ and the 
hard coefficients $\mathcal{H}_n(x)$ given in 
\cite{Bardin:1993bh,Beenakker:1994vn,Montagna:1994qu,Berends:1994pv}. 
The QED coupling constant $\alpha$ is taken at scale $Q^2$. 
By default, \Sherpa includes the hard coefficients up 
to second order ($n=2$).
However, there is some freedom in how non-leading terms are taken into account.
The choice reflects how the soft photon residue is treated, 
i.e.\ whether it is proportional to $\ln(s/m_e^2)$ or $\ln(s/m_e^2)-1$.
The corresponding options for the corresponding parameters $\beta_S$ and 
$\beta_H$ are listed in Tab.~\ref{TAB::SCHEME_PDFE}.~\footnote{
There is no such freedom of choice within the YFS resummation.
The term $\beta=\frac{\alpha}{\pi}(\ln(s/m_e^2)-1)$ is a direct result
 from the analytical integration over the entire soft-photon
phasespace and is needed to achieve the cancellation of IR singularities.}
\begin{table}[t!]
\begin{center}

\setlength{\tabcolsep}{11pt}
\renewcommand{\arraystretch}{1.5}
 \begin{tabular}{c||c|c|c}
   Scheme & $\beta_S$ & $\beta_H$ & Refs
  \\
  \hline
  Beta  & $\beta$ & $\beta$ & \cite{Bardin:1993bh}  \\
  Eta & $\eta $ & $\eta$  & \cite{Beenakker:1994vn} \\
  Mixed   & $\beta$ & $\eta$  & \cite{Montagna:1994qu} \\
  \end{tabular}
  \caption{
    The different scheme choices available in \Sherpa for the electron 
    structure function, where $\beta = \frac{\alpha}{\pi}\left(\ln(s/m_e^2)-1\right)$
    and $\eta = \frac{\alpha}{\pi}\ln(s/m_e^2)$.
  }
  \label{TAB::SCHEME_PDFE}
\end{center}

\end{table}

\subsection{Parton Shower}
The structure function can be combined with a QED parton shower~\cite{Hoeche:2009xc}
to generate exclusive multi-photon final states.
\Sherpa includes two parton showers, \CSS~\cite{Schumann:2007mg} and \Dire~\cite{Hoche:2015sya},
but only the \CSS currently supports QED radiation \cite{Hoeche:2009xc}. \footnote{%
  An extension of the \Dire shower to include
  QED radiation is straightforward and will be pursued in the future.}
It is based on Catani--Seymour dipole factorisation~\cite{Catani:1996vz,Catani:2002hc}, 
with its extension to resummation
first proposed in~\cite{Nagy:2006kb}, and implemented simultaenously in~\cite{Schumann:2007mg}
and~\cite{Dinsdale:2007mf}. 
The algorithm builds on a set of generic operators for particle emission
off (color) charged dipoles in unintegrated and spin-averaged form. Soft eikonals are matched 
to collinear splitting functions in order to eliminate double counting of soft singularities.
The splittings are ordered by their associated transverse momenta. 
In the case of QCD, the large-$N_c$ limit is used to restrict the numbers of spectators to the 
leading ones with positive definite color-correlators. 
Contributions from neglected spectators can be shown to be suppressed 
by powers of $1/N_c^2$, contributing little to a standard observables.

\Sherpa extends this algorithm to photon emissions off charged fermions, 
making use of the fact that QCD and QED splittings only differ in their 
color- or charge-correlators. 
In the case of QED, to retain positive definiteness, only opposite 
charges are considered to form valid dipoles, discarding like-signed ones. 
However, due to the lack of a $1/N_c^2$-like suppression this degrades the 
quality of the QED parton shower approximation.
Efforts are currently underway to extend this algorithm to include also 
like-sign charged dipoles, restoring the exact soft-photon limit, 
making use of the weighted Sudakov veto algorithm~\cite{Hoeche:2009xc,Hoeche:2011fd,Platzer:2011dq,Lonnblad:2012hz}. 
In a concurrent development photon splittings into fermion-antifermion pairs 
will be included in the near future. 
While photon emissions off charged quarks and leptons use the QED coupling 
constant in the Thomson limit, $\alpha(0)$, photon splittings instead evaluate 
the coupling constant at the invariant mass of the produced fermion pair.

Within \Sherpa, radiation off final state leptons will normally be handled by the YFS 
simulation module \Photons. If the QED shower is to be used, it must therefore be enabled
explicitly by the user, and the YFS based resummation needs to be disabled to avoid double counting.

\subsection{Beam spectra}
In addition to providing the electron structure function,
\Sherpa allows for a two-step definition of particles entering a hard
interaction. The simulated beam particles can differ from the collider bunch 
particles, and their energy spectra can therefore be modified.
For $e^+e^-$ colliders we anticipate that an interface to the CIRCE~\cite{Ohl:1996fi}
package, which parametrizes the $e^\pm$, and $\gamma$ beam-spectra based on the collider geometry. 
Two prominent examples for beam spectra that change the incident 
particle species are
\begin{itemize}
\item Laser backscattering, where the initial lepton beam sources highly energetic 
  photons through Compton scattering~\cite{Badelek:2001xb,Zarnecki:2002qr,Archibald:2008zzb},
\item The equivalent photon approxmination (Weizs\"acker--Williams approximation),
  where the beam particles act as quasi-classical sources of collinear quasi-real 
  photons \cite{vonWeizsacker:1934sx,Williams:1934ad,Budnev:1974de}.
\end{itemize}
The thus generated photons can then either enter the hard interaction 
directly or be further resolved using, e.g., their respective structure functions 
\cite{Gluck:1991ee,Gluck:1991jc}, upon which they enter parton evolution 
through a parton shower.

\subsection{Soft Resummation}
The YFS formalism provides
a robust method for resumming the emission of real and virtual
photons in the soft limit to all orders. This resummation can be
further improved by including exact fixed-order expression in a systematic
way. In this chapter, we will first introduce the general YFS framework
and then will concentrate on its applications to \ee collisions.

The total cross section for the production of a configuration of particles in the 
final state, $\dd\Phi_Q$, including all higher-order corrections, is given by summing over 
all real and virtual photons, $n_\gamma$ and $n_\gamma^V$, wrt.\ its leading 
order configuration, 
\beq \label{yfs:xs}
    \dd\sigma =  \sum_{n_\gamma=0}^{\infty} \frac{1}{n_\gamma!}
    \dd{\Phi_Q}
    \left[\prod_{i=1}^{n_\gamma}\dd{\Phi_i^\gamma}\right]
    \left(2\pi\right)^4
    \delta^4\left(\sum_\text{in}q_\text{in}
                  -\sum_\text{out} q_\text{out}
                  -\sum_{i=1}^{n_\gamma} k_i\right)
    \left\lvert \sum_{n_\gamma^V=0}^{\infty}  \mathcal{M}^{n_\gamma^V+\frac{1}{2}n_\gamma}_{n_\gamma}\right\rvert^2\!\!.
\eeq
$\dd\Phi_i^\gamma$ is the phase space element of photon $i$ with momentum $k_i$. 
In the notation introduced here for the matrix element $\mathcal{M}$,
the Born level contribution will be defined as $\mathcal{M}_{0}^{0}$, while
the matrix element $\mathcal{M}_{n_\gamma}^{p}$ will refer to the Born
process plus $n_\gamma$ real photons evaluated at an overall power $p$
in the electromagnetic coupling $\alpha$. While this expression
includes photon emissions to all orders in principle, only the first
few terms in the perturbative series can be computed in practice. 
Following the YFS approach the total cross section can be reformulated 
in such a way that the infrared divergences are resummed to infinite 
order yielding,
\begin{align}\label{eq:masterYFS}
    \dd\sigma = \sum_{n_\gamma=0}^\infty
        \frac{e^{Y(\Omega)}}{n_\gamma!}\,
        \dd{\Phi_Q} 
        \left[\prod_{i=1}^{n_\gamma}\dd{\Phi_i^\gamma}\,\eik{i}\right]
        \left(\tilde{\beta}_0
        + \sum_{j=1}^{n_\gamma}\frac{\tilde{\beta}_{1}(k_j)}{\eik{j}}
        +\sum_{{j,k=1}\atop{j< k}}^{n_\gamma}
        \frac{\tilde{\beta}_{2}(k_j,k_k)}{\eik{j}\eik{k}}
        + \cdots
        \right).
\end{align}
Therein,
\begin{itemize}
    \item $Y(\Omega)$ is the YFS form factor which contains contributions from 
          real and virtual photons, inside the soft domain $\Omega$, summed to 
          infinite order. 
          It can be computed using a dipole decomposition .
    \item \eikK\ is the soft YFS eikonal associated with the emission of real 
          photons and can also be decomposed in dipoles, 
          $\eikK =  \sum_{i,j}\frac{\alpha}{4\pi^2}\,Z_iZ_j\theta_i\theta_j
           \left(\frac{p_i}{p_i\cdot k}-\frac{p_j}{p_j\cdot k}\right)^2$, 
          where $i$ and $j$ label all charged particles of the process, $Z_i$ 
          and $Z_j$ are their charges, and $\theta_{i/j}=+1(-1)$ if particle 
          $i/j$ is in the final (initial) state.
    \item The remaining terms inside the brackets are the
          infrared finite residuals $\tilde{\beta}_{n_\gamma}$, 
          comprising infrared subtracted squared matrix elements 
          with $n_\gamma$ real photons. 
          They can be systematically calculated order-by-order in perturbation 
          theory, $\tilde{\beta}_{n_\gamma}=\sum_{n_\gamma^V}\tilde{\beta}_{n_\gamma}^{n_\gamma^V+n_\gamma}$. 
          For example, up to NLO, we need to include the Born amplitude 
          $\tilde{\beta}_0^0$ as well as the infrared finite renormalised 
          one-loop corrections $\tilde{\beta}_0^1$ and the infrared finite 
          hard remainder of the tree-level expression including one 
          additional photon $\tilde{\beta}_1^1$.
          In \Sherpa, the ISR corrections for \ee initial states have 
          been implemented \cite{Krauss:2022ajk} using explicit differential distributions \cite{Jadach:2000ir} 
          and for FSR corrections to $Z\to\ell^+\ell^-$ decays explicit 
          matrix elements including up to NLO EW and NNLO QED 
          \cite{Krauss:2018djz} corrections are used.
    \item The QED coupling constants associated with the resummation 
          of soft-photon emissions is taken in the Thomson limit, $\alpha(0)$. 
          All other QED couplings present in the Born-level expression $\mathcal{M}_0^0$ 
          may be evaluated at a different suitable scale, typically defined at short-distances, 
          like $\alpha_{G_\mu}$, $\alpha(m_Z)$, etc.
\end{itemize}
In the \Sherpa-2 series this resummation was implemented for the
leptonic final states in the \Photons module~\cite{Schonherr:2008av}.
This was further extended to include \NLOEW and \NNLOQED
\cite{Krauss:2018djz} effects in the decay of electroweak bosons.
More recently, in the \Sherpa-3 series the YFS resummation has been extended to include
initial state radiation for lepton-lepton collisions~\cite{price2021yfs,Price:2021ds}.
One of the advantages of the YFS method is that perturbative corrections can be included
in a systematic way order-by-order. In its current implementation these perturbative corrections
are included up to \oforder{\alpha^3L^3} by using analytical expression derived in \cite{Jadach:2000ir}.
This level of accuracy, while sufficient for LEP, still falls short of the predicted accuracy needed
for future lepton colliders~\cite{Jadach:2019huc}, such that future refinements are needed. 
The second advantage of the YFS algorithm is the explicit treatment of the photon phasespace.
The photon momenta are distributed according to the eikonal factors~\cite{Jadach:1988gb},
which allows for explicit photons in the event record.

In addition to the initial and final state soft photon resummation, a special treatment 
for QED corrections in processes with W bosons has also been implemented.
The infrared factorization of soft-photons can be easily extended to include the emission from W bosons.
This can be seen by considering the amplitude for emission of soft photons from a spin-1 boson 
with charge $Q_W$ and mass $M_W$. Due to the spin independence of the soft current,
this amplitude has the same form as in the spin-$\frac{1}{2}$ case~\cite{Jadach:1995sp},
and the known expressions can be used by substituting $m_f\rightarrow m_W$ and 
$Q_f \rightarrow Q_W$ in the respective constituent dipoles.
Another important correction to \ww production is that of the electromagnetic 
Coulomb interaction at the threshold.
This exchange leads to large loop-corrections in the threshold region due to 
the low relative velocity which enhances the cross-section by a factor 
$\approx \alpha\pi/v$, where $v$ is the relative velocity. 
The inclusion of this loop-correction was calculated to first-order in 
\cite{Bardin:1993mc,Fadin:1993kg,Fadin:1995fp,Chapovsky:1999kv} and it 
was shown to, in the threshold region, produce a positive correction.

Currently, the dedicated higher-order corrections implemented in \Sherpa 
do not include any coherence effects from interfering initial and final 
state emissions. These will enter in the infrared-finite $\tilde{\beta}$
and will be included in future improvements. In particular, it is a goal of the \Sherpa
collaboration to automatically calculate the matrix elements corrections using
the ME generators \Amegic\cite{Krauss:2001iv} and \Comix\cite{Gleisberg:2008fv}.
As well as automating the real corrections internally, we plan to include one loop
EW corrections in an automated fashion using tools such as \OpenLoops\cite{Buccioni:2019sur,Cascioli:2011va}
or \Recola\cite{Actis:2016mpe}. These corrections corrections are particular 
importance in the light of the to-be-achieved
the experimental precision of future Higgs factories.

Finally, the YFS soft photon resummation only captures collinear 
effects through its infrared finite residuals. 
Thus, an extension to incorporate photon splittings to charged particle 
pairs in the final-state is currently under development.
These splittings are at least of relative \oforder{\alpha^2} compared to the 
Born configuration, but the possibility for a photon to split into
a pair of charged particles may have important consequences for the 
dressing dressing of, e.g., 
the primary leptons who emitted the photons in the first place. 
Therefore we allow photons to split into electrons, muons, and/or 
quarks or charged hadrons (depending on the energy scale) in a 
reduced parton shower with a reconstructed starting scale for 
the independent collinear evolution of each photon.

\acknowledgments

This research was supported by the Fermi National Accelerator Laboratory (Fermilab),
a U.S. Department of Energy, Office of Science, HEP User Facility.
Fermilab is managed by Fermi Research Alliance, LLC (FRA),
acting under Contract No. DE--AC02--07CH11359.
A. Price is supported by the European Union’s Horizon 2020 research and innovation programme under the Marie
Sklodowska-Curie grant agreement No 94542
This work is supported by the Royal Society through a University Research Fellowship and an Enhancement Award
(URF\textbackslash{}R1\textbackslash{}180549, RGF\textbackslash{}EA\textbackslash{}181033 and
 CEC19\textbackslash{}100349: M.~Sch\"onherr) as well as the STFC under grant agreement ST/P001246/1 (L.\ Flower, M.~Sch\"onherr).





\section{The {\sc Whizard} event generator for future lepton 
lolliders\label{sec:whiz}}
\begin{center}
\emph{J\"urgen Reuter}
\end{center}

\subsection{Introduction to {\sc Whizard}}
\label{sec:whiz_intro}

{\sc Whizard} is a multi-purpose event generator for lepton, hadron
and lepton-hadron collider physics, originally developed for
applications in lepton collider physics for the planned linear
collider TESLA, and has also been applied to fixed-target
experiments. From its early application originates its now deprecated
synonym, ``{\em W}, {\em Hi}ggs, {\em Z}, {\em A}nd {\em R}espective
{\em D}ecays''. {\sc Whizard} is a complete framework for very general
high-energy physics simulations in a modular form that allows a high
level of flexibility~\cite{Kilian:2007gr}. The program core is written
in modern, object-oriented Fortran, while its matrix-element
generator, {\sc O'Mega},  is written in the function programming
language OCaml. Dating back to the late 1990s, the program was
rewritten as version 2.0.0, released in April 2010 for LHC physics,
with the most recent version as of January 2022 being 3.0.2. Version
3.0.0 marked the milestone of automated NLO calculations.

The purpose of this contribution is, after a short 
introduction into the tool, to give an overview of {\sc Whizard}'s
capabilities for future $e^+e^-$ colliders, together with its short-
and mid-term plans. On the technical side, {\sc Whizard} uses a
modern, object-oriented software framework with a highly modular
structure. Most parts of the program core are endowed with unit tests,
and the majority of features is tested by means of core and extended
functional tests in a continuous-integration system. These tests cover
all aspects of the code, even Ward identities and supersymmetric Ward
identities are used~\cite{Ohl:2002jp} to protect against
regressions. Git is used as a version-control system for development. 

\subsection{Hard matrix elements: leading order and BSM}
\label{sec:whiz_me_bsm}

For hard-scattering processes, {\sc Whizard} provides its own
(tree-level) matrix-element generator, {\sc O'Mega}~\cite{Moretti:2001zz},
which produces matrix elements in a recursive way as directed
acyclical graphs. For QCD, {\sc Whizard} and {\sc O'Mega} use the
color-flow formalism~\cite{Kilian:2012pz}. 
This framework supports as hard-coded models the SM and some very
common BSM models like supersymmetry, composite models and effective
field theory setups. Beyond that, it has a very general
interface to FeynRules~\cite{Christensen:2010wz} and
UFO~\cite{Degrande:2011ua} which allows to use almost all possible
external Lagrangian models. Of particular importance for the physics
program at Higgs factories is the support for the complete SM
effective field theory (SMEFT); to just list a few examples of
additional models, extended Higgs sectors like two-Higgs doublet
models, inert doublet models, Higgs-singlet extensions, Higgs portal
models, twin-Higgs models, dark photons, supersymmetric and composite
Higgs models are supported. Les Houches accord (SLHA-like) files can
be used for inputs e.g. in parameter scans. {\sc Whizard} also
supports the definition of customized propagators from the UFO file.
Loop-matrix elements for next-to-leading
(NLO) QCD and EW processes (cf. Sec.~\ref{sec:whiz_nlo} can be used by
dedicated interfaces to one-loop providers (OLP) like Openloops, GoSam
and Recola. 

For all processes within all models, Feynman diagram selection is
possible, however should be treated with care 
due to possible gauge-invariance violations. Processes can be
factorized into production and decays by using the decay features of
{\sc Whizard}, also consecutively in decay chains. All decays of a
resonance at a given final-state multiplicity can be auto-generated.
In resonant processes, intermediate resonances can be specified as
polarizied, e.g. for the study of specific longitudinal or transversal
EW gauge bosons. This is especially important for multi-EW boson
production and vector-boson scattering~\cite{Brass:2018hfw}. The
program also allows to calculate unitarity limits and to generate
bin-per-bin maximal numbers of event consistent with
unitarity~\cite{Alboteanu:2008my,Kilian:2014zja,Fleper:2016frz}.

{\sc Whizard} ships with its 
one scripting language, SINDARIN, which supports a almost completely
general framework of analysis syntax: this allows arbitrary cuts to be
defined on the hard matrix elements, as well as selections for the
event generation. The hard matrix elements can be convoluted with a
large number of parton distribution functions, structure functions or
beam spectra: these include hadron-collider proton PDFs via
LHAPDF~\cite{Whalley:2005nh,Buckley:2014ana}
externally or from a selected set of internally shipped PDFs,
effective $W/Z$ approximation for the content of EW bosons inside
quarks or leptons, the effective-photon approximation
(EPA)~\cite{vonWeizsacker:1934nji,Williams:1934ad,Budnev:1975poe}, also 
known as Weizs\"acker-Williams approximation. The latter is provided
in different variants, depending on the application for low-energy
$\gamma\gamma \to$ hadron background simulations or for high-energy
particle production.

For lepton collisions, electron/muon PDFs are
provided which sum up photons in collinear factorization (augmented
by soft non-collinear photon emission to all orders at the lowest
logarithmic accuracy~\cite{Gribov:1972rt}) and hard-collinear
emissions up to third order. They are implemented and 
well tested for leading logarithmic (LL)~\cite{Kuraev:1985hb,Skrzypek:1990qs} accuracy
and are being 
implemented at the moment at NLL~\cite{Frixione:2019lga,Bertone:2019hks}. This is
expected to be completed 
within 2022. {\sc Whizard} supports several options for structured
lepton beams: Gaussian beam spreads individually adjustable for each
beam, beam spectrum files read in as tables of pairs of energy values
and an approximation to Guinea-Pig spectra via its beamstrahlung
generator {\sc Circe}~\cite{Ohl:1996fi}. This supports for backwards
compatibility a 6- to 7-parameter fit to the beam spectra and a
smoothened, adaptive bin fit that can very well describe also
wake-field machines like CLIC, photon collisions from Compton
backscattering, muon colliders or plasma wake-field machines. Ongoing
work will also include a proper simulation of the $z$-dependence along
the beam axis of the beamstrahlung; this is at the moment being
included in {\sc Circe} and has then to be consistently included in
the infrastructure of the {\sc Whizard} core. This is expected within
the year 2022. {\sc Whizard} allows initial state particle of all
kinds to be polarized, both for scattering as well as  for decay
processes. For scattering processes, the polarization of both beams
can be arbitrarily correlated by specifying a spin density matrix. In
addition, polarization fractions can be given. Electron PDFs are at
the moment inclusive in polarization, i.e. the polarization from the
beam is carried through to the hard matrix element. Work has started
to investigate the effects from polarized versions of the EPA.
Finally, initial state beams can be asymmetric (like in flavor
factories or HERA), and crossing angles can be defined.

\subsection{Phase space integration, performance, event formats}
\label{sec:whiz_misc}
  
{\sc Whizard} can write out events in all major event formats: it
supports a lot of ASCII formats for backwards compatibility, it
supports LHA and LHE formats (in all three different version
variants), it also ships StdHEP, and it provides HepMC2 and HepMC3 as
well as LCIO interfaces. Using the ROOT classes of HepMC3, {\sc
  Whizard} is able to directly write out ROOT (tree) files.

As the program has always had a strong focus on weak production
processes of multi-fermion final states, at lepton colliders it
contains a phase-space parameterization that is flexible enough to
adapt to many different (interfering) resonant production
channels. This builds upon an adaptive multi-channel Monte Carlo
integration encoded in the {\sc VAMP}
subpackage~\cite{Ohl:1998jn,Lepage:1980dq}. In the recent years there
have been several attempts to make the MC integration much more
efficient by using highly parallelized computing architectures: using
Multi-Processing Interface (MPI), speed-ups of several tens up to
roughly a hundred can be achieved~\cite{Brass:2018xbv}. Also, {\sc
O'Mega} can produce very small expressions for the matrix elements in
the form of a so-called virtual machine, which are as efficient as
compiled code~\cite{ChokoufeNejad:2014skp}.

\subsection{Higher order calculations and matching to parton showers}
\label{sec:whiz_nlo}

Turning to high-precision calculations at next-to-leading order in the
SM, {\sc Whizard} has a complete implementation for NLO QCD and
electroweak (as well as mixed) corrections for hadron and lepton
colliders, based on its implementation of the FKS
subtraction~\cite{Frixione:1995ms,Frixione:1997np,Frederix:2009yq}
scheme. In the same setup as for scattering processes, NLO decays can
be calculated. As mentioned above, virtual amplitudes can be used from
OpenLoops~\cite{Cascioli:2011va,Buccioni:2019sur},
GoSam~\cite{Cullen:2011ac,Cullen:2014yla} or 
Recola~\cite{Actis:2016mpe,Denner:2017wsf}. The automated
implementation has been  validated for dozens of processes for LHC and
several energy stages of 
$e^+e^-$ colliders. The parallelized adaptive MC integration discussed
above allows to calculate processes at NLO (QCD) for rather high
final-state multiplicites, e.g. $e^+e^- \to \ell\ell\nu\nu b\bar{b} H$
or $e^+e^- \to jjjjjj$~\cite{Rothe:2021sml,Brass:xxx}. In addition to the
standard FKS, {\sc Whizard} also since several years has a general
implementation of a resonance-aware FKS subtraction scheme. The
program provides a default framework to produce arbitrary differential
distributions, using the conecpt of event groups with an analysis like
e.g. Rivet~\cite{Buckley:2010ar,Bierlich:2019rhm}. Electroweak
corrections for lepton colliders can be 
calculated already for massive leptons, while the infrastructure for
corrections for massless leptons is being validated right now. The NLL
electron PDFs are being implemented, and first results for this are
expected within 2022. Also under validation are QCD corrections to
hadron collider processes for sub-leading QCD coupling powers, where
overlapping QCD and electroweak infrared divergencies appear, and
which cannot be separated. To defne infrared-safe quantities,
{\sc Whizard} provides jet clustering with a
FastJet~\cite{Cacciari:2011ma} interface as well 
as photon isolation~\cite{Frixione:1998jh}. To define fixed-order NLO
cross sections in a 4- or 3-flavor scheme, the user can specify
$b$-jet, $c$-jet and light jet clustering.

In order to generate complete events, {\sc Whizard} provides its own
parton shower, specifically $k_T$-ordered parton shower as well as an
analytic, virtuality-order (QCD) shower. In addition, {\sc Whizard}
ships with the last Fortran Pythia version
(6.427)~\cite{Sjostrand:2006za} which is specifically tuned to LEP2
hadron data, and with a dedicated interface to Pythia
8~\cite{Sjostrand:2014zea}; the latter allows to directly communicate
between the event records of {\sc Whizard} and {\sc Pythia}.

For inclusive jet samples with tree-level matrix elements, {\sc
Whizard} can apply the MLM merging~\cite{Mangano:2001xp} algorithm,
while for NLO matching between matrix elements and the 
parton shower, a general, process-indepen\-dent version of POWHEG
matching~\cite{Frixione:2007vw} is available for NLO QCD
corrections~\cite{ChokoufeNejad:2015kpc}. 
This has been developed for colored final states in $e^+e^-$ and is
under validation for LHC processes at the moment. POWHEG matching for
EW corrections is in preparation, while for mid-term planning also
alternative matching and merging schemes are envisioned. In order to
reconcile of multi-fermion final states, e.g. $e^+e^- \to jjjj$ with
parton showering and hadronization, {\sc Whizard} is able to find and
quantify contributions from underlying resonant subprocesses, and then
provide pseudo-shower histories to the final state weighted by the
relative cross sections of these underlying processes. This is
important to correctly hadronic correlations in the final states, and
also inclusive observables like the total number of final-state
neutral and charged hadrons as well as photons.

One of the most important tasks for precision physics at future Higgs
factories will be a modelling of QED and photon radiation as precise
as possible. For the normalization of QED cross section, a resummation
based on collinear factorization gives very precise results. As
mentioned above, this is possible in {\sc Whizard} at LL level and
will be available at NLL level soon. Specifically for the assessment
of experimental/systematic uncertainties, an exclusive simulation of
photon radiation is necessary. {\sc Whizard} can of course produce
explicit photons from matrix elements; on the other hand, the energy
loss from LL electron PDF in the collinear approximation is collected
in a single photon per beam. A heuristic approximation is possible to
generate a $p_T$ distribution for these photons, which can also be
applied to recoiling charged leptons from
Weizs\"acker-Williams/EPA. The photon distribution is generated using
a logarithmic scaling, while the event of the hard scattering (with
subsequent parton shower and hadronization) is boosted accordingly. It
has been shown that for both signal and background in monophoton dark
sector searches where this distribution matters a lot, this heuristic
approach agrees very well with exact matrix-element
calculations~\cite{Kalinowski:2020lhp,1859727}. {\sc Whizard} has
different options to chose the value of $\alpha$, which first of all
depends on the electroweak scheme; the native scheme for tree-level
calculations is the $G_F-M_Z-M_W$ scheme, for NLO calculations the
complex-mass scheme. There is also the possibility to take $\alpha$
directly from the OLP provider. For the interal $\alpha$, {\sc
Whizard} provides a running $\alpha$ as well, either at the one- or
two-loop level.

While a collinear resummation results in very accurate results for
cross sections of (inclusive) processes with initial-state radiation,
the highest possible precision of resummed calculations has to be
combined with exclusive photon radiation in the events. There are
different approaches to do that, e.g. using a POWHEG or MC@NLO-type
matching for QED or eikonal-based resummation in the form of the YFS 
formalism~\cite{Yennie:1961ad}. The {\sc Whizard} team pursues both
approaches, the first by extended the extended the NLO POWHEG matching
from QCD to QED, and for the second it is starting 
the implementation of the YFS formalism for general processes. First
results could be expected end of 2022 or early in 2023. Furthermore,
{\sc Whizard} will also get its own QED parton shower, which will be
available as a relatively slim stand-alone shower for pure QED, and
potentially also as an interleaved shower together with the QCD
shower.

\subsection{Special applications: Top and $WW$ threshold}

{\sc Whizard} has a special support for the top threshold to model the
process $e^+e^- \to t\bar{t} \to bW^-\bar{b}W^+$. This process allows
to determine the top mass with the highest possible theoretical
precision. Analytic calculations using an NRQCD EFT approach can
include NNNLO corrections, which allows to reduce the error to 30-70
MeV. This calculation, however, is completely inclusive. To study
experimental efficiencies and systematic uncertainties, Monte Carlo
samples resembling event weights close to the true top threshold cross 
sections are needed. {\sc Whizard} includes a special treatment of a
matched calculation between the continuum NLO off-shell fixed-order
calculation and the NLL-threshold resummed calculation avoiding double
counting~\cite{Bach:2017ggt}, fully off-shell and exclusive in the
top decay products. This allows to study fully exclusive distributions.

In a similar framework, work has started to match resummed
QED-Coulombic corrections to the QED NLO/NNLO fixed-order calculation
for the $WW$ threshold~\cite{Beneke:2007zg,Actis:2008rb}. Like the top
threshold, this will be available as a dedicated process within a
specialized model. These simulations  allow experimental studies with
the desired accuracy to support measurements of the $W$ mass from a
threshold scan with a precision of 1-2 MeV or below.

\vspace{0.5cm}

In summary, the main ongoing work inside {\sc Whizard} is the
completion of fully automatized NLO SM corrections (QCD/EW/mixed) for
any kind of collider, for arbirtrary differential distributions at
fixed order, and also matched to QCD/QED/EW parton showers. 

\acknowledgments

JRR acknowledges the support by the Deutsche Forschungsgemeinschaft
(DFG, German Research Association) under Germany's Excellence
Strategy-EXC 2121 ``Quantum Universe"-39083330.



\section{Lepton collisions in MadGraph5\_aMC@NLO\label{sec:mg5amc}}
\begin{center}
\emph{Giovanni Stagnitto, Marco Zaro}
\end{center}

In this Section we will report on the functionalities of \aNLO\ related to the
simulation of lepton colliders.
We remind the reader that \aNLO~\cite{Alwall:2014hca, Frederix:2018nkq} is a
computer program for the automatic computation of LO- and NLO-accurate cross
sections (the latter both in the QCD and in the EW coupling) for scattering
processes. While \aNLO\ is widely used in the context of LHC simulations, it can
also be employed for lepton collisions. Indeed, many results for leptonic
collisions where already provided in Ref.~\cite{Alwall:2014hca}, including
NLO-QCD corrections but limited to the case of a strictly fixed centre-of-mass
energy.
The extension to the case with Initial-State Radiation (ISR) at leading
logarithmic accuracy (LL) and possibly beamstrahlung is more recent, and has
documented in Ref.~\cite{Frixione:2021zdp}.
Developments are in progress for the inclusion of NLO EW corrections to the
short distance cross section and a next-to-leading logarithmic (NLL) accurate
treatment of ISR, allowing for the computation of NLL+NLO observables.
Here we will briefly review the changes in the code which are necessary in order
to deal with leptonic collisions, following the discussion in
Ref.~\cite{Frixione:2021zdp}, and expanding it to the case of NLL ISR and NLO EW
corrections.
An in-depth phenomenological study of NLL+NLO effects on physical observables
will be the subject of a forthcoming paper~\cite{epdf-pheno}.\\

Following the notation of Refs.~\cite{Frixione:2019lga,Frixione:2021zdp}, if we
start from two colliding beams of electrons and positrons with momenta
$P_{e^\pm}$, and we define the corresponding cross-section for the reaction
\begin{equation}
    e^+(P_{e^+})  e^-(P_{e^-}) \to X
\end{equation}
as $d\Sigma_{e^+ e^-}$,
the following steps happen:
\begin{enumerate}
    \item A pair $(k,l)$ of particles emerge from beam dynamics, which carry a
      fraction $y^\pm$ of longitudinal momentum of the two incoming beams. The
      beam-level cross section factorises as a convolution of particle-level
      cross section $d \sigma_{kl}$ and the beamstrahlung function $\mathcal
      B_{kl}$
\begin{equation}
    d\Sigma_{e^+ e^-} \left(P_{e^+},P_{e^-}\right) = 
\sum_{kl}\int dy_+ dy_- \, \mathcal B_{kl} (y_+, y_-) \, d \sigma_{kl}(y_+ P_{e^+}, y_- P_{e^-}) \,.
\end{equation}
    \item Particles $k, l$ undergo a hard collision, where ISR effects are
      included by writing $d \sigma_{kl}$ as yet another convolution of a
      parton-level cross section $d \hat \sigma_{ij}$ and QED parton
      distribution functions (PDFs) $\Gamma_{i/k}$
\begin{equation}\label{eq:fact}
    d \sigma_{kl} (p_k,p_l) = \sum_{ij} \int dz_+ dz_+\, \Gamma_{i/k}(z_+,\mu,m) \, \Gamma_{j/l}(z_-,\mu,m)
    \, d \hat \sigma_{ij}(z_+ p_k, z_- p_l, \mu)\,,
\end{equation}
    where $z_{\pm}$ are the longitudinal momentum fractions carried by the
    partons w.r.t. their mother particle, $\mu$ is the factorisation scale and
    $m$ the lepton mass, which is neglected in the parton-level cross section.
    In the following, we will mostly focus on the PDFs relevant to an incoming
    unpolarised electron particle, $\Gamma_{i/e-}$; the PDFs of an incoming
    positron are trivially related by charge conjugation. We will refer to
    $\Gamma_{e^\pm/e^\pm}$ as electron PDF, and to $\Gamma_{\gamma/e^\pm}$ as
    photon PDF.
\end{enumerate}

Eq.~\eqref{eq:fact} recalls the standard QCD factorisation formula at hadron
colliders. However, at variance with hadronic PDFs, QED PDFs are entirely
calculable with perturbative techniques. Their role is to resum to all order the
large contributions stemming from photon collinear emissions in the initial
state, which appear as logarithms of some hard physical scale $E$ over the mass
of the electron $m$, $\log^k(E^2/m^2)$.
The collinear terms present in the PDFs are universal, and their resummation by
means of QED DGLAP evolution
equations~\cite{Altarelli:1977zs,Gribov:1972ri,Lipatov:1974qm,Dokshitzer:1977sg}
(see Ref.~\cite{deFlorian:2016gvk} for explicit expressions of the two-loop QED
splitting kernels) is a process-independent procedure.
Let us stress that within the PDF formalism, we are taking into account only the
logarithms related to (hard or soft) collinear radiation off initial-state
particles. In principle, by means of fragmentation functions (FFs), which are
the time-like analogue of PDFs, it would be possible to also account for
collinear radiation off final-state particles (as it is usually done in QCD when
heavy quarks are present in the final state~\cite{Mele:1990cw}). Soft logarithms
and interference terms can instead be resummed by means of other resummation
techniques~\cite{Yennie:1961ad,Jadach:2000ir,Anlauf:1991wr,Munehisa:1995si,CarloniCalame:2000pz},
which are usually tailored to specific class of processes though.

In practice, both beamstrahlung and ISR effects are included in \aNLO\ by
means of the definition of suitable partonic densities. The relevant formulas
are reported in Sec.~3-5 of Ref.~\cite{Frixione:2021zdp} and will not be
repeated here.
As for ISR, the current public release of \aNLO\ includes,
for lepton collisions, the long-known LL analytical expressions~\cite{Skrzypek:1990qs,Skrzypek:1992vk,Cacciari:1992pz}, which resum the tower of $(\aem
\log(E^2/m^2))^k$ terms.
Such LL analytical expressions are built out of an additive matching between a
recursive solution up to some order in $\aem$, typically $\mathcal{O}(\aem^3)$,
and an all-order $\aem$ solution valid in the region $z \to 1$ (where the bulk
of the cross section is), as usually done in the literature, see
e.g.~\cite{Denner:2000bj}.
Note that, in the case of NLO EW (QED) corrections to the short-distance cross
section with LL PDFs, a scheme of change term is needed in the short-distance
cross section in order to avoid overcounting. The peculiar structure of the PDFs,
which feature an integrable divergence for $z\to 1$, requires a suitable
re-parameterization of the phase-space, as described in 
Ref.~\cite{Frixione:2021zdp}.

Recently in Ref.~\cite{Bertone:2019hks}, QED PDFs have been extended to NLL
accuracy i.e.\ resumming also the $\aem (\aem \log(E^2/m^2))^k$ terms; they have
been obtained by solving the NLO evolution equations with NLO initial conditions
(derived in Ref.~\cite{Frixione:2019lga}) by means of both analytical and
numerical methods.
By working at the NLL accuracy, the mixing between the electron/positron (and
possibly other fermion families) and the photon PDFs is taken into account. Note
that NLL PDFs not only provide a NLL correction to processes with incoming
electrons, but also allow to treat photon-initiated hard processes in the same
framework.
A public code, \ePDF, has been developed in the context of
Ref.~\cite{Bertone:2019hks}: it provides the numerical solution of the evolution
equations, as well as the implementation of the analytical large-$z$ solutions,
both in the $\MSb$ and in the $\Delta$~\cite{Frixione:2021wzh} factorisation
scheme\footnote{Neither $\Gamma_{i/k}$ nor $d\hat\sigma_{ij}$ in
eq.~\eqref{eq:fact} are physical quantities; at NLL and beyond, they depend on
the choice of the factorisation scheme and on the value of the mass scale $\mu$
(it is sensible to choose a value of $\mu \sim E$, with $E$ a scale of the order
of the hardness of the process). The $\Delta$ scheme is similar to the DIS
scheme usually adopted in QCD, in which one maximally simplifies the PDFs'
initial conditions. We refer the interested reader to
Ref.~\cite{Frixione:2021wzh}, where the $\Delta$ scheme is discussed in more detail.
}.

The key point to stress is that, both at LL and at NLL, the electron PDF
features an integrable divergence as $z_\pm \to 1$, which has to be handled in
the proper way when convoluting the short-distance cross section with the PDFs.
This is why it is crucial to have an analytic control of the electron PDF in the
large-$z$ region.
As for the photon PDF, it does not pose any problem in the large-$z$ region: in
the $\MSb$ scheme it features at most a logarithmic divergence, which can be
handled by means of a proper change of integration variables, whereas in the
$\Delta$ scheme does not diverge.

The default way adopted in \ePDF\ in order to build an electron PDF well-behaved
in the whole $z$-range is a multiplicative matching between the numerical
solution $\Gamma_{\rm num}$ and the associated analytical large-$z$ solution
$\Gamma_{\rm asy}$. 
We introduce a switching point $z_0$ and we define the matched solution as
\begin{equation}\label{eq:msol}
  \Gamma_{\rm mtc}(z) =
  \begin{cases}
    \Gamma_{\rm num}(z)\,, & z < z_0\,, \\
    \Gamma_{\rm asy}(z)\,\dfrac{\Gamma_{\rm num}(z_0)}{\Gamma_{\rm asy}(z_0)}\,, & z > z_0\,. \\
  \end{cases}
\end{equation}
The switching point $z_0$ is chosen in such a way at $z_0$ the solutions
$\Gamma_{\rm num}(z_0)$ and $\Gamma_{\rm asy}(z_0)$ are essentially identical,
and the numerical solution $\Gamma_{\rm num}(z_0)$ has not yet lost
accuracy. Typical values for $z_0$ are between $1-10^{-6}$ and $1-10^{-7}$.
Since in the $\MSb$ scheme the analytical recursive solutions up to
$\mathcal{O}(\aem^3)$ are also available~\cite{Bertone:2019hks}, in such a
scheme it is also possible to use a fully analytical additively-matched
solution, in keeping with what was done in the literature at LL.

In practice, a runtime evaluation of the numerical solution is too slow, so it
is convenient to use grids, as it is customary in the hadronic PDFs case.
However, as said above, at the same time it is important to have an analytic
control of the electron PDF in the $z \to 1$ region, since during the phase-space
integration values of $z$ extremely close to 1 are probed (note that one should
avoid to explicitly evaluate the difference $1-z$, so as to not incur in
overflow errors).
Hence, in order to build the matched solution eq.~\eqref{eq:msol}, a two-stage
approach is adopted. First, we save the numerical solution as a grid in
\LHAPDF~\cite{Buckley:2014ana} format. Then, at runtime, we read the grid and we
call the appropriate large-$z$ solution, and eventually we perform the matching
between the two solutions.
The whole procedure has been implemented in a (still) private version of \ePDF,
which has been interfaced to \aNLO.
Note that \ePDF\ is a standalone code which can in principle be linked to any
other partonic generator: the user is just required to call a high-level
function of \ePDF.\\

So far, we have not considered the case of NLO EW (QED) corrections in the
short-distance cross section. While the generation of the relevant
matrix-elements and counterterms does not require any change in the code, it is
again the structure of $\Gamma_{e^\pm/e^\pm}(z_\pm,\mu,m)$ which poses
efficiency problems in the phase-space integration. We remind the reader that
\aNLO\ employs the FKS subtraction scheme~\cite{Frixione:1995ms}, as automated
in {\sc MadFKS}~\cite{Frederix:2009yq}, and that in such a scheme sectors are
defined in terms of pair of particles $(i,j)$, which can be seen as the emitted
parton and its sister.  In the case $j$ belongs to the
initial state, the standard way the real-emission kinematic is generated in
\aNLO\ relies on the so-called event projection: in the real emission, the sum
of all final-state momenta excluding parton $i$ has the same invariant mass and
rapidity as the sum of the final-state momenta in the Born kinematics (see
Sec.~5 of Ref.~\cite{Frixione:2007vw} for the explicit construction). This is a
necessary feature in order to be able to match NLO computations with current
parton showers, as it mimics what the latter do in the case of initial-state
backward evolution. However, this leads to the fact that real-emission and
subtraction terms (the latter have Born kinematics) have different Bjorken
$x$'s. While this is not a problem in the case of hadron collisions, the
structure of the ISR density creates serious efficiency issues if this
phase-space parameterisation is adopted.

In order to cope with this issue, a new parameterisation for the real emission
has been adopted, to be applied in the FKS sectors where particle $j$ is in the
initial state. Before sketching it, we stress that employing this
parameterisation forbids to match with QED parton showers, unless the latter are
modified in a consistent manner.

The new generation for the real emission works as follows:
\begin{enumerate}
    \item The kinematics of the real-emission parton ($i$ in the FKS notation)
      is generated from the variables
    \begin{equation}
        \xi=\frac{E_i}{2 \hat s}, \quad y = \cos \theta_{ij}, \quad \phi\,.
    \end{equation}
    Respectively, they correspond to the rescaled energy of parton $i$, written
    in terms of the partonic energy $\hat s$, its angle with particle $j$, and
    the azimuth, all defined in the partonic center-of-mass frame. These
    variable are in a one-to-one correspondence with the integration random
    numbers (possibly with adaptive sampling). They uniquely define the massless
    four momentum $k_{n+1}$ in the partonic c.o.m frame.
\item The kinematics of the remaining $n$ final-state momenta $\{\bar
  k_l\}_{l=1,n}$ is generated in their c.o.m.\ frame.  Their total invariant
  mass is
    \begin{equation}
    \left(\sum_{l=1,n} \bar k_l\right)^2 = (1-\xi) \, \hat s\,.
    \end{equation}
\item Finally, the momenta $\{\bar k_l\}_{l=1,n}$ are boosted in the partonic
  c.o.m.\ frame
    \begin{equation}
        k_l = \mathbf B \bar k_l
    \end{equation}
    in order to ensure momentum conservation, i.e. in the same frame we have
    \begin{equation}
        \sum_{l=1,n+1} k_l =  \left(\sqrt{\hat s}, \vec 0 \right)\,.
    \end{equation}
\end{enumerate}

Thanks to these modifications, a stable and efficient evaluation of NLO EW
corrections can be performed also for lepton collisions within \aNLO, as it will
be documented in Ref.~\cite{epdf-pheno}.

\acknowledgments

M.Z. is supported by the “Programma per Giovani Ricercatori Rita Levi Montalcini” granted by the Italian Ministero dell’Università e della Ricerca (MUR).



\section{BabaYaga\label{sec:BY}}
\begin{center}
\emph{Carlo M. Carloni Calame,
Guido Montagna,
Oreste Nicrosini,
Fulvio Piccinini
}\end{center}

\subsection{Introduction}
\label{introduction}
The knowledge of the luminosity ${\cal L}$ is an important ingredient for any
measurement at $e^+ e^-$ machines. The common strategy is to calculate it through
the relation ${\cal L} = N_{obs} /\sigma_{th}$ , where $\sigma_{th}$ is the theoretical
cross section of a QED process, namely $e^+ e^- \to e^+ e^-$
(Bhabha~\footnote{Bhabha scattering events are used in two different
kinematical regimes: 
large angle (typically charged particles are required to lie in the
interval $[20,160]$ or $[15,165]$ degrees) for flavour factories with center of mass energies up to $\sim 10$~GeV and 
small angle (typically one charged particle is required within a few  degrees in
the forward direction and the other one in the backward direction) for LEP and future $e^+ e^-$ machines. In the latter regime Bhabha scattering is less sensitive than
the former to potential New Physics effects at high energy collisions and therefore more suited for
luminometry.}),
$e^+ e^- \to \mu^+ \mu^-$ or $e^+ e^- \to \gamma \gamma$, and
$N_{obs}$ is the number of observed events. QED processes are the best choice
because of their clean signal, low background and the possibility to push the
theoretical accuracy up to very high precision. The latter requires
the inclusion of the relevant radiative corrections (RCs) in the cross sections
calculation and their implementation into Monte Carlo (MC) event generators (EGs)
in order to easily account for realistic event selection criteria. Present
EGs used for luminometry rely on exact fixed-order QED corrections and Leading Logarithmic
(LL) approximation of higher order effects, together with their consistent matching.
The following sections describe how this is achieved within the \babayaga\ EG,
targeting NLOPS accuracy. It was originally developed for the precise simulation of
large-angle Bhabha scattering at low energy $e^+ e^-$ colliders, 
with center of mass energy up to 10 GeV~\cite{CarloniCalame:2000pz,CarloniCalame:2001ny}, 
and later extended~\cite{Balossini:2008xr} to simulate also
$\mu^+ \mu^-$ and $\gamma \gamma$ final states in the same energy regime.
In its first version~\cite{CarloniCalame:2000pz,CarloniCalame:2001ny},
the generator relied upon a QED Parton Shower (PS) to 
account for the LL photonic corrections, resummed up to all 
orders in perturbation theory. 
The PS algorithm implemented in \babayaga\ is described in Section~\ref{sec:ps},
while the matching with the NLO calculation is described in Section~\ref{sec:matching}.
Last Section~\ref{sec:high-energy} discusses the current work in progress
to update the generator to match the precision requirements of future
high-energy $e^+ e^-$ machines.

\subsection{QED Parton Shower algorithm}
\label{sec:ps}
The PS is a MC algorithm which gives an exact iterative numerical
solution of the Dokshitzer-Gribov-Lipatov-Altarelli-Parisi (DGLAP)
evolution equation~(\cite{Gribov:1972ri,Gribov:1972rt,Dokshitzer:1977sg,Altarelli:1977zs})
in QED for the non-singlet QED structure function (SF)\footnote{The
term ``Structure Functions'' applies strictly only when a LL evolution
is considered, which is here the case. Beyond LL, the more appropriate term ``Parton
Distribution Function'' should be used.}
$D(x,Q^2)$, which allows to describe the effects of multiple emission
of photons in the collinear limit to all orders in perturbation theory. 
The SF represents the probability density of finding ``inside'' a parent electron
an electron with momentum fraction $x$ and virtuality $Q^2$. The evolution
equation reads~\footnote{For definiteness in the following, the QED coupling
constant $e$ is chosen to be renormalized in the on-shell scheme,
{\it i.e.} it is connected with the fine structure constant $\alpha$ by
the relation $e^2 = 4\pi\alpha$.} 
\begin{equation}  
Q^2\frac{\partial}{\partial Q^2}D(x,Q^2)=
\frac{\alpha}{2\pi}\int_x^{1}\frac{dy}{y}P_+(y) D(\frac{x}{y},Q^2)  .
\label{eq:ap} 
\end{equation}
In the previois equation, $P_+(x)$ is the regularized $e \to e + \gamma$ 
splitting function 
\begin{equation} 
P_+(x)=\frac{1+x^2}{1-x}-\delta(1-x)\int_0^1 dt P(t) ,
\label{eq:vertex}
\end{equation}
where $P(x) = (1+x^2)/(1-x)$ is the unregularized $e\to e+\gamma$
splitting vertex. Equation~(\ref{eq:vertex}) is symbolic and its numerical
evaluations require the introduction of an infrared regulator
$x_+\equiv1-\epsilon$ with $\epsilon \ll 1$, which changes it into the form
\begin{equation} 
P_+(x)=\frac{1+x^2}{1-x}\Theta(x_+-x)-\delta(1-x)\int_0^{x_+} dt P(t) ,
\label{eq:vertexir}
\end{equation}
where $\Theta(x)$ is the Heaviside step function.

The energy scale $Q^2$ entering the 
SF is, in general, dependent on the specific process under study. 
The QED SFs account for photon radiation emitted by 
both initial-state and final-state charged fermions. 
An important feature of the PS solution is that the kinematics of the emitted photons
can be recovered (within some approximation) and hence an exclusive event generation
can be performed, i.e. all the momenta of the final state particles (fermions
and an indefinite number of photons) can be reconstructed.
Within the SF approach the corrected cross section can be written as 
\begin{equation}
\sigma(s)=\int dx_- dx_+ dy_- dy_+ \int d\Omega 
D(x_-,Q^2)D(x_+,Q^2)
D(y_-,Q^2)D(y_+,Q^2) \frac{d\sigma_0(x_- x_+ s)}{d\Omega_{cm}} \Theta(cuts) , 
\label{eq:sezfs}
\end{equation}   
where $D(x_\pm,Q^2)$ and $D(y_\pm,Q^2)$ refer to radiation from initial and final
state charged legs, respectively~\footnote{For the process
$e^+ e^-\to \gamma \gamma$, $D(y_\pm, Q^2) = 1$ is understood.}, and
$d\sigma_0 / d\Omega$ is the Born-like differential cross section.

The starting point of the PS algorithm is the Sudakov form
factor~\cite{Sudakov:1954sw}:
\begin{equation} 
\Pi (s_1,s_2) = \exp \left[-\frac{\alpha}{2 \pi} 
\int_{s_2}^{s_1} \frac{d s'}{s'} \int_0^{x_+} dz P(z)  \right]  ,   
\label{eq:sudakov} 
\end{equation}
which represents the probability that an electron 
evolves from virtuality $s_2$ to virtuality $s_1$ with no 
emission of photons of energy fraction 
greater than $\epsilon = 1-x_+$, where 
$\epsilon$ is an infrared regulator. In terms of 
the factor $\Pi (s_1,s_2)$ and with the initial condition $D(x,m^2)=\delta(1-x)$, 
the DGLAP equation can be written in 
iterative form as: 
\begin{eqnarray} 
D(x,Q^2)&=&\Pi(Q^2,m^2)\delta(1-x) \nonumber\\
&+& \int_{m^2}^{Q^2}\Pi (Q^2,s')\frac{d s'}{s'}\Pi (s',m^2)\frac{\alpha}{2\pi}
\int_0^{x_+} dy P(y) \delta (x-y) \nonumber\\
&+& \int_{m^2}^{Q^2}\Pi (Q^2,s')
\frac{ds'}{s'}\int_{m^2}^{s'}\Pi (s',s'')\frac{ds''}{s''}
\Pi (s'',m^2)\times \nonumber \\
&&\bigg(\frac{\alpha}{2\pi}\bigg)^2\int_0^{x_+} dx_1\int_0^{x_+}
dx_2 P(x_1)P(x_2) \delta (x-x_1x_2) + \cdots    
\label{eq:alpha2}             
\end{eqnarray} 
Equation~(\ref{eq:alpha2}) suggests the steps to compute $D(x,s)$
by means of a MC algorithm detailed in Ref.~\cite{CarloniCalame:2000pz}.  
In this way the emission of a shower of photons by an electron is simulated, and 
the $x$ distribution of the PS event sample reproduces 
$D(x,Q^2)$. Equation~(\ref{eq:alpha2}) accounts for soft plus virtual and real photon
radiation up to all orders in perturbation theory in LL approximation. It is
worth noticing that, setting the scale $Q^2$ equal to $s t / u$, the Sudakov form
factor exponentiates the LL contribution of the ${\cal O}(\alpha)$ soft plus virtual
cross section as well as the dominant contribution coming from the infrared cancellation
between the virtual box and the initial-final state interference of bremsstrahlung
diagrams~\footnote{The above applies for charged fermion pair production, while for
$e^+ e^- \to \gamma \gamma$ only the choice $Q^2 = s$ is physically meaningful.}. 

The PS algorithm offers the possibility to go naturally 
beyond the strictly collinear treatment of the electron evolution, 
by generating the transverse momentum $p_\perp$ of electrons and 
photons at each branching. In Ref.~\cite{CarloniCalame:2001ny} an improvement
has been introduced to include the coherence effects, due to radiation
from different charged legs, by generating the angular spectrum of the $l^{th}$
photon as
\begin{equation}
  \cos \vartheta_l \propto - \sum_{i,j}^N \eta_i \eta_j
  \frac{1 - \beta_i \beta_j \cos\vartheta_{ij}}{(1-\beta_i \cos\vartheta_{il})
    (1 - \beta_j \cos\vartheta_{jl})},
  \label{eq:YFSangular}
\end{equation}
where $N$ is the number of the involved charged fermions, $\beta_i$ is the
speed of the $i^{th}$ emitter, $\vartheta_{ij}$ is the relative angle between
particles $i$ and $j$, $\eta_i$ is a charge factor, which is $+1$ for incoming
lepton or outgoing antilepton and $-1$ for incoming antilepton or outgoing lepton.
Equation~(\ref{eq:YFSangular}) is inspired to the expression of the differential
cross section for a generic process with emission of additional photons in the
soft limit~\cite{Yennie:1961ad}. The prescription of Equation~(\ref{eq:YFSangular}) 
has been shown to generate exclusive radiation in nice agreement with exact
${\cal O}(\alpha)$ results for Bhabha scattering and $e^+ e^- \to \mu^+ \mu^-$. 

Despite its nice features, the PS described above is intrinsically accurate at the LL level
and a precision in integrated or differential cross sections better
than 0.5-1\% can not be expected. In order to account also for missing ${\cal O}(\alpha)$ non-log
contributions to exclusive and inclusive observables, a matching with
exact NLO RCs is mandatory, in such a way that the features of the PS
are preserved (i.e. exclusive event generation and resummation of LL corrections
up to all orders) while avoiding the double counting of the ${\cal O}(\alpha)$
LL corrections, present both in the PS approach and in the NLO calculation.

\subsection{Matching NLO corrections with Parton Shower: \babayagaNLO} 
\label{sec:matching}
In this Section we discuss the matching algorithm as implemented in
\babayagaNLO~\cite{Balossini:2006wc}. It is worth mentioning that
PS/NLO matching algorithms have been firstly put forward in
the context of QCD simulations~\cite{Frixione:2002ik,Nason:2004rx}: although our
procedure certainly shares some common ideas with them, we never carried out a
thorough comparison of general similarities and differences with our
algorithm, which would be extremely interesting to perform in the
future.

The fully differential cross section implicit in Equation~(\ref{eq:sezfs}) 
can be recast in the following form:
\begin{equation}
d\sigma^{\infty}_{LL}=
{\Pi}(Q^2,\varepsilon)~
\sum_{n=0}^\infty \frac{1}{n!}~|{\cal M}_{n,LL}|^2~d\Phi_n ,
\label{generalLL}
\end{equation}
where ${\Pi}(Q^2,\varepsilon)$ is the Sudakov form-factor accounting for the
soft-photon (up to an energy equal to $\varepsilon$ in units of the
incoming fermion energy $E$) and virtual emissions, $\varepsilon$ is
an infrared separator 
dividing soft and hard radiation and $Q^2$ is
related to the energy scale of the process.
$|{\cal M}_{n,LL}|^2$ is the squared amplitude in LL
approximation describing the process with the emission of $n$ hard
photons, with energy larger than $\varepsilon$ in units of $E$.
$d\Phi_n$ is the exact
phase-space element of the process (divided by the incoming flux
factor), with the emission of $n$ additional photons
with respect to the Born-like final-state configuration. 
The cross section $d\sigma^{\infty}_{LL}$ of Equation~(\ref{generalLL}) 
is numerically independent of the infrared separator
$\varepsilon$. More precisely, only power-suppressed $\epsilon$ terms
might be present, which can be made vanishingly small.


According to the factorization theorems of soft and/or collinear
singularities, the squared amplitudes in LL approximation can be written in a factorized
form. In the following, for the sake of clarity and without loss of
generality, we write photon emission formulas as if only one external
fermion radiates. We are aware that it is a completely unphysical
case, but it allows to write more compact formulas, being the
generalization to the real case straightforward when including the
suited combinatorial factors. With this in mind,
the one-photon emission squared amplitude in LL approximation can be
written as
\begin{equation}
|{\cal M}_{1,LL}|^2=\frac{\alpha}{2\pi}\frac{1+z^2}{1-z}I(k)
~|{\cal M}_0|^2~
\frac{8\pi^2}{E^2 z (1-z)} ,
\label{onegammaLL}
\end{equation}
where $1-z$ is the fraction of the fermion energy $E$ carried by the
photon, $k$ is the photon four-momentum,
$I(k)$ is a function describing the angular spectrum of the photon
and $P(z)=(1+z^2)/(1-z)$ is the Altarelli-Parisi 
${e} \to {e} + \gamma$ splitting function. 
In Equation~(\ref{onegammaLL}) we observe the factorization of the Born squared
amplitude and that the emission factor $\frac{\alpha}{2\pi}P(z)I(k)
\frac{8\pi^2}{E^2z(1-z)}$
can be iterated for each photon emission, up to all orders,
to obtain $|{\cal M}_{n,LL}|^2$.
It is worth noticing that ${d^3\vec{k}}/{k^0} = (1-z)E^2d\Omega_\gamma dz$
and that in the collinear limit the cross
section of Equation~(\ref{generalLL}) reduces to the cross section calculated
by means of the QED PS algorithm described in Refs.~\cite{CarloniCalame:2000pz,CarloniCalame:2001ny}.

The Sudakov form factor $\Pi(Q^2,\varepsilon)$ reads explicitly
\begin{equation}
\Pi(Q^2,\varepsilon)=
\exp\left(
-\frac{\alpha}{2\pi}~I_+~L^\prime
\right),~~~~L^\prime=\log\left(\frac{Q^2}{m^2}\right),~~~~
I_+\equiv
\int_0^{1-\varepsilon}
dz P(z) .
\end{equation}
The function $I(k)$ has the property that
$\int d\Omega_{\gamma}I(k)=\log (Q^2/m^2)$
and allows the cancellation of the infrared logarithms.

The cross section calculated in Equation~(\ref{generalLL}) has the
advantage that the photonic corrections, in LL approximation, are
resummed up to all orders of perturbation theory. On the other side,
the weak point of the formula~\ref{generalLL} is that its
expansion at ${\cal O}(\alpha)$ does not coincide with an exact ${\cal O}(\alpha)$ (NLO) result, being
its LL approximation. In fact
\begin{eqnarray}
d\sigma^{\alpha}_{LL}&=&
\left[
1-\frac{\alpha}{2\pi}~I_+~\log\frac{Q^2}{m^2}
\right] |{\cal M}_0|^2 d\Phi_0+
|{\cal M}_{1,LL} |^2 d\Phi_1\nonumber\\
&\equiv&
\left[
1+C_{\alpha,LL}
\right] |{\cal M}_0|^2 d\Phi_0
+
|{\cal M}_{1,LL} |^2 d\Phi_1 ,
\label{LL1}
\end{eqnarray}
whereas an exact NLO cross section can be always cast in the form
\begin{equation}
d\sigma^{\alpha}
=
\left[
1+C_{\alpha}
\right] |{\cal M}_0|^2 d\Phi_0
+
|{\cal M}_{1} |^2 d\Phi_1 .
\label{exact1}
\end{equation}
The coefficient $C_{\alpha}$ contains the complete virtual
${\cal O}(\alpha)$ and the ${\cal O}(\alpha)$ soft-bremsstrahlung squared matrix elements, in units of
the Born squared amplitude,
and $|{\cal M}_1|^2$ is the exact squared matrix element with the
emission of one hard photon.
We remark that $C_{\alpha,LL}$ has the same logarithmic structure as
$C_\alpha$ and that $|{\cal M}_{1,LL}|^2$ has the same singular
behaviour of $|{\cal M}_1|^2$.

In order to match the LL and NLO calculations, we introduce
the correction factors, which are by construction
infrared safe and free of collinear logarithms,
\begin{equation}
F_{SV}~=~
1+\left(C_\alpha-C_{\alpha,LL}\right),~~~~~~
F_H~=~
1+\frac{|{\cal M}_1|^2-|{\cal M}_{1,LL}|^2}{|{\cal M}_{1,LL}|^2}
\label{FSVH}
\end{equation}
and we notice that the exact ${\cal O}(\alpha)$ cross section can be expressed, up
to terms of ${\cal O}(\alpha^2)$,
in terms of its LL approximation as
\begin{equation}
d\sigma^\alpha~=~
F_{SV} (1+C_{\alpha,LL} ) |{\cal M}_0|^2 d\Phi_0
~+~
F_H |{\cal M}_{1,LL}|^2 d\Phi_1
\label{matchedalpha}
\end{equation}
Driven by Equation~(\ref{matchedalpha}), Equation~(\ref{generalLL})
can be improved by writing the resummed cross section as
\begin{equation}
d\sigma^{\infty}_{matched}=
F_{SV}~\Pi(Q^2,\varepsilon)~
\sum_{n=0}^\infty \frac{1}{n!}~
\left( \prod_{i=0}^n F_{H,i}\right)~
|{\cal M}_{n,LL}|^2~
d\Phi_n ,
\label{matchedinfty}
\end{equation}
The extension of the matching formula Equation~(\ref{matchedinfty}) to the
realistic case, where every charged particle radiates photons,
is almost straightforward. Equation~(\ref{matchedinfty}) is the master
formula according to which event generation and cross secton
calculation are performed in \babayagaNLO. It is worth stressing that
the $F_{SV}$ and $F_{H,i}$ correction factors are applied at differential
level on an event-by-event basis and that the ${\cal O}(\alpha)$
expansion of Equation~(\ref{matchedinfty}) coincides with Equation~(\ref{exact1}).

 We would like to remark also that
the LL cross section of Equation~(\ref{generalLL}) is by construction
positively defined in every point of the phase space, whereas the
correction factors of Equation~(\ref{FSVH}) can in principle make the
differential cross section of Equation~(\ref{matchedinfty}) negative in
some point, namely where the PS approximation is less accurate
(e.g. for hard photons at large angles). Nevertheless, we verified
that this never happens when considering typical event selection criteria for
luminosity at flavour factories.

It is useful to present, in the realistic case, the expression of the
function $I(k)$, which describes the leading behaviour of the angular spectrum
of the emitted photons, accounting also for interference of radiation
coming from different charged particles:
\begin{equation}
I(k)~=~
\sum_{i,j=1}^4~
\eta_i \eta_j~
\frac{p_i\cdot p_j}{(p_i\cdot k)(p_j\cdot k)}~
E_\gamma^2
\label{idik}
\end{equation}
where $p_l$ is the momentum of the external fermion $l$, 
$\eta_l$ is the charge factor introduced in Equation~(\ref{eq:YFSangular}), 
$k$ is the photon momentum, $E_\gamma$ is its energy
and the sum runs over all the external fermions. The function $I(k)$
does not depend on the photon energy and it is analogous to
Equation~(\ref{eq:YFSangular}). Given Equation~(\ref{idik}), and considering for instance Bhabha scattering, one can
quite easily convince that, after integrating over photon angles, the
scale $Q^2$ in the Sudakov form
factor takes the form $L^\prime=\log\frac{Q^2}{m^2} = \log\frac{st}{u m^2} - 1\equiv L-1$ where $s$, $t$ and $u$ are the
Mandelstam variables of the process and $m$ is the electron mass.

The exact ${\cal O}(\alpha)$ soft plus virtual corrections to the Bhabha scattering
have been taken from Ref.~\cite{Altarelli:1989hv,Greco:1987uw}. The soft plus virtual
cross section reads
\begin{eqnarray}
d\sigma^\alpha_{SV} &=&
d\sigma^{\alpha,s}_{SV}+d\sigma^{\alpha,t}_{SV}
+d\sigma^{\alpha,st}_{SV}\nonumber\\
d\sigma^{\alpha,i}_{SV}&=&d\sigma_0^i [2 (\beta + \beta_{int})
\log\varepsilon+C_F^i]
\label{exactsv}
\end{eqnarray}
where $i$ is an index for $s$, $t$ and $s$-$t$ subprocesses 
contributing to the 
Bhabha cross section, $\beta = \frac{2\alpha}\pi[\log(s/m^2)-1]$, $\beta_{int} =
\frac{2\alpha}\pi\log(t/u)$ and the explicit expression for $C_F^i$
can be found in Refs.~\cite{Altarelli:1989hv,Greco:1987uw}. We notice that in
Equation~(\ref{exactsv}) the terms coming from $s$, $t$ and $s$-$t$
interference diagrams are explicitly given.

The above discussion concerns the subset of photonic RCs. In \babayaga, any leptonic
RCs due to photon vacuum polarization effects are included
in the building-block matrix elements at tree-level and one-loop order,
paying attention to the correct cancellation of infra-red divergencies,
as described in more detail in Ref.~\cite{Balossini:2006wc}.

\subsection{Future improvements and upgrades for high energies}
\label{sec:high-energy}
As stressed in the previous sections, \babayagaNLO\ is
tailored to energies up to 10 GeV, where the bulk of RCs is due to
QED radiation. We give here a list of needed and desirable
upgrades of the code which will improve its theoretical accuracy
and make it suitable for simulations at higher energies:
\begin{enumerate}
\item running at higher energies will require the inclusion of $Z$
exchange diagrams for
$e^+e^-$ and $\mu^+\mu^-$ final states~\footnote{These are already included in the
tree-level matrix elements, but not in higher-order corrections.},
together with the full set of NLO electroweak corrections. A
first step towards this goal has been taken in
Ref.~\cite{CarloniCalame:2019dom}, where the process
$e^+e^-\to\gamma\gamma$ has been considered as a luminosity monitoring
process at FCC-ee energies and a phenomenological study of the effects
induced by the complete one-loop RCs in the Standard Model
is presented. We remark that the framework of the matching algorithm described in
Sec.~\ref{sec:matching} has been succesfully adopted also for charged-
and neutral-current Drell-Yan processes in
Refs.~\cite{CarloniCalame:2006zq,CarloniCalame:2007cd} and for Higgs decay
into four leptons~\cite{Boselli:2015aha}, 
including NLO electroweak corrections. The same theoretical basis can be in
the future applied to Bhabha scattering and $e^+e^-\to\mu^+\mu^-$ for simulations at
high-energy $e^+e^-$ machines with \babayagaNLO;
\item a possible improvement to better control higher-order
  corrections is the implementation into a Parton Shower algorithm of
  the NLO initial conditions discussed in
  Ref. ~\cite{Frixione:2019lga} and NLL evolution as detailed in Refs.~\cite{Bertone:2019hks,Frixione:2021wzh};
\item one of the developments foreseen for the future is the generalization
of the matching algorithm to include also exact NNLO RCs in QED. This
will allow to improve the theoretical accuracy of the generator,
presently at the ${\cal O}(0.1\%)$ level, to meet the precision requirements for
measurements at future machines (see for instance
Refs.~\cite{TLEPDesignStudyWorkingGroup:2013myl,Proceedings:2019vxr,Blondel:2018mad}). The
parallel efforts on the MUonE
project~\cite{Alacevich:2018vez,Banerjee:2020tdt,CarloniCalame:2020yoz,Budassi:2021twh}
can help towards the achievement of the goal.  
\end{enumerate}


\section{Theoretical predictions for $\Pe^+\Pe^-\to \PW^+\PW^-\to 4f$
\label{sec:racoon}}
\noindent
\begin{center}
\emph{Ansgar Denner, Stefan Dittmaier}
\end{center}

\subsection{Introduction}
\label{sec:introdd}

Future $\Pe^+\Pe^-$ colliders, either realized as linear
or circular collider, will offer fantastic opportunities to perform
high-precision measurements to challenge the validity of the
Standard Model (SM) and to identify possible deviations from SM predictions
even if the Large Hadron Collider (LHC) does not discover any new particles
(see, e.g., \cite{EuropeanStrategyforParticlePhysicsPreparatoryGroup:2019qin} 
and references therein).
The investigation of W-boson pairs is one of the cornerstones in this
vision of ``discovery via precision''.
For instance, from a scan of the total WW~production cross section
over its threshold, the W-boson mass $\MW$ could be determined with a
precision of 3\,MeV and even of 0.5--1\,MeV at the linear and circular
collider options ILC~\cite{LCCPhysicsWorkingGroup:2019fvj} and FCC-ee~\cite{TLEPDesignStudyWorkingGroup:2013myl,dEnterria:2016sca}, respectively.
Moreover, the large number of $\sim10^7$--$10^8$ W~pairs
expected to be produced at these colliders would turn the investigation of differential WW cross sections at higher energies into high-precision physics.
In particular, such analyses would enormously tighten
the limits on anomalous gauge-boson self-interactions,
which are nowadays quoted in terms of limits on Wilson coefficients of
effective dimension-6 operators in SM Effective Field Theory.
On the theory side, the full exploitation of this physics potential is
a great challenge as well (see, e.g., \cite{Freitas:2019bre} and references therein).

The purpose of this contribution is to describe the salient steps of
increasing sophistication in the
theoretical predictions for $\Pe^+\Pe^-\to \PW^+\PW^-\to 4f$ as, 
starting from the status of predictions for the Large Electron--Positron collider (LEP),
in which radiative corrections were taken into account in the form of
resonance expansions, and leading to the current state of the art,
where differential next-to-leading-order (NLO) predictions based on full
matrix elements exist for all relevant scattering energies and 
dedicated cross-section predictions near threshold
via an effective field theory (EFT) for including corrections at and beyond NLO.
Finally, we specify further improvements that are necessary to account for the
precision expected for future $\Pe^+\Pe^-$ colliders.

\subsection{Predictions based on the double-pole approximation}

In the course of phase~2 of LEP,
the theoretical description of charged-current four-fermion production in $\Pe^+\Pe^-$ collisions
turned from pure lowest-order predictions into systematic precision calculations
with a proper inclusion of the W-boson decay subprocesses and NLO corrections
to all stages of the full resonance process.
A major step towards percent-level accuracy in the cross-section
 prediction for the LEP2 energy range
was made by the concept of a systematic expansion about the two W-boson resonances,
leading to the so-called {\it double-pole approximation} (DPA).
The DPA decomposes, in a gauge-invariant way, the full resonance process into production
and decay subprocesses, which each receive their own {\it factorizable corrections} and
are linked by {\it non-factorizable corrections} due to soft photon
exchange.

In more detail, the contributions of the non-factorizable corrections in DPA, which
result from the presence of IR singularities, were systematically
studied in \cite{Denner:1997ia,Beenakker:1997bp,Beenakker:1997ir}
and generalized to other processes in
\cite{Accomando:2004de,Dittmaier:2015bfe}.
The factorizable corrections consist of the corrections to the
subprocesses of on-shell W-pair production~\cite{Bohm:1987ck,Fleischer:1988kj}
and on-shell W~decay~\cite{Denner:1990tx}, properly taking into account W~spin correlations.
The DPA was formulated and evaluated by different groups in different
variants~\cite{Beenakker:1998gr,%
Jadach:1996hi,%
Denner:2000bj,Denner:2002cg,%
Kurihara:2001um},
as e.g.\ reviewed in \cite{Grunewald:2000ju},
but only the two Monte Carlo programs {\sc YFSWW}~\cite{Jadach:1996hi}
and {\sc RacoonWW}~\cite{Denner:1999gp,Denner:2000bj,Denner:2002cg}
were used in the final LEP2 WW cross-section analyses~\cite{Schael:2013ita}.
Theoretical arguments as well as comparisons between the different DPA variants suggested
that sufficiently above the WW threshold a $\sim0.5\%$ accuracy could be achieved
for LEP2 energies ($\sqrt{s}\sim165{-}210\,$GeV). 

In {\sc RacoonWW} only the virtual one-loop corrections are treated in
the DPA approximation, while the NLO real corrections are taken into
account exactly. Real and virtual corrections are combined using
either two-cutoff phase-space slicing or the dipole subtraction method
for photon radiation \cite{Dittmaier:1999mb}. 
Corrections from initial-state radiation are resummed to all orders
 at the leading-logarithmic accuracy
via the structure-function approach
\cite{Berends:1987ab,Beenakker:1996kt}, which is based on the
collinear approximation improved by soft-photon exponentiation.
Moreover, off-shell effects of the Coulomb
singularity \cite{Fadin:1993kg,Bardin:1993mc},
resulting from the exchange of virtual photons between the produced
W~bosons, are included.

The Monte Carlo program {\sc YFSWW}~\cite{Jadach:1996hi} is based on
NLO corrections to on-shell W-pair production, improved by
YFS exponentiation for the electromagnetic corrections including also
LL final-state W-decay radiative effects
\cite{Jadach:1996hi,Jadach:1998tz}. 
The NLO cross section for on-shell
W-pair production after subtraction of photonic corrections is dressed
by photon radiation in the YFS approach. Leading logarithmic
corrections to W~decays are included via PHOTOS and non-factorisable
corrections only approximately via a screened Coulomb ansatz.
Spin correlations are fully included at leading order (LO) and approximatively at NLO.

Predictions based on the DPA have limitations both at low and high energies.
Directly around the WW threshold the DPA is not applicable, leaving
predictions with a $\sim2\%$ accuracy from some {\it improved Born
  approximation} (IBA) \cite{Dittmaier:1991np,Denner:2001zp}, which is based
on universal corrections that can be easily incorporated into the
structure of the off-shell LO matrix elements.
For higher energies relevant for future linear colliders
($\sqrt{s}>500\,$GeV) a deterioration of the DPA was expected due to
the increasing relevance of so-called (singly-resonant) background
diagrams. Finally, the IBA suffers
from the lack of electroweak logarithms at high energies.

\subsection{Predictions based on full matrix elements}

The situation significantly improved with the completion of the NLO
calculation~\cite{Denner:2005es,Denner:2005fg} for the full $2\to4$ process 
$\Pe^+\Pe^-\to \PW^+\PW^-\to4f$ and its implementation in {\sc Racoon4f}, 
supporting NLO accuracy both in resonant and non-resonant regions.
The precision of this NLO calculation in integrated
cross sections was estimated to be $\sim0.2\%$ in the whole LEP2
energy range and to be $\sim0.5\%$ up to energies in the TeV range.
Moreover, the uncertainty estimate of the DPA was nicely confirmed by
this more precise calculation (see Fig.~\ref{fig:XSeeWW} below).

The full NLO calculation required significant technical and conceptual
progress in two directions: The first concerns the gauge-invariant
treatment of resonances in NLO accuracy in resonance and off-shell
regions and in the region in between.  To this end, the so-called {\it
  complex-mass scheme} (CMS) was suggested in \cite{Denner:2005fg}
(more details can be found in \cite{Denner:2006ic,Denner:2019vbn}),
which introduces complex W- and Z-boson masses and analytically
continues couplings in such a way that all gauge-invariance relations
are maintained (Ward identities, gauge-parameter cancellation). In the
CMS perturbative calculations proceed as usual, i.e.\ amplitudes are
calculated order by order, but with complexified renormalization
prescriptions and with complex masses in loop integrals. The fast and
numerically stable evaluation of the multi-leg one-loop diagrams (with
up to six-point amplitudes) was the second type of complication that
required significant technical progress.  This problem was solved by
the introduction of new reduction techniques for one-loop tensor
integrals~\cite{Denner:2005nn} that avoid the appearance of dangerous
inverse kinematical determinants (mostly of Gram type) of external
momenta.  To achieve this, the traditional 
Passarino--Veltman tensor reduction~\cite{Passarino:1978jh}
was complemented by dedicated expansions of tensor coefficients
about exceptional phase-space points where those dangerous
determinants vanish. Further optimizations and improvements of this
technique lead to the development of the public {\sc Fortran} library
{\sc Collier}~\cite{Denner:2016kdg} which evaluates arbitrary one-loop
tensor integrals with complex masses and various regularizations up to
high tensor ranks (without any hard limit).
More details on these one-loop concepts and techniques can also be found in the
review~\cite{Denner:2019vbn}.

A comparison of predictions for the inclusive
W-pair production cross section for centre-of-mass energies ranging from the LEP2 energy
range up to $2\,$TeV is shown in Fig.~\ref{fig:XSeeWW}.
\begin{figure}
  \centering
     \includegraphics[scale=.6]{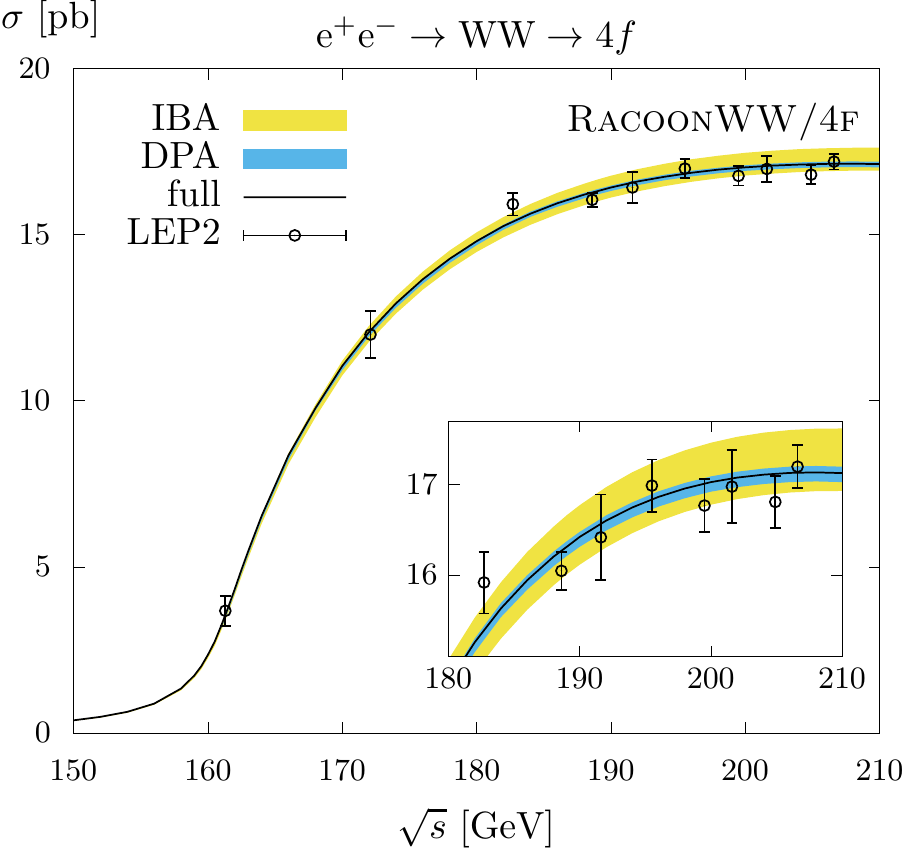}
\hspace*{1em}
     \includegraphics[scale=.6]{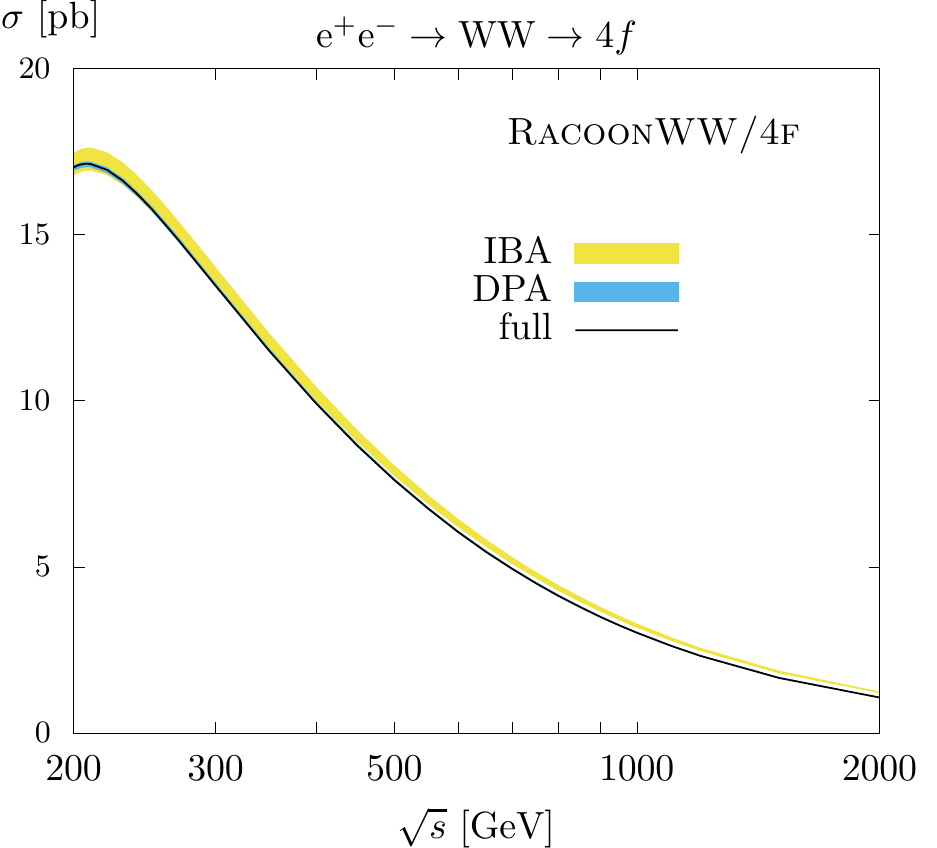} \\
     \includegraphics[scale=.6]{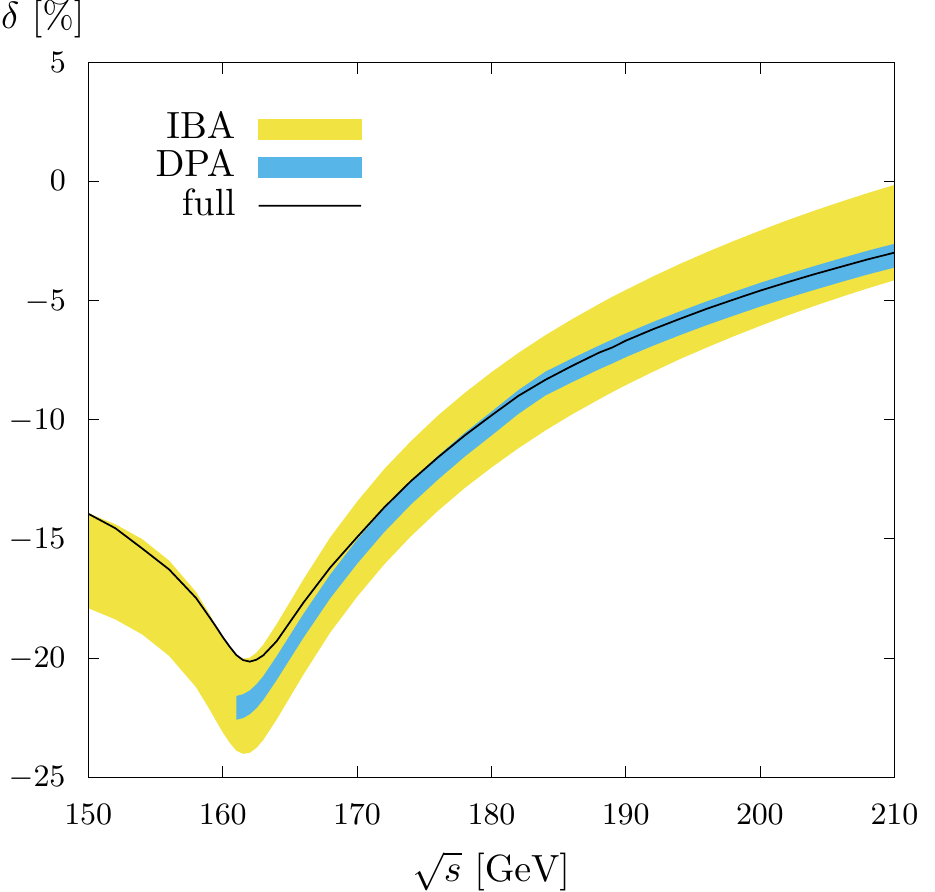}
\hspace*{1em}
     \includegraphics[scale=.6]{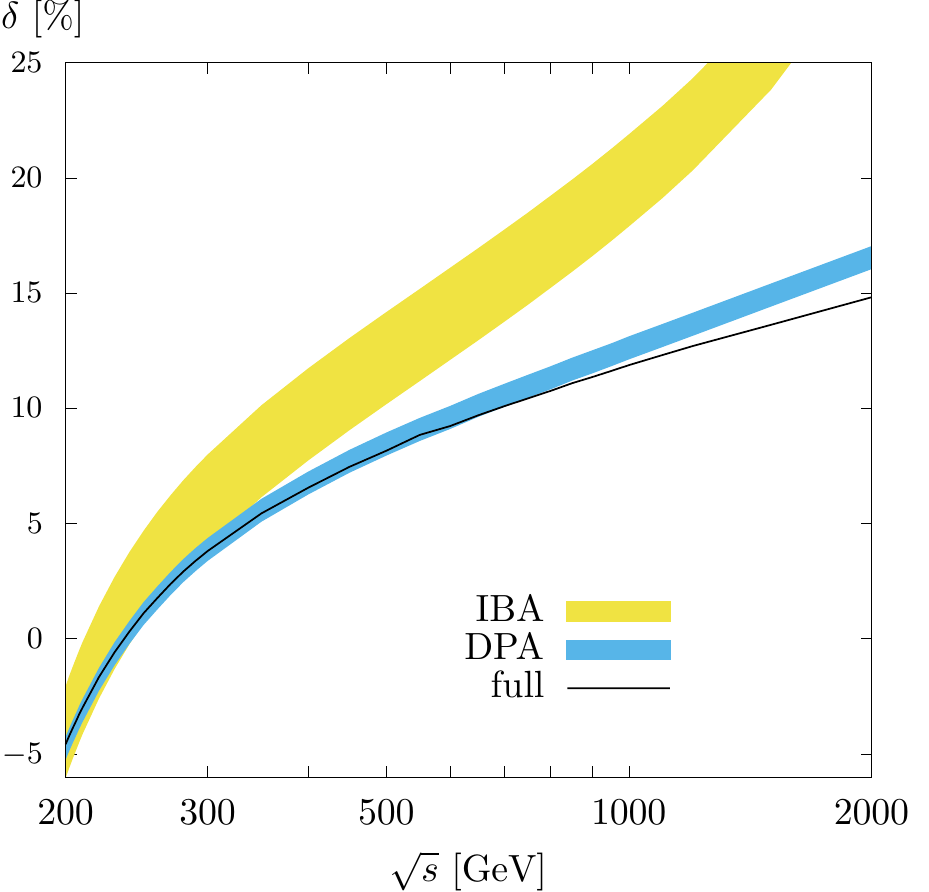} \\
\vspace*{-.5em}
  \caption{Inclusive cross section  for $\Pe^+\Pe^-\to \PW^+\PW^-\to 4f$ 
    as obtained from {\sc RacoonWW/4f}, including NLO EW corrections in
    IBA, in DPA~\cite{Denner:2000bj}, 
    and fully off shell in the CMS~\cite{Denner:2005es,Denner:2005fg}. 
    The IBA and DPA bands illustrate the
    uncertainty of $\pm2\%$ and $\pm0.5\%$, as assessed for the
    LEP2 energy range.  LEP2 cross-section measurements
    \cite{Schael:2013ita} are shown as data points.
    (Plots taken from \cite{Denner:2019vbn}.)}
\label{fig:XSeeWW}
\end{figure}
The results include NLO EW corrections based on an IBA, the DPA of
{\sc RacoonWW}, and the full off-shell calculation of
\cite{Denner:2005es,Denner:2005fg} as implemented in {\sc Racoon4f}. 
All cross sections are obtained in the $G_\mu$ scheme
\cite{Denner:2019vbn}, where the electromagnetic coupling is
determined from the Fermi constant. Leading higher-order effects from
initial-state radiation (ISR) are incorporated in all versions.  In the LEP2
energy range, the IBA, the DPA, and the full off-shell calculations
agree within the expected uncertainties.%
The agreement with LEP2 data~\cite{Schael:2013ita}, which are included
in the plot, with theory predictions required the inclusion of
non-universal EW corrections, as provided by the DPA.  For energies
above the LEP2 energy range, the error assessments of $\pm2\%$ and
$\pm0.5\%$ for IBA and DPA, respectively, fail.  For
$\sqrt{s}\gtrsim300{-}400\,$GeV, logarithmically enhanced negative
high-energy EW corrections appear, which are not incorporated in the
IBA, but still by the DPA. Those corrections reach $\sim-10\%$ and
more in the TeV range for total cross sections and are even larger in
differential distributions.  The difference of some percent between
DPA and full off-shell result, on the other hand, is due to the
increasing contribution of background diagrams, which are included
only at LO but not in the DPA calculation of the virtual corrections.

\subsection{Near-threshold predictions from EFT}

In \cite{Beneke:2007zg,Actis:2008rb} an EFT approach was devised to
efficiently compute the $\Pe^+\Pe^-\to4\,$fermions cross section close
to the W-pair production threshold, where $\sqrt{s}-2 \MW \sim
\GamW$. 
The EFT exploits the scale hierarchy in the threshold regime,
$\MW \gg \MW v \gg \MW v^2 \sim \GamW \sim \MW\alpha$, where $v$ is the typical velocity of the W bosons in the centre-of-mass
frame, $\GamW$ is the W decay width, and $\alpha$ represents the electroweak
coupling constant.
It builds on non-relativistic QED (NRQED)~\cite{Pineda:1997bj} to describe the dynamics of the (slightly) off-shell W bosons, on soft-collinear effective theory (SCET)~\cite{Bauer:2000yr} to describe photonic ISR, and on unstable-particle EFT~\cite{Beneke:2003xh} to describe the W decay.

In the EFT the cross-section calculation is organized as a consistent
(``threshold'') expansion in the small parameter $\delta \sim \alpha
\sim v^2$. This corresponds to a simultaneous expansion in the small parameters $v$ and $\alpha$.
The parametric scaling of the 
LO cross section
in the EFT is $\sigma_\eff^\mathrm{LO} \sim \alpha^2 v \sim \delta^{5/2}$.
In accordance with the (N$^n$LO, $n=1,2,3,\ldots$) terminology of the full-theory perturbative expansion in $\alpha$,
the higher-order corrections in the threshold expansion with relative scaling factors $\delta^{n/2}$ are named  ``N$^{n/2}$LO''.

The relevant field modes, i.e.\ the resonant degrees of freedom, (and their typical four-momenta $k^\mu$) of the EFT are:
``hard'' ($k^\mu \sim \MW$), ``soft'' ($k^\mu \sim \delta^{1/2}\MW$),
``ultrasoft'' ($k^\mu \sim \delta \MW$), ``potential'' ($k_0 \sim \delta \MW$, $|\vec{k}| \sim \delta^{1/2} \MW$), or ``collinear'' ($k_0 \sim \MW$, $k^2 \sim \delta \MW^2$).
The non-relativistic W bosons are, for example, represented in the EFT by potential fields, while the incoming electrons and positrons have collinear momenta pointing in opposite directions.
Every (potential) interaction due to exchange of a ``Coulomb'' photon between the non-relativistic W bosons yields a relative correction of order $\alpha/v \sim \delta^{1/2}$ to the cross section near threshold.

In \cite{Beneke:2007zg} the EFT calculation of the total cross section was carried out to NLO in the threshold expansion and compared to the full-theory NLO result of \cite{Denner:2005es,Denner:2005fg}. The observed deviations are in agreement with the expected size of the higher-order corrections in the $v$ expansion missing in the EFT result. Still, a more detailed comparison including further refinements on the full-theory side is desirable
(see also \cite{Schwinn:2019qbp}).

As a first step beyond full NLO the ``dominant NNLO'' corrections,
i.e.\ the terms of order N$^{3/2}$LO in the threshold expansion that
are contained in the full NNLO (and N$^3$LO), were computed in
\cite{Actis:2008rb}. 
Individual contributions to these corrections were found to affect the
outcome of the future W-mass measurement from a threshold scan by
about 5\,MeV. They combine to an overall effect on the measured W-boson
mass in the same ballpark.
It was also pointed out that the QED ISR encoded in the electron/positron structure functions has a large effect. The ISR-related large logarithms must therefore be resummed at least at next-to-leading logarithmic (NLL) accuracy, while the state-of-the art by the time of \cite{Actis:2008rb} was only LL precision.
On the other hand, the effect of loose
invariant-mass cuts on the decay products of the W bosons, as may be required by experiment, was argued to be negligible.

To consolidate the findings of \cite{Actis:2008rb} and to further decrease the theoretical 
uncertainties, all NNLO threshold corrections that are not already contained in the full NLO result of the total cross section have to be calculated.
Moreover, the treatment of ISR effects should be improved 
to a level of precision sufficient for future W-mass measurements.

\subsection{Outlook}
Addressing the fantastic precision expected from the next generation
of $\Pe^+\Pe^-$ colliders on the theory side poses a great challenge to
predict W-pair production at the highest possible precision in a
twofold way.  Firstly, the prediction of the total cross section near
threshold that is required for the $\MW$ determination has to be
precise within about $0.01$--$0.05\%$ while the current uncertainty is
$\sim0.2\%$ resulting from a NLO calculation for the full $2\to4$
particle process, or alternatively, from a dedicated higher-order
calculation at the WW~threshold within a non-relativistic EFT for
unstable particles.  

Going beyond this current precision is a major
effort and requires the calculation of various corrections beyond NLO.
On the one hand, the EFT approach for a cross-section prediction at
the WW threshold has to be pushed to an unprecedented level of
precision, forcing us to significantly extend the level of general
understanding of this EFT.  
On the other hand, various refinements are
necessary in current Monte Carlo generators for the process class
$\Pe^+\Pe^-\to4\,$fermions, such as {\sc Racoon4f}, which is the successor
of the public Monte Carlo program {\sc
  RacoonWW}~\cite{Denner:1999gp,Denner:2000bj,Denner:2002cg}, that
incorporates the current state-of-the-art NLO
prediction~\cite{Denner:2005es,Denner:2005fg}.  
While the calculation of \cite{Denner:2005es,Denner:2005fg} makes use
of ``generic'' leptonic, semileptonic, and hadronic $4f$~final states,
more refined predictions should make use of all individual $4f$~final states,
properly taking into account all interferences 
with $t$-channel diagrams in channels with  $\Pe^\pm$ in the final states
and with Z-pair production channels.
Further improvements are needed on the side of QCD corrections at the
differential level and beyond NLO, which is more challenging.

Finally, for a unified
approach for the full energy region ranging from threshold to the TeV range,
it will be necessary to implemented the higher-order EFT
corrections for the improved threshold prediction in Monte Carlo
generators and to refine and improve the calculation of differential
cross sections above threshold to increase the achieved accuracy at
higher energies.

\acknowledgments
Max Stahlhofen is gratefully acknowledged for discussions about
the EFT approach for threshold predictions.



\section{CEEX and EEX Realizations of the YFS MC approach to precision 
theory for $e^+e^-$ Colliders\label{sec:YFS}}
\begin{center}
\emph{S. Jadach, B.F.L. Ward}
\end{center}

\subsection{The YFS Approach to Precision Theory for $e^+e^-$ Colliders}
The exact amplitude-based  CEEX/EEX YFS MC approach to EW higher order corrections is pioneered in Refs.~\cite{Jadach:1993yv,Jadach:1999vf,Jadach:2000ir,Jadach:2013aha}. Here, EEX denotes exclusive exponentiation as originally formulated by Yennie, Frautschi and Suura (YFS) in Ref.~\cite{yfs:1961} and it is effected at the squared amplitude level. CEEX denotes the coherent exclusive exponentiation developed in Refs.~\cite{Jadach:1999vf,Jadach:2000ir,Jadach:2013aha} in which IR singularities are resummed at the level of the amplitude. We, with our collaborators, have developed and implemented several MC event generators which realize the YFS MC approach for EW higher order corrections in precision physics for $e^+e^-$ colliding beam devices: KORALZ(EEX 2f production in $e^+e^-$ annihilation)~\cite{Jadach:1993yv}, BHLUMI(EEX Bhabha scattering at small angles)~\cite{bhlumi2:1992,bhlumi4:1996}, BHWIDE(EEX Bhabha scattering at wide angles)~\cite{bhwide:1997}, KKMC-ee(CEEX/EEX 2f production in $e^+e^- annihilation)$~\cite{Jadach:1999vf,Jadach:2013aha}, YFSWW3(EEX W pair production and decay in $e^+e^-$ annihilation)~\cite{yfsww3:2001}, KORALW(EEX 4f production in $e^+e^-$ annihilation) ~\cite{koralw:1998}, KORALW\&YFSWW3(EEX Concurrent 4f background processes and 1st order W pair production in $e^+e^-$ annihilation))~\cite{kandy-2001}, and YFSZZ(EEX Z pair production and decay in $e^+e^-$ annihilation)~\cite{yfszz:1997}-- see Ref.~\cite{MC2000} for succinct descriptions of these MC's. All of these MC's, except perhaps for KORALZ, which has been superseded by KKMC-ee, are candidates for appropriate upgrades to meet the new precision requirements for the future $e^+e^-$ colliding beam devices. We now give a brief review of the physics in the YFS MC approach in this context.

Since it is still not generally used, we first recall the master formula for the CEEX/EEX realization of the higher corrections to the SM~\cite{SM1,SM2,SM3,SM4} EW theory. 
For the purpose of illustration, let us consider the prototypical process 
$e^+e^-\rightarrow f\bar{f}+n\gamma, \; f = \ell, q, \; \ell=e,\mu,\tau,\nu_e,\nu_\mu,\nu_\tau, \; q = u,d,s,c,b,t.$  For this process, we have the cross section formula
\begin{equation}
\sigma =\frac{1}{\text{flux}}\sum_{n=0}^{\infty}\int d\text{LIPS}_{n+2}\; \rho_A^{(n)}(\{p\},\{k\}),
\label{eqn-yfsmth-1}
\end{equation}
where $\text{LIPS}_{n+2}$ denotes Lorentz-invariant phase-space for $n+2$ particles, $A=\text{CEEX},\;\text{EEX}$, the incoming and outgoing fermion momenta are abbreviated as $\{p\}$ and the $n$ photon momenta are denoted by $\{k\}$.
Thanks to the use of conformal symmetry, full $2+n$ body  phase space is covered without any
approximations. The respective algorithm's details are covered in Ref.~\cite{Jadach:1999vf}.
Specifically, we have from Refs.~\cite{Jadach:2000ir,Jadach:1999vf} that 
\begin{equation}
\rho_{\text{CEEX}}^{(n)}(\{p\},\{k\})=\frac{1}{n!}e^{Y(\Omega;\{p\})}\bar{\Theta}(\Omega)\frac{1}{4}\sum_{\text{helicities}\;{\{\lambda\},\{\mu\}}}
\left|\Meu\left(\st^{\{p\}}_{\{\lambda\}}\st^{\{k\}}_{\{\mu\}}\right)\right|^2.
\label{eqn-yfsmth-2}
\end{equation}
(See Refs.~\cite{Jadach:2000ir,Jadach:1999vf} for the corresponding formula for the $A=\text{EEX}$ case.) Here,  $Y(\Omega;\{p\})$ is the YFS infrared exponent. The respective infrared integration limits are specified by the region $\Omega$ and its characteristic function
$\Theta(\Omega,k)$ for a photon of energy $k$, with $\bar\Theta(\Omega;k)=1-\Theta(\Omega,k)$ and $$\bar\Theta(\Omega)=\prod_{i=1}^{n}\bar\Theta(\Omega,k_i).$$
See Refs.~\cite{Jadach:1999vf,Jadach:2000ir,Jadach:2013aha} for the definitions of the latter functions as well as the CEEX amplitudes~$\{\Meu\}$. 
\KK MC 4.22~\cite{Jadach:2013aha} featured the exact
${\cal O}(\alpha)$ EW corrections implemented using the  DIZET {6.21,...,6.42} EW libraries from the semi-analytical
program ZFITTER~\cite{zfitter1,zfitter6:1999,zfitter:2006,dizet642}.
Recent public versions of \KK{MC} include also the DIZET 6.45 version,
see \cite{Arbuzov:2020coe}.
The respective implementation is described in Ref.~\cite{Jadach:2000ir} so that we do not repeat it here.
In {\KK}MC-ee, the CEEX amplitudes $\{\Meu\}$ in
Eqs.~(\ref{eqn-yfsmth-1},\ref{eqn-yfsmth-2}) are exact 
in ${\cal O}(\alpha^2 L^2, \alpha^2L)$ in the sense that all terms in the respective cross section at orders ${\cal O}(\alpha^0),\;{\cal O}(\alpha),\; {\cal O}(\alpha L),\;{\cal O}(\alpha^2 L), \; \text{and}\; {\cal O}(\alpha^2 L^2)$ are all included in our result for that cross section.
Here the big log is $L=\ln\frac{Q^2}{m^2}$ where
$Q$ is the respective hard 4-momentum transfer. In our case, the charged lepton masses and the quark masses will determine $m$, depending on the specific process
under consideration.
We follow Ref.~\cite{mstw-mass} and use the current quark masses~\cite{PDG:2016} $m_u = 2.2 \text{MeV}.\; m_d = 4.7 \text{MeV}, \;  m_s = 0.150\text{GeV}, \; m_c = 1.2 \text{GeV},\;  m_b = 4.6\text{GeV} \text{and}\; m_t = 173.5 \text{GeV}$\footnote{See Ref.~\cite{kkmchh1} for a relevant discussion of the uncertainty of our results due to realistic uncertainties on our values of the current quark masses - we find the uncertainties due to the choices of the quark masses are approximately equal to 10\% of the overall corrections.}. We note for completeness that in our MC's all real and soft virtual photonic corrections have
$\alpha=\alpha(0)=\frac{1}{137.035999...}$, since photons are massless.
For hard QED corrections, we use $\alpha=\alpha(Q),$\footnote{This is the analogue of $\alpha(M_Z)$ but defined at the scale $Q$.} with the attendant hadronic vacuum polarization taken after Ref.~\cite{fjeger-fccwksp2019}.

We note that the EEX realization in {\KK}MC-ee includes as well the exact ${\cal O}(\alpha^3 L^3)$ corrections. The user always has the option to switch on this correction as needed.
For an overview on the other programs listed above which feature EEX see Refs.~\cite{Kobel:2000aw} and \cite{Grunewald:2000ju}. As we noted,
most of the latter programs may serve as a starting point for the future development. However, as an introductory step they have to 
be translated 
from Fortran77 to a modern programming language.
Presently {\KK}MC-ee is already fully translated into C++ and the BHLUMI translation is in the process.

It is worth to stress that the \KK{MC} program is unique 
concerning the spin polarization treatment. 
It is the only one which implements the full spin density matrix formalism
(transverse and longitudinal) for both incoming beams and outgoing fermions
in the presence of higher order corrections,
any number of photons, all over the phase space.

The anticipated new precision requirements of the future $e^+e^-$ colliding beam devices,
as summarized in Ref.~\cite{Jadach:2019bye}, are at least a of factor 10 and often a factor 100 higher
than any present MC programs can provide.
The path of the development of new calculations in the form of the MC event generators was
already outlined, see Refs.~\cite{Jadach:2019bye}.
The improvement plan for the low angle Bhabha process 
is outlined Refs.~\cite{Jadach:2018jjo,Jadach:2021ayv,Blondel:2018mad}.
A new variant of the CEEX scheme for the production and decay of the charged
particles implementable in the MC programs, like what one needs for WW pair production, was the scenario outlined in Ref.~\cite{Jadach:2019wol}.

Generally we expect to move most of these MC's to the CEEX paradigm to take advantage of its efficiency in simulation interference effects especially near resonance thresholds.
The other feature of the CEEX scheme underlined in Ref.~\cite{Blondel:2018mad}
is that it provides a flexible and consistent scheme of combining 
electroweak (EW) and 
non-soft QED corrections in different orders. This feature is critical for
achieving precision goals of the future electron colliders where typically EW corrections
have to be completed to second order while QED corrections are needed up to third or fourth order
plus soft photon resummation\footnote{While it is true that the CEEX methodology is mainly for/on QED, it 
includes a very valuable thing concerning the so-called genuine EW corrections:
so far it is the only technique which offers a consistent (gauge 
invariant) and practical method of building matrix elements (spin 
amplitudes) for the complete (photonic) QED corrections up to 
${\cal O}(\alpha^2)$ and EW corrections up to ${\cal O}(\alpha^1)$ only.
As stressed and described in Ref.~\cite{Blondel:2018mad}, this 
methodology extends to complete QED corrections up to ${\cal O}(\alpha^3)$ (with resummation) and  EW corrections up to ${\cal O}(\alpha^2)$.
}.

As we have already anticipated in Ref.~\cite{radcor2021-bw}, with the appropriate CEEX realizations, we expect to be able to reach the desired precision expectations for
all of the respective processes addressed by the aforementioned MC's. 
The above plans and expectations assume the availability of the proper financial support.

\subsection{Comparison with the Collinear Factorisation Approach}

The YFS MC approach which we presented above should be compared with the collinear factorisation approach~\cite{fad-kur:1985,altarelli-mart:1986,nicro-trent:1987,fad-khz:1987,berends-neerver-burgers:1988,bluemlein-freitas-vnNeervn:2011,frixione-2019,bertone-2019,frixione-2021} which for Z physics has been reviewed also in Ref.~\cite{bardin-zplep1:1989}. Indeed, in Refs.~\cite{bhabhacern-lep2:1996}, detailed comparisons were made 
between BHLUMI, featuring the EEX YFS MC approach to the low angle Bhabha scattering process which was used to determine the luminosity at LEP, and SABSPV~\cite{sabspv:1995} which also treated the same process to a comparable over-all normalization accuracy using the collinear factorisation approach. A related set of comparisons at low energy were made featuring BHWIDE and BaBaYaga~\cite{babayaga-2019} in Ref.~\cite{Actis:2010gg}. From these comparisons we can draw some
general conclusions as we now discuss\footnote{Since the calculation of any specific process involves some components that are generic to the $e^+e^-$ annihilation environment and some components that are specific to the process under study, there is no contradiction when one says that they will draw general conclusions about the comparison of the two approaches from studies done with a specific process.}.

For definiteness, let us write down the generic collinear factorisation representation of the same process that we 
considered above: $e^+e^-\rightarrow f\bar{f}+n\gamma, \; f = \ell, q, \; \ell=e,\mu,\tau,\nu_e,\nu_\mu,\nu_\tau, \; q = u,d,s,c,b,t.$ We have
\begin{equation}
\sigma =\int dx_1dx_2\sum_{ij} f_i(x_1)f_j(x_2)\sigma_{ij}(Q^2)\delta(Q^2-x_1x_2s),
\label{eqn-yfsmth-3}
\end{equation}
where the sub-process cross section for the $ij$ parton-parton interaction, where $i$($j$) denotes a parton in
the incoming $e^-$($e^+$) beam, with $\hat{s}=Q^2$ when the $e^+e^-$ cms energy squared is $s$,
is denoted in a conventional notation for parton densities $\{f_j\}$. Here, $\sigma_{ij}(Q^2)$ is given by the 
appropriate expression as dictated by the respective QED electron collinear factorisation calculus as described in Refs.~\cite{fad-kur:1985,altarelli-mart:1986,nicro-trent:1987,fad-khz:1987,berends-neerver-burgers:1988}.

Already by comparing Eq.(\ref{eqn-yfsmth-1}) and Eq.(\ref{eqn-yfsmth-3}) we see one important advantage of the YFS MC approach: the exact phase space for the multiple photon radiation is realized on an event by event basis to all orders in 
$\alpha$ whereas in the collinear factorisation approach the radiation transverse degrees of freedom that have been integrated to reach the 1-dimensional QED PDF's have to be restored and this restoration is inherently approximate, as it was illustrated in Ref.~\cite{bhabhacern-lep2:1996},for example\footnote{In other words, the distributions which the QED collinear factorisation approach produces are not exact for the transverse degrees of freedom which were integrated out to arrive at the collinear limit represented by the QED PDF's while our distributions are exact in these degrees of freedom. We have seen in the LEP studies~\cite{bhabhacern-lep2:1996} that the detailed measurements of the exclusive photon distributions show this deviation from exactness.}.

Application of Eq.~(\ref{eqn-yfsmth-3}) is limited to ``academic observables'' 
with a cut-off on the total photon energy $E_{\max} = (sx_1x_2)^{1/2}$\footnote{We stress that in any real observable there is always multiple photon radiation to all orders in $\alpha$. Any fixed-order calculation thus is necessarily academic, and its usefulness has to determined on a case-by-case basis. In many cases, the effects of the multi-photons missing from the fixed-order result are small enough that the fixed-order result can be used to assess the data. Even in the latter cases, the lack of exactness of the treatment of the transverse degrees of freedom in Eq.~(\ref{eqn-yfsmth-3}) limits its applicability.}.
All realistic experimental observables select events using acollinearity
and other similar cuts depending on photon momenta in a complicated way.
Only a Monte Carlo with full multiple photon phase space can 
provide predictions for the real experiments.
On the other hand, variants of Eq.~(\ref{eqn-yfsmth-3}) with added subleading
corrections in $f_l$ and $\sigma_{ll'}$ are quite useful in testing/calibrating
Monte Carlo programs.
For instance the BHLUMI program includes the QED collinear factorisation
based program LUMLOG, while \KK{MC} provides the KKsem and KKfoam auxiliary programs,
which serve for testing/calibrating the main multiphoton generator,
albeit for academic observables.

\subsection{Summary}
We conclude with the following observation: {\KK}MC-ee is the unique $e^+e^-$ annihilation MC which features CEEX exact ${\cal O}(\alpha^2 L)$ corrections and control over the corresponding EW initial-final interference (IFI) effects as well the exact ${\cal O}(\alpha)$ pure weak corrections and complete 
spin effects both for the initial beams and outgoing fermions.
This uniqueness is important: it proves that we have a unique set of the tools to extend the CEEX method to the other important processes in the future $e^+e^-$ colliders' precision physics programs in the effort to reach the new required precision for these processes.

\vskip 2 mm
\centerline{\bf Acknowledgments}
\vskip 2 mm

The authors thank Prof. G. Giudice for the support 
and kind hospitality of the CERN TH Department. 
S.J. acknowledges funding from the European Union’s Horizon 2020 research and innovation programme under grant agreement No 951754 and support of the National Science Centre, Poland, Grant No. 2019/34/E/ST2/00457.



\newcommand\mcmule{{\sc McMule}}
\section{NNLO QED calculations with \mcmule{}\label{sec:mcm}}
\begin{center}
\emph{Tim Engel, Marco Rocco, Adrian Signer, Yannick Ulrich}
\end{center}














\subsection{Introduction}

While a huge effort is ongoing to tackle next-to-next-to-leading order
(NNLO) corrections in QCD, fixed-order calculations in QED have
received comparatively little attention in the literature. From a
high-energy point of view this is understandable due to the numerical
importance of QCD corrections, driven by a coupling that is much larger
than the QED coupling $\alpha$. Hence, NNLO QED calculations are
typically considered in low-energy processes involving mainly
leptons. However, for electron-positron colliders initial-state QED
corrections are of utmost importance. Often such
corrections are taken into account to high orders in $\alpha$ but with
a soft and/or collinear approximation in the kinematics.

In this contribution we take a slightly different approach in that we
start from complete fixed-order NNLO QED calculations. We will
describe a modern approach for such calculations, based partly on the
developments made for QCD, and explore how these calculations can be
adapted to the case of a high-energy electron-positron collider. In
particular, we regularise all singularities with dimensional
regularisation in $d=4-2\epsilon$ dimensions and use a subtraction
method for the phase-space integration. This is a notable difference
to many QED calculations in the literature where a photon-mass
regularisation as well as a slicing method, separating photon
radiation into hard and soft, is used. We stress that the fact that
we have thus far considered only fixed-order calculations and neglected
all-order resummation effects does not imply that the latter cannot be
matched with the former -- in fact, we are now working towards this
very goal; we shall briefly comments on this in Section~\ref{sec:mcmoutlook}.

In contrast to most QCD calculations, for QED calculations it is
essential to keep non-vanishing fermion masses $m$. Since QED with
massive fermions has only soft singularities, this leads to a
tremendous simplification of the infrared (IR) structure. In
Section~\ref{sec:fksl} we will describe how this can be exploited to
devise an efficient subtraction scheme for phase-space integrations at
all orders in $\alpha$.

However, this simplification comes at a cost. The presence of fermion
masses introduces additional scales to the problem and leads to a
complication in the calculation of virtual amplitudes. In fact, often
it is not possible to obtain analytic expressions for two-loop
amplitudes with $m\neq 0$. Fortunately, in many cases fermion masses
are much smaller than the other kinematic variables $Q^2$. As
described in Section~\ref{sec:mass}, we can then use a procedure
dubbed massification. It allows to recover all not polynomially
suppressed mass effects of two-loop virtual amplitudes from the
corresponding amplitudes with a massless fermion.

Another technical difficulty related to small but non-vanishing
fermion masses is the presence of large logarithms $\ln(m^2/Q^2)$.
These are the remnants of collinear singularities regulated by $m\neq
0$. In massless QCD, only observables that are inclusive in
final-state collinear emission are physical, where these collinear
singularities cancel as $1/\epsilon$ poles.  Initial-state
collinear emissions are absorbed into the parton distribution
functions.  In QED, the $\ln(m^2/Q^2)$ are physical in that they can
be measured. Being less inclusive leads to numerical issues in
particular in the phase-space integration of the real-virtual
contributions. Rather than using pure numerical techniques or brute
force to address this issue, we use properties of radiative QED
amplitudes. More precisely, we approximate these amplitudes in the
soft limit at next-to-leading power to enable a stable numerical
integration. This is described in Section~\ref{sec:nts}.

The techniques outlined in this contribution are being implemented in
the publicly available Monte Carlo framework
\mcmule{}~\cite{mcmule:website}. One of the main motivations for these
developments is to obtain a full NNLO prediction of $\mu$-$e$
scattering. This process has gained a lot of
interest~\cite{Banerjee:2020tdt} in the context of the MUonE
experiment~\cite{MUonE:LoI}. Already some important contributions have
been calculated at NNLO~\cite{Banerjee:2020rww, CarloniCalame:2020yoz,
Fael:2018dmz, Fael:2019nsf, Budassi:2021twh} and the two-loop matrix
element is known with $m_\mu \gg m_e= 0$~\cite{Bonciani:2021okt}.
However, these techniques can also be used for processes at higher
energies.  Indeed, they form the basis for our exploration in
Section~\ref{sec:mcmoutlook} where we discuss first steps towards using
\mcmule{} at medium- and high-energy electron-positron colliders.

\subsection{The FKS\texorpdfstring{$^\ell$}{l} subtraction scheme}\label{sec:fksl}

As with all cross-section calculations beyond leading order in gauge
theories, also in QED with massive fermions IR singularities appearing
in the phase-space integration have to be isolated and properly
combined with those of virtual loop integration. What is particularly
simple in our case is that no collinear $1/\epsilon$ are present and
the soft singularities exponentiate. In other words, the $\ell$-loop
matrix element (squared) with $n$ external particles $\M{n}{\ell}$
satisfies the YFS formula~\cite{Yennie:1961ad}
\begin{align}
\sum_{\ell = 0}^\infty \M{n}{\ell} = e^{-\ieik}\, 
\sum_{\ell = 0}^\infty \fM{n}{\ell}
\label{eq:yfsnew}
\end{align}
where all $1/\epsilon$ IR poles of $\M{n}{\ell}$ are contained in the
integrated eikonal factor $\ieik$ and the remainders of the matrix elements,
$\fM{n}{\ell}$, are finite. This simple structure can be exploited to convert the 
FKS subtraction scheme~\cite{Frixione:1995ms,Frederix:2009yq} initially developed for
NLO QCD calculations to FKS$^\ell$~\cite{Engel:2019nfw} for QED
calculations at all orders. At NNLO for example, the double virtual,
real-virtual and double-real corrections read
\begin{subequations}
\label{eq:nnlo:4d}
\begin{align}
\begin{split}
\sigma^{(2)}_n(\xc) &= \int\!
   \D\Phi_n^{d=4}\,\bigg(
    \M n2
   +\ieik(\xc)\,\M n1
   +\frac1{2!}\M n0 \ieik(\xc)^2
\bigg) =
\int\! \D\Phi_n^{d=4}\, \fM n2(\xc)
\,,
\end{split}\label{eq:nnlo:n}
\\
\sigma^{(2)}_{n+1}(\xc) &= \int\!
 \D\Phi^{d=4}_{n+1}
  \cdis{\xi_1} \Big(\xi_1\, \fM{n+1}1(\xc)\Big)\label{eq:nnlo:n1}
\,,\\
\sigma^{(2)}_{n+2}(\xc) &= \int\!
  \D\Phi_{n+2}^{d=4}
   \cdis{\xi_1}\,
   \cdis{\xi_2}\,
     \Big(\xi_1\xi_2\, \fM{n+2}0\Big) \label{eq:nnlo:n2}\, .
\end{align}
\end{subequations}
where $\xi_i$ are the (scaled) energies of the radiated photons,
$\D\Phi_n^{d=4}$ is the $n$-parton phase space in 4 dimensions, and
the distribution $(1/\xi_i)_c$ acts on a test function $f(\xi_i)$ as
\begin{align}
  \label{subcond}
\int_0^1\D\xi_i\, \cdis{\xi_i}\, f(\xi_i)
&\equiv
\int_0^1\D\xi_i\,\frac{f(\xi_i)-f(0)\theta(\xc-\xi_i)}{\xi_i}
\,.
\end{align}
All three terms in \eqref{eq:nnlo:4d} are individually finite and the
dependence on the unphysical parameter $\xc$ cancels in the sum. This
cancellation is exact and serves as a test for a correct and
numerically stable implementation. The FKS$^\ell$ procedure does not
require to split photon radiation into a soft and a hard part, nor
does it introduce a photon mass as regulator. If all matrix elements
are known, numerical integration of \eqref{eq:nnlo:4d} combined with a
measurement function to define the IR-safe observable(s) results in a
fully-differential Monte Carlo code. This is the basis of the
\mcmule{} framework~\cite{Banerjee:2020rww}.

\subsection{Massification}\label{sec:mass}

The statement ``if all matrix elements are known'' is always easy to
make, but often impossible to realise. Of course, tree-level matrix
elements pose no problem. For one-loop matrix elements we often use
OpenLoops~\cite{Buccioni:2017yxi,Buccioni:2019sur}. However, two-loop
matrix elements are a bottleneck.

To alleviate the problem it is possible to start with the two-loop
matrix element computed with massless fermions,
$\mathcal{M}_n^{(2)}(m\!=\!0)$.  This matrix element contains poles up
to $1/\epsilon^4$ due to double soft-collinear singularities
associated with the massless fermion and photon. The corresponding
matrix element with massive fermions, $\mathcal{M}_n^{(2)}(m)$, instead
only contains $1/\epsilon^2$ soft poles and $\ln^2(m^2/Q^2)$ as
remnants of the collinear singularities. For small masses $m^2 \ll
Q^2$ the idea is tempting to use the massless matrix element as an
approximation of the massive one. This is, however, not possible due
to the different singularity structure. Instead, the massive all-order
result can be related to the massless one via the factorisation
formula~\cite{Penin:2005eh, Becher:2007cu, Engel:2018fsb}
\begin{align}\label{eq:massification}
  \mathcal{M}_n(m)&= \Big(\prod_j {Z(m)}\Big) \times S
   \times \mathcal{M}_n(0) + \mathcal{O}(m) \, .
\end{align}
The product is over all external fermion legs with a small mass $m$.
The process-independent factor ${Z(m)}$ is given in (3.12)
of~\cite{Engel:2018fsb} to NNLO and converts collinear $1/\epsilon$
poles into $\ln(m^2/Q^2)$. However, it is important to stress that
\eqref{eq:massification} does not only reproduce the logarithmically
enhanced terms but in fact all terms that are not polynomially
suppressed. The soft part $S$ is process dependent, starts at NNLO,
and receives contributions from only fermion loops. Conceptually this
is the most complicated contribution due to the factorisation
anomaly~\cite{Engel:2018fsb}. In practice, it is often advantageous to
compute the fermion-loop contributions -- including the hadronic ones
-- of two-loop matrix elements semi-numerically. This has the added
advantage of eliminating $S$ in \eqref{eq:massification} and rendering
massification completely universal.

To obtain a perfect cancellation of the $1/\epsilon$ singularities in
\eqref{eq:nnlo:4d} the divergent part of $\mathcal{M}_n^{(2)}(m)$ has to
be taken without approximation using~\eqref{eq:yfsnew}. We also use
the full $m$ dependence in the generation of the momenta and the phase
space.  Parametrically, the error induced by \eqref{eq:massification}
at NNLO is of the order $(\alpha/\pi)^2\, m^2/Q^2$, potentially
multiplied by a $\ln(m^2/Q^2)$.  For example, in the case of the
electron energy spectrum of the muon decay, massification reproduces
the NNLO correction with full electron-mass dependence at the level of
a few percent~\cite{Engel:2019nfw}. Going to higher energies this
approximation is expected to perform even better.

\subsection{Next-to-soft stabilisation}\label{sec:nts}

Even though all parts in \eqref{eq:nnlo:4d} are finite and can in
principle be evaluated numerically in 4 dimensions, in practice
numerical instabilities are often encountered. In particular
$\sigma^{(2)}_{n+1}(\xc)$, \eqref{eq:nnlo:n1}, is delicate as it
requires a numerically stable evaluation of the real-virtual one-loop
amplitude also in corners of phase space where large hierarchies of
scales are present. OpenLoops shows a remarkable stability and has
been used successfully for the real-virtual corrections of many QCD
NNLO calculations. However, the presence of (small) fermion masses
exacerbates the situation. The problem can be solved by replacing the
full one-loop radiative matrix element with a sufficiently precise
approximation in the numerically problematic soft-photon region. To
this end, the next-to-leading-power expression for the limit where the
photon momentum $k$ becomes soft, i.e. $k\to 0$, is needed. This
next-to-soft stabilisation has been used to compute the photonic
corrections to Bhabha scattering~\cite{Banerjee:2021mty} as well as
the full corrections to M{\o}ller scattering~\cite{Banerjee:2021qvi}
at NNLO. The precision of the numerical evaluation of the radiative
matrix elements used in these calculations is better than $10^{-5}$
for all photon energies.

The leading term in the  soft expansion has the celebrated form
$\mathcal{M}_{n+1}^{(\ell)}=\mathcal{E}\,\mathcal{M}_{n}^{(\ell)}$
where the eikonal factor $\mathcal{E}$ scales as $1/k^{2}$ (and upon
integration over the photon phase space yields $\hat{\mathcal{E}}$).
This approximation is not sufficient for our purpose and has to be
improved by including the subleading-power contributions scaling as
$1/k$. At tree-level, the Low-Burnett-Kroll (LBK)
theorem~\cite{Low:1958sn, Burnett:1967km, Adler:1966gc} provides a
universal structure for these next-to-soft contributions. The LBK
theorem has been extended to one loop for massive QED
in~\cite{Engel:2021ccn} and according to its (3.31) is given by the
sum of the hard and soft contributions
\begin{subequations}
    \label{eq:lbk_oneloop}
\begin{align}
    \label{eq:lbk_oneloop_hard}
    \mathcal{M}_{n+1}^{(1),\text{hard}}
    &\simeq \sum_l\sum_i Q_i Q_l \Big(
    - \frac{p_i\cdot p_l}{(k\cdot p_i)(k\cdot p_l)}
    + \frac{p_l\cdot \tilde{D}_i}{k\cdot p_l}
    \Big)
    \mathcal{M}_n^{(1)}, \\
    \label{eq:lbk_oneloop_soft}
    \mathcal{M}_{n+1}^{(1),\text{soft}}
    &\simeq \sum_l \sum_{i \neq j} Q_i^2 Q_j Q_l
    \Big( \frac{p_i\cdot p_l}{(k\cdot p_i)(k\cdot p_l)}
    - \frac{p_j\cdot p_l}{(k\cdot p_j)(k\cdot p_l)} \Big) \,
    2 S(p_i,p_j,k)\, \mathcal{M}_{n}^{(0)},
\end{align}
\end{subequations}
where the sums run over the external legs with charges $Q_i$ and
momenta $p_i$. The first term on the r.h.s. of
\eqref{eq:lbk_oneloop_hard} scaling as $1/k^2$ is just the eikonal
approximation mentioned before. The second term of
\eqref{eq:lbk_oneloop_hard} scales as $1/k$ and is the expected
one-loop generalisation provided by the LBK theorem, where
$\tilde{D}_i$ is the (modified) LBK differential operator, see (3.11b)
of~\cite{Engel:2021ccn}. Taken together, these two terms correspond to
the tree-level version of the LBK theorem. Using the method of
regions~\cite{Beneke:1997zp} to perform the expansion in $k$ at the
one-loop level, they correspond to the hard
part~\cite{Bonocore:2014wua}. At one loop there is also a soft
contribution, scaling as $1/k$. The important observation
of~\cite{Engel:2021ccn} is that this soft contribution also has a
universal structure, as given in \eqref{eq:lbk_oneloop_soft}. The
function $S(p_i,p_j,k)$ scales as $k$ and can be found in (3.29)
of~\cite{Engel:2021ccn}. Using \eqref{eq:lbk_oneloop}, the
next-to-soft limits can be obtained without a detailed
process-specific computation.

There is a subtlety hidden in~\eqref{eq:lbk_oneloop_hard} in that the
non-radiative matrix element has to be evaluated with radiative
kinematics. In particular, the corresponding momenta only satisfy
momentum conservation up to the remnant $k$. A possible approach on
how to deal with this is discussed in~\cite{Engel:2021ccn}, while an
alternative is given in~\cite{DelDuca:2017twk,Bonocore:2021cbv}.

\subsection{Outlook towards higher energies} \label{sec:mcmoutlook}

The techniques described in the previous sections were developed with
low-energy processes in mind. However, they remain applicable and in
some cases become even more important when going to higher energies.
On the other hand, additional features have to be taken into account.
As possible applications we are considering $ee\to\tau\tau$ for $\sqrt
s\sim 10\,{\rm GeV}$~\cite{Kollatzsch:2022xxx} and also planning to
extend our work on Bhabha scattering~\cite{Banerjee:2021mty} and
$ee\to\gamma\gamma$ to energies relevant for luminosity determinations
at electron-positron colliders.

A first extension that is required is the inclusion of electroweak
effects. A complete NLO calculation in the Standard Model can be
achieved using OpenLoops for the matrix elements and FKS$^\ell$ for
phase-space integrations.

A second issue that will become more prominent at higher energies are
numerical issues related to collinear photon emission. At low
energies, it is sufficient to give special consideration to the soft
region, $k\ll m$. At higher energies, also the hierarchy $m^2\ll Q^2$
will require some care, especially in the case of the numerically
delicate real-virtual contribution. To this end, a similar strategy as
for the soft region can be applied. This time the limit where the
angle $\sphericalangle(k,p_i)$ becomes small has to be considered.
Since we keep $p_i^2=m^2$, but now treat $m^2\ll Q^2$, we need the
limit $k\cdot p_i \sim m^2\ll Q^2$ of one-loop radiative matrix
elements. At leading power in the small quantity and at NLO in
$\alpha$ this leads to a generalisation of massification,
\eqref{eq:massification}, that e.g. for an initial-state collinear
configuration reads
\begin{align} \label{eq:collim}
    \mathcal{M}_{n+1}(k,p_j,m) \simeq
    J_\text{ISR}(x,m) \Big( \prod_{j\neq i} Z(m) \Big)
    \mathcal{M}_n(p_i-k,p_{j\neq i}, 0),
\end{align}
where $1-x$ is the energy fraction carried off by the photon.  Thus,
for the (incoming) leg $i$ to which the photon becomes collinear, the
factor $Z(m)$ is replaced by the splitting function $J_\text{ISR}$.
The explicit expression for $J_\text{ISR}$ and the crossing related
final-state splitting function $J_\text{FSR}$ at one loop can be found
in Appendix~B of~\cite{Engel:2021ccn}. This factorisation formula
could also turn out to be useful in coping with the narrow peaks that
arise due to the collinear scale hierarchy. In particular,
\eqref{eq:collim} can be used to subtract these collinear
pseudo-singularities pointwise, in analogy to~\cite{Dittmaier:1999mb}.

Finally, \mcmule{} is currently calculating strictly at fixed order.
However, for many applications a combination with resummation, e.g.~that
achieved by a parton shower or by analytical techniques, is
required. The collinear configurations mentioned in the previous
paragraph give rise to logarithms $\ln(m^2/Q^2)$. They become more
important at higher energies and can be resummed using fragmentation
functions (for final state) or parton distribution functions 
(for initial state)\footnote{\label{ft:cllYFS}In the context of the
Snowmass 2021 proceedings, the interested reader can find a more extensive 
discussion on collinear factorisation and YFS in previous sections
and in ref.~\cite{FERMILAB-PUB-22-116-SCD-T}.}. Currently, within \mcmule{} an 
effort is ongoing to include a YFS parton shower\footnote{\ref{ft:cllYFS}}
that is similar to {\tt PHOTONS++}~\cite{Schonherr:2008av} and
matched at NNLO. For the parton shower, instead of $\ieik$ the
exponent of \eqref{eq:yfsnew} is the YFS form factor, that also
includes a virtual contribution.  The
subtraction scheme FKS$^\ell$ is well suited for this as it is already
based on YFS. This helps dealing with double counting in the matching
of the parton shower to fixed-order calculations.

With these extensions \mcmule{} will become a tool for leptonic
processes with an even broader range of applications.

\acknowledgments

TE acknowledges support by the Swiss National Science Foundation (SNF)
    under contract 200021\_178967.



\section{Collinear divergences at future $e^+e^-$ colliders\label{sec:coll}}
\begin{center}
\emph{Alessandro Vicini}
\end{center}









\subsection{Precision calculations at an $e^+e^-$ collider  }
The high-precision tests of the Standard Model (SM) at an $e^+e^-$ collider,
like the future FCC-ee or CEPC,
rely on the possibility to compute higher-order radiative corrections,
covering all the relevant phase-space regions.
Detailed studies have been performed at LEP
(cfr. \cite{TwoFermionWorkingGroup:2000nks,Grunewald:2000ju} and references therein)
and in preparation for the FCC-ee  \cite{Proceedings:2019vxr} and the CEPC \cite{CEPCStudyGroup:2018ghi},
with a systematic discussion of the available codes
and of several aspects related to the treatment of higher-order radiative corrections.
In this note we try to comment about some of the challenges set by the expected experimental
precision at $e^+e^-$ colliders and about the possible transfer of the theoretical and computational
knowledge developed for the LHC.

We assume, for the sake of discussion, that
the full set of three-loop EW corrections to the process $e^+e^-\to f \bar f+X$,
with $f$ an electrically charged fermion,
is explicitly available.
The need for such a challenging calculation
has been discussed in \cite{Freitas:2019bre,Blondel:2019qlh},
in view of high-precision measurements at the $Z$ resonance.
It has been shown that there might be corrections to the lowest-order couplings,
from virtual corrections of weak and mixed QED-weak origin, numerically relevant e.g. in the prediction of the
asymmetries.
Their role is somehow complementary to the one of the QED corrections, in the precise predictions
of the kinematical distributions.
It is known, since LEP \cite{Altarelli:1989hv},
that initial state QED radiation develops very large logarithmic corrections,
crucial to the high-precision description of the $Z$ lineshape.
The resummation to all orders of these classes of corrections is known to be mandatory,
for an accurate prediction of the central values of the observables and for a reduction
of the remaining theoretical uncertainty \cite{Montagna:1993ai,Jadach:1993yv,Bardin:1999yd}.
Few questions\footnote{
We refer to the three-loop EW case as the ultimate calculation that can be necessary at the
FCC-ee/CEPC, but the same questions could already be addressed, when the complete set
of two-loop EW will become available.}
arise:
1)  Is the three-loop calculation on its own sufficient to match the precision target,
    without the inclusion of additional higher-order contributions?
2) Since the large logarithmic effects are due to soft or collinear emissions,
  which of the two classes of corrections should receive priority in the construction
  of a framework matching fixed- and all-order results ?
3)  Does the answer to the previous question depend on the observable under study?

In order to fix some figures,
at FCC-ee the total cross section into hadrons, at the $Z$ peak, can be measured with a relative error
at the 0.01\% level \cite{Proceedings:2019vxr}.
We take this number as a illustrative reference for the needed precision of the theoretical predictions.
In the study of the $Z$ boson lineshape,
the center-of-mass energy is $\sqrt{s}\sim \mz$.
The size of the logarithmic factor,
which appears with the initial state QED radiation upon integration over the photon-emission angle
and is dominated by the phase-space region with the photon collinear to the incoming electron or positron,
is $L\equiv\log(\mzsq/m_e^2)\sim 24.18$, with $m_e$ the mass of the electron.
The structure of the cross section, putting in evidence the initial state collinear
(also dubbed mass) logarithm $L$ and the soft logarithm $\ell$,
is given, up to first order in the fine structure constant $\alpha$, by
$\sigma=\sigma_0 + \alpha \left(a_{11} L \ell + b_{10} L + c_{01} \ell + d_1 \right)$,
with $\sigma_0$ the Born cross section.
The soft logarithm $\ell=\log(\Delta E^2/\mzsq)$ appears because of the emission of soft, i.e. low-energy, photons
and its precise value depends on the resolution $\Delta E$ adopted to identify the final state particles.
We remark the presence of a double logarithmic enhancement
given by the product of a mass and a soft logarithms,
plus the contributions given by single logarithmic terms
and a subleading term without enhancement. 
In order to appreciate the need for a systematic resummation of the large logarithmic corrections,
and in particular those due to collinear emissions,
we iterate the initial state single logarithmic collinear correction factor and find:
$\alpha L \sim 0.1764,\, \alpha^2 L^2 \sim 0.0311,\, \alpha^3 L^3 \sim 0.0055,\, \alpha^4 L^4 \sim 0.0010$\,.
These factors strongly contribute to enhance the effect of the radiative corrections,
at a level relevant for the precision goal of prediction of the total cross section into hadrons.
Even if we can not draw final conclusions from this numerical inspection,
it nevertheless suggests that the answer to the first question could be negative,
and the matching of resummed and three-loop results could be necessary.
One might ask whether the resummation of terms at the third order in a logarithmic counting
is by itself sufficient to match the precision goals of the future $e^+e^-$ colliders,
without needing a full three-loop calculation and the corresponding matching.
The estimates presented in \cite{Freitas:2019bre,Blondel:2019qlh}
suggest that both sets of corrections (fixed and to all orders) will be needed, because of their different
impact on the observables.

A detailed review about the structure functions for the resummation of collinear logarithms,
and of the Yennie-Frautschi-Suura (YFS) formalisms \cite{Yennie:1961ad}
for the resummation of soft-emissions contributions,
has been presented in \cite{WorkingGrouponRadiativeCorrections:2010bjp};
we refer to Section 2.4 of that paper for technical details.
In this note we address questions analogous to those of ref. \cite{WorkingGrouponRadiativeCorrections:2010bjp}
for the matching with next-to-leading (NLO) fixed-order results,
but, at variance with that paper,
we consider now the problem of a matching which could be needed at third order.

\subsection{ The collinear singular limit of squared matrix elements   }
In QED,
the amplitude describing the emission of a photon
off an external on-shell fermion $f$
is proportional to
\begin{equation}
\frac{1}{(p+k)^2-m_f^2+i\varepsilon}=
\frac{1}{2 E_p E_k(1-\beta_f\cos\theta) + i\varepsilon}
\label{eq:propagator}
\end{equation}
where $p$ and $k$ are the fermion and photon momenta respectively,
$k^2=0$ and $p^2=m_f^2$, $m_f$ is the fermion mass,
$E_p=\sqrt{m_f^2+\vec p^2}$ is the fermion energy, $E_k=|\vec k|$
is the photon energy, $\theta$ is the emission angle of the photon with respect
to the fermion $\vec p$,
and $\beta_f=\sqrt{1-m_f^2/|\vec p|^2}$.
When we integrate the squared matrix element over the photon phase space,
the interference term of two Feynman amplitudes with the photon emitted by two charged particles
yields a factor proportional to
$
\log\left(
\frac{1+\beta}{1-\beta}
\right)
$
which diverges in the $m\to 0$ limit;
this singularity is thus called mass or collinear divergence, because the relevant phase-space region
is the one with the photon parallel to the emitting particles.

The collinear divergence depends only
on the properties (charge,mass,spin) of the emitting particles,
but it is independent of the details of the rest of the hard scattering process.
This universality feature emerges as a consequence
of the factorisation of the squared matrix element.
Beyond the first order, the correct factorisation in the collinear limit
has been explicitly demonstrated to all orders for tree level amplitudes in \cite{Catani:1999ss}.
Explicit results at second order are known \cite{Catani:1998nv},
and explicit checks at third order have been performed \cite{Anastasiou:2015vya,Duhr:2021vwj}
but a formal demonstration of collinear factorization theorems for a generic process to all orders is lacking.

The appearance, in the first QED calculations of radiative corrections  \cite{Kinoshita:1958ru}
of mass singularities,
and the remark of their cancellation in inclusive quantities,
has led to the formulation, by Kinoshita-Lee-Nauenberg (KLN) \cite{Kinoshita:1962ur,Lee:1964is},
of a theorem which states the conditions of such cancellation.
The sum over all degenerate, i.e. indistinguishable, final states
yields a finite, regular result, although the individual contributions are separately divergent.
The finiteness is achieved order by order in the perturbative coupling constant expansion,
including contributions due to virtual and real-emission corrections,
provided that both sets are degenerate with respect to the observable definition.
This theorem is fundamental in the physics of hadronic jets and more in general provides the
guiding principle to define infrared-safe and collinear-safe observables.

The emission of collinear radiation from the initial state of a scattering amplitude
modifies the kinematics of its underlying process,
because the emitted particle has in general a finite energy.
The non degeneracy of the two underlying processes,
in the virtual and real correction cases,
breaks the hypothesis of the KLN theorem.
It is thus possible that a collinear divergence
survives in the combination of the two contributions.
Thanks to the factorisation property of these divergences,
the partonic cross section can be written as the product of a finite cross section times a divergent factor.
The universality of the latter allows us to reabsorb it into the definition
of the incoming physical object starting the scattering:
the best known example is given by the proton and its parameterisation in terms of collinear
parton density functions (PDFs).
Thanks to the universality properties, this redefinition is predictive,
with a factorization-scheme ambiguity at the level of the subdominant finite corrections,
formally of a perturbative order higher with respect to the one of the calculation.
The PDFs express the probability to find a parton that carries a fraction $x$ of the proton momentum,
while probing the latter in a scattering with a momentum transfer $\mu$.
In complete analogy, we can describe a lepton scattering, like e.g. $e^+e^-$ annihilation,
as the interaction of two composite systems, the leptons,
whose internal structure is given in terms of partons
(electrons, positrons, photons, quarks, EW bosons).
The definition of the physical lepton is given, order by order in perturbation theory,
only after the factorisation of the initial state collinear divergences
and their reabsorption in the lepton PDFs.
The collinear parton densities $\vec q(x,\mu^2)$ (they can be partons inside a proton or a lepton),
satisfy the integro-differential DGLAP equations
\cite{Gribov:1972ri,Altarelli:1977zs,Dokshitzer:1977sg}
\begin{equation}
\frac{\partial}{\partial\log\mu^2}\vec q(x,\mu^2) =
\frac{\alpha(\mu)}{2\pi} \int_x^1 \frac{dz}{z}\, \mathbf{P}(z)\cdot \vec q(\frac{x}{z},\mu^2)\, ,
\end{equation}
whose solution expresses their dependence on the scale $\mu$
as the convolution of the densities themselves with the appropriate splitting functions matrix $\mathbf{P}(z)$.
The latter describe the emission responsible for the scale change,
evaluated in the collinear approximation, and can be computed in perturbation theory.
They are know up to 3-loop level \cite{Moch:2004pa,Vogt:2004mw,Ablinger:2014nga,Ablinger:2017tan} in QCD.
The solution of the DGLAP equations resums to all orders
the contributions due to the collinear emissions described by the splitting functions.
The logarithmic accuracy of this resummation is determined by
the perturbative order of the splitting function expression.

The solution of the DGLAP equations requires to fix the boundary conditions of the differential problem.
In the QCD case, since the strong interaction does not allow us to compute these conditions from first principles,
the proton structure has to be modeled in terms of PDFs and
the PDF parameters at a given scale are fit to the available experimental data.
This problem has led to a large number of studies discussing
the representation of the uncertainty on the proton structure
induced by the imperfect knowledge of the boundary conditions \cite{Butterworth:2015oua}.
The consistent determination of the latter from a fit to several experimental data is possible
only if all the relevant cross sections are known at the perturbative order under discussion. 
This latter problem represents a bottleneck in the progress towards a consistent prediction of the LHC processes
at NNLO-QCD and, a fortiori, N3LO-QCD or NNLO QCD-EW.
In summary, the DGLAP evolution of the proton PDFs can be performed
if the splitting function at a given order are known, 
but the absolute value of the parton densities can be affected
by an inconsistent knowledge of the boundary conditions.
In the lepton case, it is instead possible to compute the boundary conditions from first principles in QED
as shown at NLO-QED in \cite{Frixione:2019lga}.
An effort towards their determination at NNLO-QED and N3LO-QED is needed,
in order to apply the DGLAP formalism at a lepton collider \cite{Frixione:2021zdp} at least at second,
but hopefully also at third order in perturbation theory.

The solution of the integral-differential DGLAP equations is achieved in QCD via numerical algorithms,
with different evolution codes available in the literature \cite{Whalley:2005nh,Salam:2008qg},
developed to solve the QCD case,
including so far up to NNLO splitting kernels.
The peculiar feature, in QED, of the existence of an integrable singularity at $z=1$,
makes an analytic solution possible in this case.
While the results at NLO-QED are available \cite{Bertone:2019hks,Frixione:2021wzh},
the possibility to prepare an evolution code at NNLO or N3LO in QED
requires the two- or three-loop splitting functions, which could be derived from the QCD expressions,
but also the boundary conditions, discussed above.

\subsection{ Comparison of structure functions and YFS formalisms }
We consider the inclusive production of a final state $F$ in the scattering
of two particles $P_{1,2}$, e.g. hadron-hadron or lepton-lepton,
in the structure function approach.
We formulate the cross section as the convolution of
the parton densities describing the probability of finding partons $i,j$ inside the colliding particles,
with a partonic cross section.
\begin{equation}
\sigma(P_1P_2\to F+X) =
\sum_{i,j}\int dx_1\,dx_2\,q_i(x_1,\mu) q_j(x_2,\mu)\,\hat\sigma(ij\to F+X)
\end{equation}
A consistent result is obtained with
the partonic cross section computed at a given fixed perturbative order
and subtracted of the initial state collinear divergent factors,
convoluted with the structure functions at the same order.
The latter account, via DGLAP evolution, for the resummed contribution of the initial state collinear logarithms.
For example, we consider in QED the case of a cross section
computed with the full NLO-EW results at parton level $\hat \sigma$
and convoluted with the QED LL solution of the DGLAP equations.
The result has NLO-EW accuracy for the total cross section,
it includes to all orders terms of ${\cal O}(\alpha L)$, 
but only partially accounts for the second-order terms proportional to $\alpha^2 L$:
the fixed-order missing contributions stem from the non-factorisable two-loop corrections.
A full NNLO-EW calculation, convoluted with the corresponding second order structure functions,
will exactly include to all orders terms of ${\cal O}(\alpha^2 L)$,
but only partially subleading fixed-order terms proportional to $\alpha^3 L$.
The inclusion of terms proportional to logarithms due to soft emissions
in this formulation is restricted to those present in the fixed-order expression.
In addition, the DGLAP equations obviously include to all order the contributions
of the initial-state soft and collinear phase-space region.

A different approach to describe multiple radiation is given by the so called Parton Shower.
The latter provides an algorithmic implementation of the exponentiation of the soft LL contributions.
This description can be matched, avoiding double countings, with the exact matrix elements.
The latter improve the total cross section prediction
and the description of hard, large-transverse-momentum radiation.
\begin{equation}
d\sigma=
F_{SV}\, \Pi (Q^2,\varepsilon)\times
\sum_{n=0}^\infty \frac{1}{n!}
\left(\Pi_{i=0}^n F_{H,i}  \right)
|{\cal M}_{n,LL}|^2
d\Phi_n
\label{eq:horace}
\end{equation}
One possible formulation is given in equation \ref{eq:horace}, as implemented in the event generator
{\tt HORACE} \cite{CarloniCalame:2007cd}.
The LL Parton Shower approximation is corrected at ${\cal O}(\alpha)$
by the infrared and collinear finite factors $F_{SV}$ and $F_{H,i}$:
we have the Sudakov form factor $\Pi(Q^2,\varepsilon)$ accounting for the exponentiation
of the soft emissions; we have the real emission matrix elements ${\cal M}_{n,LL}$
written in terms of the product of $n$ eikonal currents;
the factor $F_{SV}$ takes into account the finite part of soft emissions and virtual corrections,
while the exact matrix element is applied to each hard large-transverse-momentum emission
via the $F_{H,i}$ correction.
The factorised formulation guarantees the correct semiclassical limit,
but also introduces universal higher-order terms,
which can partially account, for instance, for a subset of NLL collinear terms and in particular
for part of the subset proportional to $\alpha^2 L$,
with a significant phenomenological improvement.
Extending this kind of matching beyond NLO is a current topic of investigation,
with NNLO results matched to a LL Parton Shower \cite{Monni:2019whf}.
Also in this case, an explicit improvement is achieved for the total cross section
and for the observables sensitive to hard large-transverse-momentum radiation,
whereas the Parton Shower resummation includes only LL terms.

The YFS approach \cite{Yennie:1961ad} starts from
the resummation of the first-order logarithms in the soft approximation.
In the literature, considering the scattering of two fermions into two fermions,
not only the logarithms due to initial state radiation,
but also those stemming from the initial-final interference and the final-state emissions,
have been exponentiated and matched with fixed order expressions up to second order
\cite{Jadach:1999vf,Jadach:2000ir}.
The YFS fully exclusive formulation
is based on the factorisation of soft emissions already at the amplitude level.
Thanks to the factorisation of one-loop soft-divergent virtual corrections,
it is possible to obtain a soft-finite one-loop correction factor $Y$ which exponentiates.
The resummed fully differential expression of the cross section reads:
\begin{eqnarray}
  d\sigma&=&d\sigma_0\,\exp(Y)\,
\exp\left[
  \frac{dk\,d\cos\theta\,d\phi}{(2\pi)^2}
  k\,\Theta(k-k_{min})
  \sum_\varepsilon \alpha
  \left(\frac{\varepsilon(k)\cdot p_2}{k\cdot p_2}-
                      \frac{\varepsilon(k)\cdot p_1}{k\cdot p_1}
                \right)^2
                \right]\\
Y&=&\beta \log\frac{k_{min}}{E} +\frac14 \beta +\frac{\alpha}{\pi}\left(\frac{\pi^2}{3}-\frac12 \right)
\quad\quad
\beta=\frac{2\alpha}{\pi}\left[\log\left(\frac{s}{m_e^2}\right)-1 \right]\, .
\nonumber
\end{eqnarray}
The second exponential of the first equation is a compact notation for the sum over all possible
additional photon multiplicities $n$ of the product of eikonal emission factors, valid in the soft limit;
the presence of an integration measure is
the symbolic representation of the phase-space factors for $n$ photons, with arbitrary $n$.
The description of photons, with an energy larger than the cut-off $k_{min}$, is fully differential.
We read in the exponent $Y$ the presence of the first-order double and single logarithmic terms due to collinear
and soft-collinear contributions, evaluated in soft approximation,
integrated in the photon energy up to the cut-off $k_{min}$.
Given the soft limit approximation,
for the electromagnetic coupling the fine structure constant $\alpha$ is a customary choice.
The inclusion of higher-order contributions like the description of additional light fermion pairs
could be simulated with a running coupling,
but in this case care should be paid to avoid double countings with the rest of the hard correction factors.
The matching with the exact ${\cal O}(\alpha)$ results yields, following the original YFS algorithm,
additional corrective factors $\beta_n$ called residuals, free of any soft or collinear divergence,
expressing the impact of the exact matrix element on the $n$-th emission.
A modification of the residuals, without spoiling the formal accuracy of the calculation,
allows to improve the resummed and matched YFS formulation
to include also the missing  $\alpha^2 L$ terms. In summary, the exact exponentiation of the one-loop
soft logarithms is matched to fixed first- and partial second-order results,
covering up to second order all the relevant infrared divergent contributions.

\subsection{Discussion and conclusions}
The estimated experimental precision of future $e^+e^-$ colliders forces us to investigate the
question of the best formulation of a calculation that includes
fixed- and all-orders results in $e^+e^-$ collisions, up to three-loop level.
The various approaches can be described with respect to
$a)$ the formal inclusion of the various classes of enhanced contributions and
$b)$ the possibility of extending the resummation
beyond the exponentiation of the first-order divergent terms.
We eventually consider the dependence on the observable under study.

Any code based on a Parton Shower relies on the algorithmic exponentiation of the first-order soft emissions,
achieving LL accuracy; the angular dependence of the radiation depends on the specific implementations.
After the matching with NLO matrix elements, it is possible that
a subset of the NLL collinear contributions is correctly included,
but the full set of terms proportional to $\alpha^2 L$ is under control only with matching at NNLO level.
It is interesting to observe that, in the Parton Shower matching,
different recipes, still preserving the formal accuracy of the results,
can lead to significant numerical differences.
This remark shows the role of multiple radiation and the impact of its ``handling'',
as ruled by the matching recipe. Important examples are known from the LHC studies
and it would be interesting to check this point for an $e^+e^-$ collider,
in view of the very high-precision goal.
It is not obvis that an improvement of the accuracy of the resummation part, going beyond the LL level,
can be achieved for a general purpose tool.


The YFS formalism is based on the exclusive exponentiation of first-order contributions in the soft limit.
The inclusion of all the LL terms is achieved at amplitude level and
the full set of $\alpha^2 L$ collinear contributions can be included.
The full resummation of NLL collinear logarithms is not available.
In the public implementations of the YFS formalism,
all the radiating particles are supposed to be massive,
thus separating the issue of the collinear divergences from the discussion of the 
soft contributions. While this is a viable approach in the description of $e^+e^-$ collisions,
it would introduce a dependence on unphysical quark masses
in the description of quark-initiated processes relevant for the LHC.

In the pure structure functions approach the formulation of the physical cross section
is purely dictated by the factorisation of the initial state collinear singularities.
As a positive outcome, the extension to two- and three-loop level is conceptually
straightforward.
A technical effort to translate the N3LO splitting functions to the QED case
and to compute the boundary conditions to the DGLAP equations is needed, and is feasible.
The contributions which are resummed in this approach are the collinear logarithms,
going up to NNLL, in the case a three-loop result were available. 
In the pure structure function approach,
the soft logarithms taken into account are those appearing from the matrix elements,
and those of the soft-collinear phase-space corner.

At a future $e^+e^-$ collider,
two extremely important energy regions in the precision physics program are the $Z$-boson resonance
and the $W^+W^-$ production threshold.
The resummation of the soft logarithms clearly plays a role in the determination of the precise
position of these two points and in the description of the associated lineshapes:
these logarithms in fact control the exact amount of energy available in the center-of-mass
of the hard scattering process.
On the other hand, the measurement takes place in a fiducial volume,
with acceptance cuts also in the plane transverse to the beam, so that a formalism based
on the separation between the longitudinal and transverse components can be a natural framework
for precise predictions.

Exploring the impact of the different resummations, with the currently available tools,
can set the stage to the evaluation of higher-order corrections and the preparation of an adequate
matching formalism.

\section{Factorization beyond leading power\label{sec:pw}}
\begin{center}
\emph{Leonardo Vernazza}
\end{center}

\subsection{Factorization at next-to-leading power: state of the art}
\label{sec:introv}

Understanding gauge theory scattering amplitudes 
in the limit where external radiation becomes soft 
poses interesting challenges from a conceptual quantum 
field theory point of view and has important applications 
in the phenomenology of precision particle physics. 
Let us consider $n$-particle scattering 
processes in QED. Given the emission of an additional 
soft photon with momentum $k$, the scattering amplitude 
${\cal M}_{n+1}$ can be expressed as a power expansion 
in the energy $E = k^0$ of the photon,
\begin{equation} \label{softexpansion}
{\cal M}_{n+1} = {\cal M}_{n+1}^{\rm LP}
+ {\cal M}_{n+1}^{\rm NLP} + \ord(E)\,,
\end{equation}
where the \emph{leading power} (LP) term has scaling 
${\cal M}_{n+1}^{\rm LP} \sim 1/E$, the 
\emph{next-to-leading power} (NLP) contribution is of 
order ${\cal M}_{n+1}^{\rm NLP}\sim E^0$, and so on.

Our interest for the expansion in
\eqn{softexpansion} stems from the fact that 
the coefficients in the series \emph{factorize}, 
i.e., they can be expressed in terms of simpler,
universal objects, which relate the radiative 
amplitude to the non-radiative or \emph{elastic} 
amplitude ${\cal M}_{n}$. It is well known that 
the LP term in \eqn{softexpansion} has the 
universal form~\cite{Yennie:1961ad,Grammer:1973db}
\begin{equation}\label{factLP}
{\cal M}_{n+1}^{\rm LP}(\{p_i\},k) = S_n \,{\cal M}_n(\{p_i\})\,,
\qquad \qquad 
S_n = e \sum_{i=1}^n q_i\, \frac{p_i^{\mu} \varepsilon_{\mu}(k)}{p_i \cdot k} \,,
\end{equation}
where $p_i^\mu$ and $q_i$ represent the momentum 
and electric charge of the $i$-th hard particle, 
and $\varepsilon_{\mu}(k)$ is the polarisation 
vector of the soft photon. The soft function 
$S_n$ describes a set of eikonal interactions 
between the external particles and the soft photon, 
which are sensitive only to the direction and charge 
of the emitting particle. In general, $S_n$ can be 
calculated as the vacuum expectation value of a set 
of Wilson lines, one for each hard emitting particle.  

Relations such as the one in \eqn{factLP} are also known 
as \emph{soft} theorems. Their interest is not just purely 
theoretical, but also relevant for phenomenology. The $1/E$ 
singularity in the soft limit enhances soft radiation in 
scattering processes. Measurements that are sensitive to 
soft radiation involve a small scale, and the corresponding 
cross section contains large logarithms of the ratio of 
this small scale and the scale of the hard scattering. 
Such large logarithms potentially spoil the convergence 
of the expansion in the coupling $e$, a problem that can 
be addressed by resummation. The development of theorems 
such as the one in \eqn{factLP} led to the proofs of factorization~\cite{Bodwin:1984hc,Collins:1985ue,Collins:1988ig}, 
but also constitutes the first step towards resummation, 
as it allows one to decompose a multi-scale amplitude 
(or cross section) into the product of simpler 
single-scale functions.

The physical interpretation of \eqn{factLP} rests upon 
the long wavelength of soft radiation not being able to 
resolve the hard scattering. However, at each subsequent 
power in \eqref{softexpansion} more is revealed, and the 
factorization structure becomes more involved. The NLP 
term in \eqn{softexpansion} is still under much 
investigation. Emitted (next-to-)soft radiation 
becomes sensitive to the spin of the hard particles, 
and starts to reveal details of the internal structure 
of the hard interaction. Compared to \eqn{factLP}, 
the factorization of radiative amplitudes at NLP 
needs to take into account the following features:
\begin{itemize}
\item Emission of soft radiation from the 
hard partons beyond the eikonal approximation,
for instance involving chromo-magnetic interactions 
between the hard particle and the emitted soft photon. 
This can be taken into account by introducing 
generalizations of the soft function $S_n$ in 
\eqn{factLP} \cite{Laenen:2010uz}. 
\item Emission of soft radiation from the non-radiative 
amplitude. This is one of the mechanism which probes the
structure of the hard interaction, and was understood
a long time ago in papers by Low, Burnett and 
Kroll~\cite{Low:1958sn,Burnett:1967km}: in the case of 
massive emitting particles, the structure of the NLP 
term is dictated by gauge invariance, by means of Ward 
identities. This early formulation, now known as the 
``LBK'' theorem, was proven~\cite{DelDuca:1990gz} 
to hold only in the region $k^0 \ll m^2/Q$, with 
$Q$ the centre of mass energy. 
\item For $m^2/Q < k^0 <m$, the LBK theorem must be extended 
to account for NLP contributions arising from soft photons 
emitted from loops in which the exchanged virtual particles 
have momenta collinear to the external particles (i.e.~having 
a small virtuality, while retaining momentum components 
which are large compared to the soft radiation). 
Such type of contributions are identified as 
\emph{radiative jets}. The first of such configurations 
was investigated for Drell-Yan in~\cite{DelDuca:1990gz}.
\end{itemize}
Most of the effort in the past few years has been
devoted to the derivation of factorization theorems 
able to describe consistently such soft and collinear
radiation into well defined field-theoretical 
matrix elements. This has been done following
mainly two approaches. One, which we will refer
to as \emph{diagrammatic}, aims
at describing such matrix elements in terms of 
fields, currents and Wilson lines in QED. The second 
approach is based on Soft-Collinear Effective Theory (SCET)~\cite{Bauer:2000ew,Bauer:2000yr,Bauer:2001ct,Bauer:2001yt,Beneke:2002ph}.

SCET describes the soft and collinear limits as 
separate degrees of freedom, each with their own 
Lagrangian. The elastic amplitude ${\cal M}_n$ in 
\eqref{factLP} is encoded in terms of effective 
$n$-jet operators and their short-distance 
coefficients, which capture the contribution 
from hard loops. At LP, soft emissions from 
hard particles can be described by Wilson lines, 
as in the diagrammatic picture. Beyond LP, soft 
emissions originate from time-ordered non-local operators 
made out of soft and collinear fields, where the power 
suppression follows either from additional insertions 
of the power-suppressed soft and collinear Lagrangian, 
or from subleading operators describing the hard scattering~\cite{Larkoski:2014bxa,Moult:2019mog,Beneke:2019oqx}.
SCET provides a systematic approach, as each operator 
and Lagrangian term have by construction a definite 
power counting. When all operators consistent with 
symmetries are included, 
this completeness ensures that the resulting 
factorization is valid to all orders in the 
coupling constant. 

Both approaches present advantages and disadvantages. 
From a technical point of view, the calculation of 
soft and collinear matrix elements within SCET is 
designed such as to directly reproduce calculations 
obtained within the method of expansion by momentum 
regions \cite{Beneke:1997zp}, and can be quite 
involved already at one-loop (cf. \cite{Beneke:2019oqx,Broggio:2021fnr}). 
In this respect, the calculation of matrix 
elements within the diagrammatic approach is simpler 
(cf. \cite{Bonocore:2015esa,Bonocore:2016awd,DelDuca:2017twk}), 
although the lack of tools such as multipole 
expansion results into overlaps between soft 
and collinear matrix elements, leading to 
complicated nested subtractions beyond one 
loop.

More relevant for applications to the theory
of resummation is the fact that soft and collinear 
matrix elements involve convolutions, which 
are well defined in dimensional regularisation, 
but often divergent in $d=4$. This poses a 
challenge for a standard EFT approach, 
where one first renormalizes each function in the 
factorization theorem to derive the renormalization 
group equations needed for resummation, causing 
these convolution integrals to become divergent. 
The problem needs to be investigated case by case, 
and so far only one example is known 
\cite{Liu:2019oav,Liu:2020tzd}, 
where convergence of the convolution integrals 
can be achieved after a re-factorization process, 
which is non-trivial to be proven in general
(see also \cite{Moult:2019uhz,Beneke:2020ibj}).
In this respect, the diagrammatic approach could 
provide an easier path to resummation by exploiting 
exponentiation properties of soft radiation and 
the replica trick~\cite{Laenen:2008gt,Gardi:2010rn}, 
which can be carried out within dimensional 
regularization (cf. \cite{Bahjat-Abbas:2019fqa,vanBeekveld:2021mxn}).

The classification of soft and collinear
matrix elements contributing beyond LP is rather
straightforward in SCET, once the power-counting 
and field content of the theory is fixed 
\cite{Larkoski:2014bxa,Moult:2019mog,Beneke:2019oqx}.
Within the diagrammatic approach, the classification
of all possible soft and collinear matrix elements 
has been derived for the Yukawa theory \cite{Gervais:2017yxv} 
and more recently for QED \cite{Laenen:2020nrt}
(see also \cite{Bonocore:2021cbv}).
We will illustrate such derivation in the next 
section.

\subsection{Factorization of QED amplitudes: 
diagrammatic approach}
\label{sec:fact}

Understanding the factorization structure of 
${\cal M}_{n+1}^{\rm NLP}$ requires in turn 
to determine the factorization of the elastic 
amplitude ${\cal M}_{n}$. This can be easily 
seen by recalling the LBK theorem and 
its failure to include virtual collinear 
configurations: one starts by separating 
the radiative amplitude in two contributions: 
one in which the radiation is emitted from the 
external legs, plus another in which the radiation 
is emitted from a particle within the hard scattering 
kernel: 
\begin{equation}\label{LBKeq}
{\cal M}_{n+1} = {\cal M}_{n+1}^{\rm ext} 
+ {\cal M}_{n+1}^{\rm int}\,.
\end{equation}
The amplitude ${\cal M}_{n+1}^{\rm int}$ 
can be obtained by means of the Ward identity from 
${\cal M}_{n+1}^{\rm ext}$. Focusing on the latter, 
consider the emission from an outgoing
fermion $i$: ${\cal M}_{n+1}^{\rm ext}$ 
takes the form 
\begin{equation}\label{LBKwrong}
{\cal M}_{n+1}^{\rm ext} = \bar u(p_i) (i e q_i \gamma^{\mu})
\, \frac{i(\slashed{p}_i+\slashed{k}+m)}{(p_i+k)^2 - m^2} 
\, {\cal M}_n(p_1,\ldots, p_i+k, \ldots p_n)\,,
\end{equation}
where ${\cal M}_n$ represents the elastic amplitude
(stripped off the spinor $\bar{u}(p_i)$). Within the 
LBK theorem, one expands the amplitude in the soft 
momentum $k$. As discussed, this expansion 
gives correct results only in the regime 
$p_i\cdot k/Q \ll m^2/Q$. But for parametrically small 
masses $m^2/Q < p_i\cdot k/Q <m$, naively expanding the 
elastic amplitude in the soft momentum $k$ misses a 
contribution in which the soft photon is emitted from 
internal particles collinear to the external leg, which 
is thus included neither in ${\cal M}_{n+1}^{\rm ext}$, 
nor in ${\cal M}_{n+1}^{\rm int}$. \Eqn{LBKwrong}
tells us that in order to understand the factorization 
properties of the radiative amplitude ${\cal M}_{n+1}$ we 
also need to obtain the correct factorization structure 
of the elastic amplitude ${\cal M}_{n}$, in presence of 
a small off-shellness $p_i\cdot k \sim m^2 \ll Q^2$. 
At leading power the factorization structure is known, 
see for instance~\cite{Collins:1989bt,Dixon:2008gr}, 
and it takes the following schematic form:
\begin{equation}\label{factLP_nonrad}
{\cal M}_n = H_n\times S_n\times\prod_{i=1}^n \, \frac{J_{i}}{{\cal J}_i}\,.
\end{equation}
In this equation the jet and soft functions $J_j$ and $S$
describe long-distance collinear and soft virtual radiation 
in ${\cal M}_n$. These functions are universal, i.e. they 
depend only on the colour and spin quantum numbers of the 
external states, and determine also the structure of 
collinear and soft singularities of the elastic amplitude. 
Note that one must divide each jet by its eikonal counterpart 
${\cal J}_i$, to avoid the double counting of soft and 
collinear divergences.

\begin{figure}[t]
\begin{center}
	\includegraphics[width = .3\textwidth]{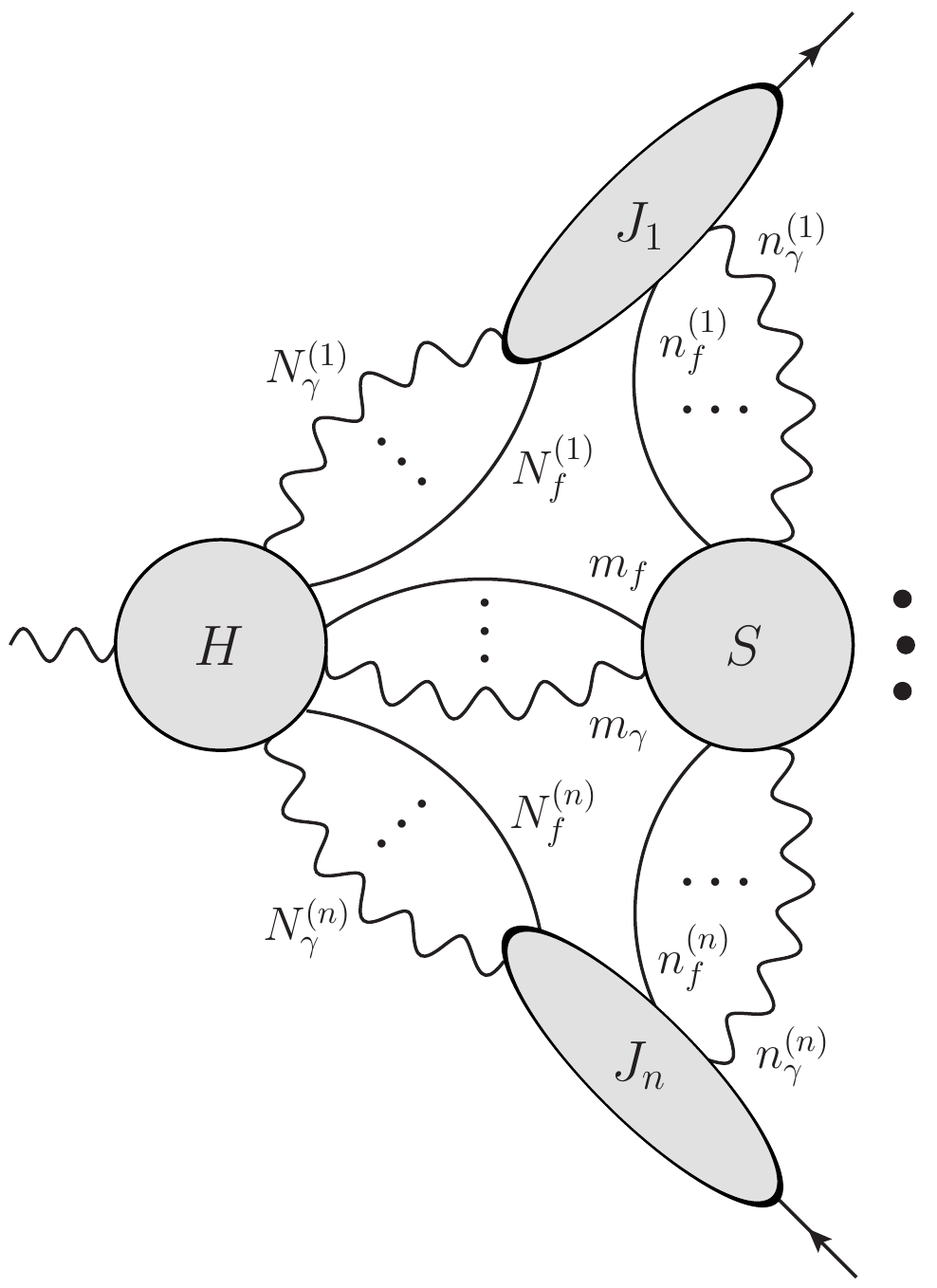}
	\caption{The reduced diagram for a vector boson 
		induced QED process with $n$ hard particles 
		in the final state. 
		Ellipses denote the presence of an arbitrary 
		number of photons/(anti-)fermions.}
	\label{fig:reduced_diagram}
\end{center}
\end{figure}

The corresponding factorization structure
at NLP for QED has been determined in 
\cite{Laenen:2020nrt}. In short, the task 
consists in obtaining a classification of 
the jet-like structures, consisting of virtual 
radiation collinear to any of the $n$ external 
hard particles, contributing at subleading power. 
This is obtained by power counting the 
\textit{pinch surfaces} that underlie 
the collinear (and soft) contributions one wishes 
to describe in terms of jet (and soft) functions, 
for a general QED scattering amplitude. The pinch 
surfaces are the solutions of the Landau 
equations~\cite{Landau:1959fi} and are 
represented by reduced diagrams in the 
Coleman-Norton picture~\cite{Coleman:1965xm}. 
In these diagrams, all off-shell lines are shrunk 
to a point, while the on-shell lines are kept and may be 
organised according to the nature of the singularity they 
embody, be it soft or collinear. This gives the general 
reduced diagram of fig. \ref{fig:reduced_diagram}, in which 
one distinguishes a soft ``blob" containing all lines carrying 
solely soft momentum, $n$ jets $J_i$ comprised of lines with 
momenta collinear to the respective external parton and 
lastly, a hard blob $H$ collecting all contracted, off-shell 
lines. 
The power counting is carried in light-cone 
coordinates: a generic momentum $k$ is decomposed 
as $k^\mu = (k^+,\vec{k}_\perp,k^-)$, where 
the light-cone components $k^{\pm}$ lie along 
the direction of two light-like vectors $n_i$ 
and $\bar n_i$, normalized such that 
$n_i \cdot \bar{n}_i = 1$, with $n_i$
along the direction of $p_i$. Soft 
and collinear momenta are taken to 
scale as 
\begin{align} 
\text{Soft:} \hspace{20pt} k^\mu\sim Q\left(\lambda^2,\lambda^2,\lambda^2 \right)\, ,
\qquad  \quad
\text{Collinear:} \hspace{20pt} k^\mu\sim Q\left(1,\lambda, \lambda^2\right) \,.  
\label{softcol} 
\end{align} 
One focuses on virtual corrections to a hard scattering 
configuration, for which all invariants $s_{ij} = (p_i+p_j)^2 
\sim Q^2$ involving external momenta are large compared 
to the energy of the radiated soft photon in 
${\cal M}_{n+1}$. Requiring the soft momentum 
to be of order $\lambda^2$ rather than $\lambda$ 
guarantees that the photon is soft with respect 
to all particles in the elastic amplitude. 
Within this power counting, a NLP 
quantity is suppressed by two powers in 
$\lambda$ with respect to the leading 
power contribution.

Using the momentum scaling in \eqn{softcol}, 
one first derives the \textit{superficial degree of divergence} 
of a particular reduced diagram $\mathcal{G}$ contained 
in fig. \ref{fig:reduced_diagram}, which is simply the 
$\lambda$-scaling of this diagram, 
$\mathcal{G}\sim \lambda^{\gamma_\mathcal{G}}$. 
In practice, $\gamma_\mathcal{G}$ can be expressed 
as function of the number of fermion and photon connections 
between the hard, soft and collinear subgraphs and, 
in presence of massive fermions, on the internal structure 
of the soft subgraph. Such a formula expresses, at any 
perturbative order, which pinch surfaces contribute 
up to NLP, which allows one to set up a consistent 
and complete NLP factorization framework for 
QED. The approach is analogous to the one taken for 
Yukawa theory in \cite{Gervais:2017yxv}, and it is a 
direct application of the well-known method first 
developed for deriving factorisation of the elastic 
amplitude at LP in Ref.~\cite{Sterman:1978bi}.
In general, one has that $\gamma_\mathcal{G} \geq 0$, 
independent of the number of hard particles in the 
final state. The $\gamma_\mathcal{G} = 0$ diagrams 
contain at most logarithmic singularities, while the 
$\gamma_\mathcal{G} > 0$ are finite and give a vanishing 
contribution in the $\lambda\rightarrow 0$ limit. For 
small but non-zero values of $\lambda$, the 
$\gamma_\mathcal{G} = 0$ diagrams form LP 
contributions, with the $\gamma_\mathcal{G} > 0$ 
diagrams acting as power corrections.

\begin{figure}[t]
	\centering
	\subfloat[ $\gamma=0$]{
		\makebox[.24\textwidth][c]{\includegraphics[width=.15\textwidth]{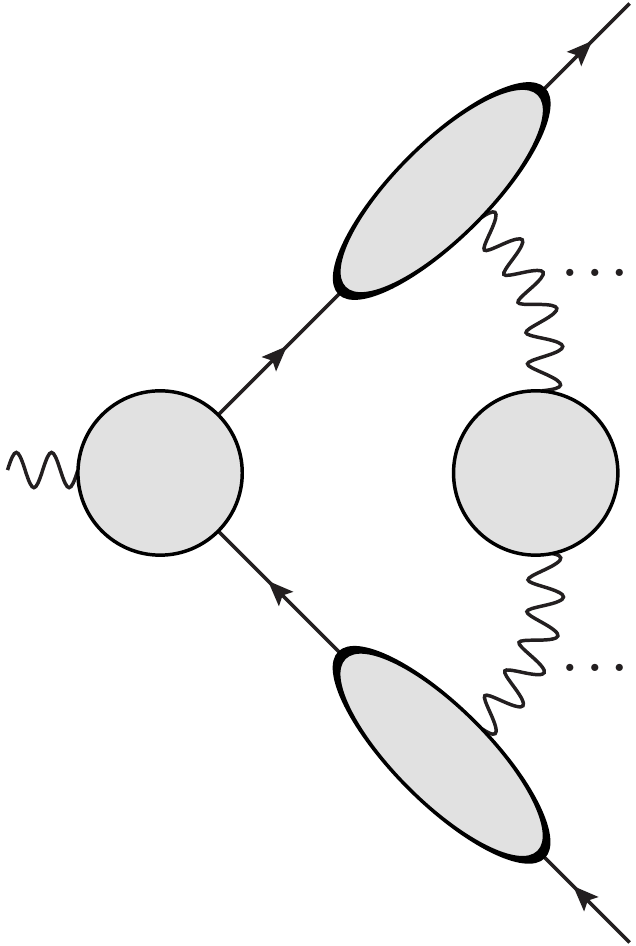}}
		\label{fig:red_diag_g0}}
	\subfloat[ $\gamma=1$]{
		\makebox[.24\textwidth][c]{\includegraphics[width=.15\textwidth]{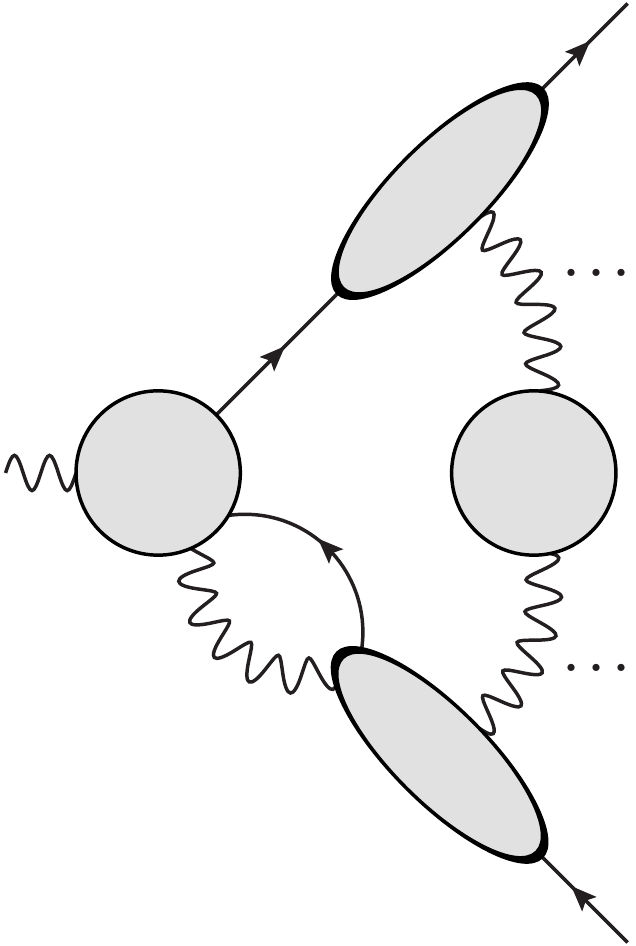}}
		\label{fig:red_diag_g1}}  
	\caption{Reduced diagrams for the process $\gamma \to f \bar{f}$. (a) $\gamma = 0$ (b) $\gamma = 1$, where a similar configuration exists with the double jet connection on the upper leg instead.}
	\label{fig:red_diag01}
\end{figure}
\begin{figure}[t]
	\centering
	\subfloat[]{
		\makebox[.24\textwidth][c]{\includegraphics[width=.16\textwidth]{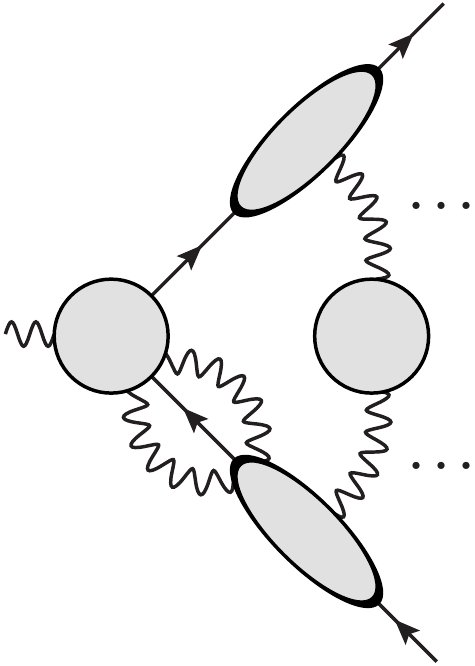}}
		\label{fig:red_diag_HJ_fgammagamma}
	}
	\subfloat[]{
		\makebox[.24\textwidth][c]{\includegraphics[width=.16\textwidth]{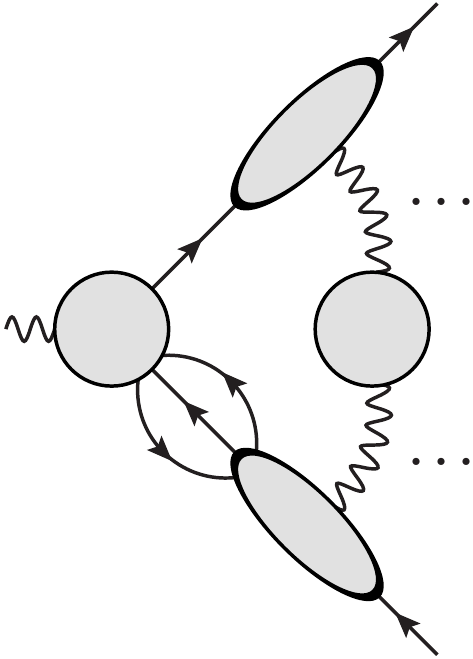}}
		\label{fig:red_diag_HJ_fff}
	}
	\subfloat[]{
		\makebox[.24\textwidth][c]{\includegraphics[width=.16\textwidth]{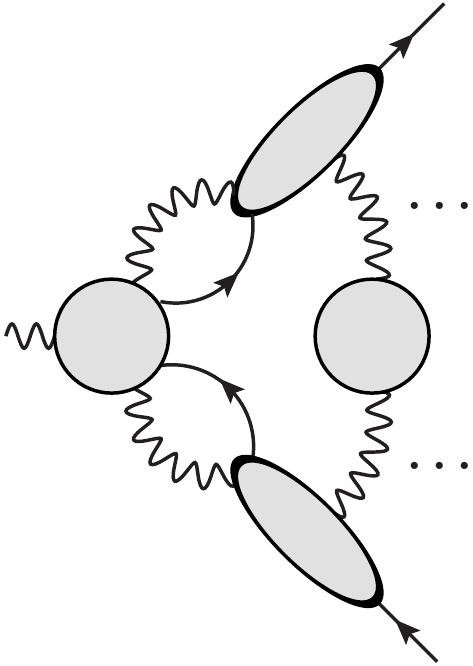}}
		\label{fig:red_diag_HJ_fgamma_x2}
	}
	\vspace{-5pt}
	\subfloat[]{
		\makebox[.24\textwidth][c]{\includegraphics[width=.16\textwidth]{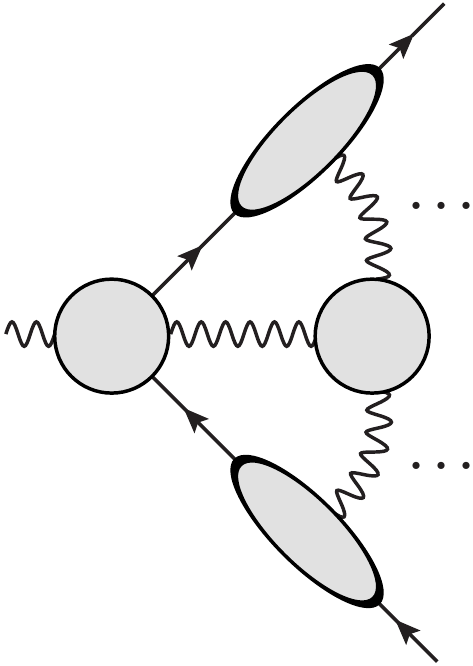}}
		\label{fig:red_diag_HS_gamma}
	}
	\subfloat[]{
		\makebox[.24\textwidth][c]{\includegraphics[width=.16\textwidth]{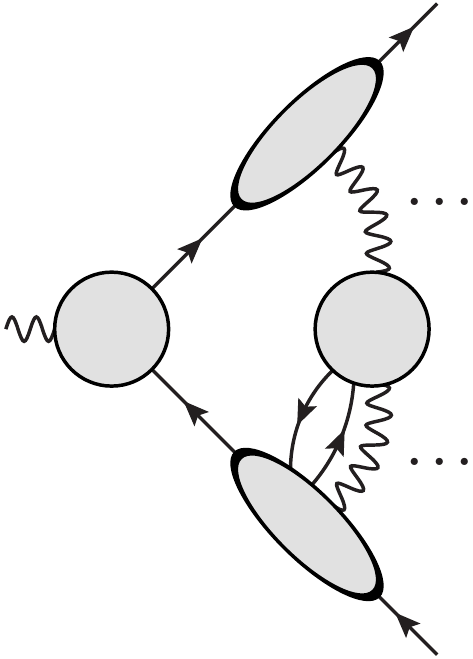}}
		\label{fig:red_diag_JS_ffgamma}
	}  
	\subfloat[]{
		\makebox[.24\textwidth][c]{\includegraphics[width=.16\textwidth]{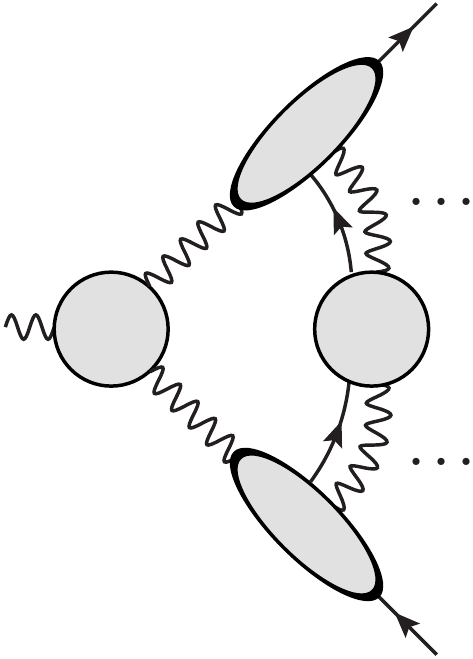}}
		\label{fig:red_diag_JS_fgamma_x2}
	}  
	\caption{Reduced diagrams for $\gamma = 2$ and $m = 0$. For $m \sim \lambda Q$, (e) and (f) contribute beyond NLP. (a) Two fermions and a photon connecting a collinear blob to the hard blob. (b) Triple fermion connection. (c) Two collinear blobs connected to the hard blob by a fermion and photon each. (d) A single photon connecting the hard and soft blobs. (e) Two fermions connecting a collinear blob to the soft blob. (f) Two collinear blobs connected to the soft blob by a fermion each.}
	\label{fig:red_diag2}
\end{figure}

After some work, one finds that the relevant 
diagrams contributing up to NLP are represented 
in figures \ref{fig:red_diag01} and \ref{fig:red_diag2}, 
where diagram (a) in fig.~\ref{fig:red_diag01} represents
the LP contribution, diagram (b) contributes starting at
$\sqrt{\rm N}$LP, and the diagrams in fig.~\ref{fig:red_diag2}
contribute at NLP. We see that they involve collinear 
jets with up to three collinear partons in the 
same collinear sector, and interactions between the 
collinear and the soft sector mediated by a soft quark,
or a soft quark-antiquark pair in addition to soft gluons. 
Restricting to the hard-collinear sector, up to 
NLP the content of figures \ref{fig:red_diag01} and 
\ref{fig:red_diag2} translates into the factorization 
formula 
\begin{align} \label{NLPfactorization}
\mathcal{M}_{\rm coll}^{\NLP} &= 
\bigg(\prod_{i=1}^n J^i_{(f)}\bigg) H \, S\, +
\sum_{i=1}^n \bigg(\prod_{j\neq i}J_{(f)}^j\bigg)
\Big[J^i_{(f\gamma)} \otimes H^i_{(f\gamma)}
+J^i_{(f\partial \gamma)} \otimes H^i_{(f\partial \gamma)}\Big]\,S 
\nonumber \\
&+\, \sum_{i=1}^n \bigg(\prod_{j\neq i}J_{(f)}^j\bigg)
J^i_{(f\gamma\gamma)} \otimes H^i_{(f\gamma\gamma)}\, S 
+ \sum_{i=1}^n \bigg(\prod_{j\neq i}J_{(f)}^j\bigg)
J^i_{(f\!f\!f)} \otimes H^i_{(f\!f\!f)}\, S \nonumber \\
&+ \sum_{1\leq i \leq j \leq n} \bigg(\prod_{k\neq i,j}J_{(f)}^k \bigg)
J^i_{(f\gamma)}J^j_{(f\gamma)} \otimes H^{ij}_{(f\gamma)(f\gamma)}\, S\,,
\end{align}
where the tensor product $\otimes$ denotes 
convolution and a contraction of spinor indices, 
we suppress the arguments of the functions and 
introduce the indices $i,j$, labelling the collinear 
sectors. For simplicity, we assume that the potential 
overlap between the soft and collinear regions 
has already been accounted for in a redefinition 
of the jet functions. 

The first term represents the LP contribution
from \eqn{factLP_nonrad}. The second term describes 
the effect of fig.~\ref{fig:red_diag_g1} and starts 
contributing at order $\lambda$. This implies 
that at order $\lambda^2$ we may expect a 
dependence of the hard function on the 
perpendicular momentum component of the collinear 
photon emerging from it, which can be re-expressed 
in terms of the $H_{(f\partial \gamma)}^i$ function. 
The third and fourth terms describe diagrams $(\rm a)$ 
and $(\rm b)$ in fig.~\ref{fig:red_diag2}, while 
diagram $(\rm c)$ corresponds to the last term. 
These contributions, as well as the $f\partial\gamma$-term, 
are strictly $\ord(\lambda^2)$, which implies that the 
soft function appearing in those terms is the same 
appearing at LP. 
\Eqn{NLPfactorization} has been obtained both 
for the case of massless and massive fermions, 
with a small parametric mass $m \sim \lambda Q$, 
and we refer to \cite{Laenen:2020nrt} for a 
discussion highlighting differences between 
the two cases. The functions in \eqn{NLPfactorization}
can be defined as expectation values of fields, 
currents and Wilson lines in QED, in analogy 
to the LP jet and soft functions.

Let us conclude by noting that approaches based on 
SCET \cite{Larkoski:2014bxa,Beneke:2019oqx} contain 
similar functions describing amplitudes at the 
next-to-leading power. The $J_{f \gamma}$, $J_{f\partial\gamma}$, 
and $J_{f\!f\!f}$ jets in \eqn{NLPfactorization} 
are related to matrix elements of the operators 
$J^{B1}$, $J^{B2}$, $J^{C1}$ in the position-space 
SCET in \cite{Beneke:2019oqx}, and 
to matrix elements of the part of $N$-jet operators 
${\cal O}_{N}^{(1,X)}$, ${\cal O}_{N}^{(2,X\delta)}$, 
${\cal O}_{N}^{(2,X^2)}$ corresponding to a specific 
collinear direction in the label formulation of SCET 
\cite{Larkoski:2014bxa}.

\subsection{Outlook}
\label{sec:outlook}

The study of scattering amplitudes in the limit 
where external radiation becomes soft is relevant 
both for obtaining a deeper understanding 
of quantum field theory and for phenomenology. 
In presence of soft radiation, scattering amplitudes 
are naturally written as a power expansion in the 
energy of the emitted radiation. The coefficients 
of this expansion factorize. Determining the 
factorization structure constitutes the first 
step toward the resummation of large logarithms 
in physical observables which are sensitive to the 
energy scale of soft radiation, but it useful 
for other applications, too. For instance, 
determining the factorization properties of
soft and collinear radiation can be useful 
to derive analytic expression for amplitudes 
in soft and collinear limits, which can be used 
to stabilize and speed up numerical evaluations 
of amplitudes in such limits \cite{Engel:2021ccn}.
In this context, current efforts are devoted to 
understand the factorization of amplitudes at 
next-to-leading power. This can be done following 
a diagrammatic approach, or by means of effective 
field theories. 

In this review we have discussed recent 
developments within the diagrammatic 
approach applied to QED, which will be 
useful for the phenomenological applications
at $e^+ e^-$ colliders. Among various 
contributions, the description of soft 
radiation beyond leading power requires to 
be able to take into account the emission 
of soft radiation from clusters of collinear 
virtual particles, which flows in or out 
from the hard scattering process, and are 
known as \emph{radiative jets}. 
Such contributions are absent at leading power, 
where one needs to take into account only the 
eikonal emission of soft photons/gluons from 
the hard particles of the process. The first 
radiative jets was identified long ago 
\cite{DelDuca:1990gz}, and studied at one 
loop in Drell-Yan in \cite{Bonocore:2015esa,Bonocore:2016awd}. 
More recently, a classification of all 
possible jets contributing at NLP in QED 
has been derived in \cite{Laenen:2020nrt}, 
and we have sketched this derivation 
in section \ref{sec:fact}. 

This classification constitutes also 
the first step toward the development 
of a theory for the resummation of NLP 
(threshold) logarithms. Within a diagrammatic 
approach, one envisage to study the combinatorial 
properties of soft photons/quarks emitted 
from the various functions in the 
factorization formula by means of 
tools such as the replica trick, 
to understand the exponentiation 
structure of soft radiation at 
cross section level.

\acknowledgments

This work is supported by Fellini Fellowship for 
Innovation at INFN, funded by the European Union's Horizon 2020 
research programme under the Marie Sk\l{}odowska-Curie Cofund Action, 
grant agreement no. 754496.





\begin{thebibliography}{100}

\bibitem{Campbell:2022qmc}
J.M.~Campbell et~al., \emph{{Event Generators for High-Energy Physics
  Experiments}},  in \emph{{2022 Snowmass Summer Study}}, 3, 2022
  [\href{https://arxiv.org/abs/2203.11110}{{\ttfamily 2203.11110}}].

\bibitem{Sjostrand:2006za}
T.~Sjostrand, S.~Mrenna and P.Z.~Skands, \emph{{PYTHIA 6.4 Physics and
  Manual}}, \href{https://doi.org/10.1088/1126-6708/2006/05/026}{\emph{JHEP}
  {\bfseries 05} (2006) 026}
  [\href{https://arxiv.org/abs/hep-ph/0603175}{{\ttfamily hep-ph/0603175}}].

\bibitem{Sjostrand:2014zea}
T.~Sj\"ostrand, S.~Ask, J.R.~Christiansen, R.~Corke, N.~Desai, P.~Ilten et~al.,
  \emph{{An introduction to PYTHIA 8.2}},
  \href{https://doi.org/10.1016/j.cpc.2015.01.024}{\emph{Comput. Phys. Commun.}
  {\bfseries 191} (2015) 159}
  [\href{https://arxiv.org/abs/1410.3012}{{\ttfamily 1410.3012}}].

\bibitem{Sjostrand:1985xi}
T.~Sj{\"o}strand, \emph{{A model for initial state parton showers}},
  {\emph{Phys. Lett.} {\bfseries B157} (1985) 321}.

\bibitem{Sjostrand:2004ef}
T.~Sj{\"o}strand and P.Z.~Skands, \emph{{Transverse-momentum-ordered showers
  and interleaved multiple interactions}}, {\emph{Eur. Phys. J.} {\bfseries
  C39} (2005) 129} [\href{https://arxiv.org/abs/hep-ph/0408302}{{\ttfamily
  hep-ph/0408302}}].

\bibitem{Norrbin:2000uu}
E.~Norrbin and T.~Sj{\"o}strand, \emph{{QCD radiation off heavy particles}},
  {\emph{Nucl. Phys.} {\bfseries B603} (2001) 297}
  [\href{https://arxiv.org/abs/hep-ph/0010012}{{\ttfamily hep-ph/0010012}}].

\bibitem{Kleiss:1989de}
R.~Kleiss et~al., \emph{{MONTE CARLOS FOR ELECTROWEAK PHYSICS}},  in \emph{{LEP
  Physics Workshop}}, 1989.

\bibitem{Nicrosini:1986sm}
O.~Nicrosini and L.~Trentadue, \emph{{Soft Photons and Second Order Radiative
  Corrections to e+ e- ---\ensuremath{>} Z0}},
  \href{https://doi.org/10.1016/0370-2693(87)90819-7}{\emph{Phys. Lett. B}
  {\bfseries 196} (1987) 551}.

\bibitem{Miu:1998ju}
G.~Miu and T.~Sj{\"o}strand, \emph{{W production in an improved parton-shower
  approach}}, \href{https://doi.org/10.1016/S0370-2693(99)00068-4}{\emph{Phys.
  Lett.} {\bfseries B449} (1999) 313}
  [\href{https://arxiv.org/abs/hep-ph/9812455}{{\ttfamily hep-ph/9812455}}].

\bibitem{Brooks:2020upa}
H.~Brooks, C.T.~Preuss and P.~Skands, \emph{{Sector Showers for Hadron
  Collisions}}, \href{https://doi.org/10.1007/JHEP07(2020)032}{\emph{JHEP}
  {\bfseries 07} (2020) 032}
  [\href{https://arxiv.org/abs/2003.00702}{{\ttfamily 2003.00702}}].

\bibitem{Kleiss:2017iir}
R.~Kleiss and R.~Verheyen, \emph{{Final-state QED Multipole Radiation in
  Antenna Parton Showers}},
  \href{https://doi.org/10.1007/JHEP11(2017)182}{\emph{JHEP} {\bfseries 11}
  (2017) 182} [\href{https://arxiv.org/abs/1709.04485}{{\ttfamily
  1709.04485}}].

\bibitem{Skands:2020lkd}
P.~Skands and R.~Verheyen, \emph{{Multipole photon radiation in the Vincia
  parton shower}},
  \href{https://doi.org/10.1016/j.physletb.2020.135878}{\emph{Phys. Lett. B}
  {\bfseries 811} (2020) 135878}
  [\href{https://arxiv.org/abs/2002.04939}{{\ttfamily 2002.04939}}].

\bibitem{Kuhn55thehungarian}
H.W.~Kuhn and B.~Yaw, \emph{The hungarian method for the assignment problem},
  {\emph{Naval Res. Logist. Quart} (1955) 83}.

\bibitem{Munkres1957Assignment}
J.R.~Munkres, \emph{{Algorithms for the Assignment and Transportation
  Problems}}, {\emph{Journal of the Society for Industrial and Applied
  Mathematics} {\bfseries 5} (1957) 32}.

\bibitem{10.1007/BF02278710}
R.~Jonker and A.~Volgenant, \emph{A shortest augmenting path algorithm for
  dense and sparse linear assignment problems},
  \href{https://doi.org/10.1007/BF02278710}{\emph{Computing} {\bfseries 38}
  (1987) 325–340}.

\bibitem{Brooks:2019xso}
H.~Brooks and P.~Skands, \emph{{Coherent showers in decays of colored
  resonances}}, \href{https://doi.org/10.1103/PhysRevD.100.076006}{\emph{Phys.
  Rev. D} {\bfseries 100} (2019) 076006}
  [\href{https://arxiv.org/abs/1907.08980}{{\ttfamily 1907.08980}}].

\bibitem{Hoche:2015sya}
S.~H{\"o}che and S.~Prestel, \emph{{The midpoint between dipole and parton
  showers}}, \href{https://doi.org/10.1140/epjc/s10052-015-3684-2}{\emph{Eur.
  Phys. J.} {\bfseries C75} (2015) 461}
  [\href{https://arxiv.org/abs/1506.05057}{{\ttfamily 1506.05057}}].

\bibitem{Catani:1996vz}
S.~Catani and M.H.~Seymour, \emph{{A general algorithm for calculating jet
  cross sections in NLO QCD}}, {\emph{Nucl. Phys.} {\bfseries B485} (1997) 291}
  [\href{https://arxiv.org/abs/hep-ph/9605323}{{\ttfamily hep-ph/9605323}}].

\bibitem{Dittmaier:1999mb}
S.~Dittmaier, \emph{{A General approach to photon radiation off fermions}},
  \href{https://doi.org/10.1016/S0550-3213(99)00563-5}{\emph{Nucl. Phys. B}
  {\bfseries 565} (2000) 69}
  [\href{https://arxiv.org/abs/hep-ph/9904440}{{\ttfamily hep-ph/9904440}}].

\bibitem{Catani:2002hc}
S.~Catani, S.~Dittmaier, M.H.~Seymour and Z.~Trocsanyi, \emph{{The dipole
  formalism for next-to-leading order QCD calculations with massive partons}},
  {\emph{Nucl. Phys.} {\bfseries B627} (2002) 189}
  [\href{https://arxiv.org/abs/hep-ph/0201036}{{\ttfamily hep-ph/0201036}}].

\bibitem{Hoeche:2011fd}
S.~H{\"o}che, F.~Krauss, M.~Sch{\"o}nherr and F.~Siegert, \emph{{A critical
  appraisal of NLO+PS matching methods}}, {\emph{JHEP} {\bfseries 09} (2012)
  049} [\href{https://arxiv.org/abs/1111.1220}{{\ttfamily 1111.1220}}].

\bibitem{Prestel:2019neg}
S.~Prestel and M.~Spannowsky, \emph{{HYTREES: Combining Matrix Elements and
  Parton Shower for Hypothesis Testing}},
  \href{https://doi.org/10.1140/epjc/s10052-019-7030-y}{\emph{Eur. Phys. J. C}
  {\bfseries 79} (2019) 546}
  [\href{https://arxiv.org/abs/1901.11035}{{\ttfamily 1901.11035}}].

\bibitem{Gellersen:2021caw}
L.~Gellersen, S.~Prestel and M.~Spannowsky, \emph{{Coloring mixed QCD/QED
  evolution}},  \href{https://arxiv.org/abs/2109.09706}{{\ttfamily
  2109.09706}}.

\bibitem{Schonherr:2017qcj}
M.~Sch{\"o}nherr, \emph{{An automated subtraction of NLO EW infrared
  divergences}},
  \href{https://doi.org/10.1140/epjc/s10052-018-5600-z}{\emph{Eur. Phys. J.}
  {\bfseries C78} (2018) 119}
  [\href{https://arxiv.org/abs/1712.07975}{{\ttfamily 1712.07975}}].

\bibitem{Kniehl:1992ra}
B.A.~Kniehl and L.~Lonnblad, \emph{{Renormalization scales in electroweak
  physics: and Photon radiation in the dipole model and in the Ariadne
  program}},  in \emph{{Workshop on Photon Radiation from Quarks}}, 2, 1992.

\bibitem{Lonnblad:2012hz}
L.~L{\"o}nnblad, \emph{{Fooling Around with the Sudakov Veto Algorithm}},
  \href{https://doi.org/10.1140/epjc/s10052-013-2350-9}{\emph{Eur.Phys.J.}
  {\bfseries C73} (2013) 2350}
  [\href{https://arxiv.org/abs/1211.7204}{{\ttfamily 1211.7204}}].

\bibitem{Catani:1992ua}
S.~Catani, L.~Trentadue, G.~Turnock and B.R.~Webber, \emph{{Resummation of
  large logarithms in $e^+ e^-$ event shape distributions}},
  \href{https://doi.org/10.1016/0550-3213(93)90271-P}{\emph{Nucl. Phys.}
  {\bfseries B407} (1993) 3}.

\bibitem{Marchesini:1987cf}
G.~Marchesini and B.R.~Webber, \emph{{Monte Carlo Simulation of General Hard
  Processes with Coherent QCD Radiation}},
  \href{https://doi.org/10.1016/0550-3213(88)90089-2}{\emph{Nucl. Phys.}
  {\bfseries B310} (1988) 461}.

\bibitem{Gieseke:2003rz}
S.~Gieseke, P.~Stephens and B.~Webber, \emph{{New formalism for QCD parton
  showers}}, {\emph{JHEP} {\bfseries 12} (2003) 045}
  [\href{https://arxiv.org/abs/hep-ph/0310083}{{\ttfamily hep-ph/0310083}}].

\bibitem{Bahr:2008pv}
M.~Bahr et~al., \emph{{Herwig++ Physics and Manual}},
  \href{https://doi.org/10.1140/epjc/s10052-008-0798-9}{\emph{Eur. Phys. J.}
  {\bfseries C58} (2008) 639}
  [\href{https://arxiv.org/abs/0803.0883}{{\ttfamily 0803.0883}}].

\bibitem{Forshaw:2019ver}
J.R.~Forshaw, J.~Holguin and S.~Pl\"atzer, \emph{{Parton branching at amplitude
  level}}, \href{https://doi.org/10.1007/JHEP08(2019)145}{\emph{JHEP}
  {\bfseries 08} (2019) 145}
  [\href{https://arxiv.org/abs/1905.08686}{{\ttfamily 1905.08686}}].

\bibitem{Richardson:2018pvo}
P.~Richardson and S.~Webster, \emph{{Spin Correlations in Parton Shower
  Simulations}},
  \href{https://doi.org/10.1140/epjc/s10052-019-7429-5}{\emph{Eur. Phys. J. C}
  {\bfseries 80} (2020) 83} [\href{https://arxiv.org/abs/1807.01955}{{\ttfamily
  1807.01955}}].

\bibitem{Olsson:2019wvr}
J.~Olsson, S.~Pl\"atzer and M.~Sj\"odahl, \emph{{Resampling Algorithms for High
  Energy Physics Simulations}},
  \href{https://doi.org/10.1140/epjc/s10052-020-08500-y}{\emph{Eur. Phys. J. C}
  {\bfseries 80} (2020) 934}
  [\href{https://arxiv.org/abs/1912.02436}{{\ttfamily 1912.02436}}].

\bibitem{vonWeizsacker:1934nji}
C.F.~von Weizsacker, \emph{{Radiation emitted in collisions of very fast
  electrons}}, \href{https://doi.org/10.1007/BF01333110}{\emph{Z. Phys.}
  {\bfseries 88} (1934) 612}.

\bibitem{Williams:1934ad}
E.J.~Williams, \emph{{Nature of the high-energy particles of penetrating
  radiation and status of ionization and radiation formulae}},
  \href{https://doi.org/10.1103/PhysRev.45.729}{\emph{Phys. Rev.} {\bfseries
  45} (1934) 729}.

\bibitem{Budnev:1975poe}
V.M.~Budnev, I.F.~Ginzburg, G.V.~Meledin and V.G.~Serbo, \emph{{The Two photon
  particle production mechanism. Physical problems. Applications. Equivalent
  photon approximation}},
  \href{https://doi.org/10.1016/0370-1573(75)90009-5}{\emph{Phys. Rept.}
  {\bfseries 15} (1975) 181}.

\bibitem{Bertone:2019hks}
V.~Bertone, M.~Cacciari, S.~Frixione and G.~Stagnitto, \emph{{The partonic
  structure of the electron at the next-to-leading logarithmic accuracy in
  QED}}, \href{https://doi.org/10.1007/JHEP03(2020)135}{\emph{JHEP} {\bfseries
  03} (2020) 135} [\href{https://arxiv.org/abs/1911.12040}{{\ttfamily
  1911.12040}}].

\bibitem{Frixione:2019lga}
S.~Frixione, \emph{{Initial conditions for electron and photon structure and
  fragmentation functions}},
  \href{https://doi.org/10.1007/JHEP11(2019)158}{\emph{JHEP} {\bfseries 11}
  (2019) 158} [\href{https://arxiv.org/abs/1909.03886}{{\ttfamily
  1909.03886}}].

\bibitem{Altarelli:1977zs}
G.~Altarelli and G.~Parisi, \emph{{Asymptotic Freedom in Parton Language}},
  \href{https://doi.org/10.1016/0550-3213(77)90384-4}{\emph{Nucl. Phys. B}
  {\bfseries 126} (1977) 298}.

\bibitem{Gribov:1972ri}
V.N.~Gribov and L.N.~Lipatov, \emph{{Deep inelastic e p scattering in
  perturbation theory}}, {\emph{Sov. J. Nucl. Phys.} {\bfseries 15} (1972)
  438}.

\bibitem{Lipatov:1974qm}
L.N.~Lipatov, \emph{{The parton model and perturbation theory}}, {\emph{Yad.
  Fiz.} {\bfseries 20} (1974) 181}.

\bibitem{Dokshitzer:1977sg}
Y.L.~Dokshitzer, \emph{{Calculation of the Structure Functions for Deep
  Inelastic Scattering and e+ e- Annihilation by Perturbation Theory in Quantum
  Chromodynamics.}}, {\emph{Sov. Phys. JETP} {\bfseries 46} (1977) 641}.

\bibitem{Skrzypek:1990qs}
M.~Skrzypek and S.~Jadach, \emph{{Exact and approximate solutions for the
  electron nonsinglet structure function in QED}},
  \href{https://doi.org/10.1007/BF01483573}{\emph{Z. Phys. C} {\bfseries 49}
  (1991) 577}.

\bibitem{Hoeche:2009xc}
S.~H{\"o}che, S.~Schumann and F.~Siegert, \emph{{Hard photon production and
  matrix-element parton-shower merging}},
  \href{https://doi.org/10.1103/PhysRevD.81.034026}{\emph{Phys. Rev.}
  {\bfseries D81} (2010) 034026}
  [\href{https://arxiv.org/abs/0912.3501}{{\ttfamily 0912.3501}}].

\bibitem{Yennie:1961ad}
D.R.~Yennie, S.C.~Frautschi and H.~Suura, \emph{{The infrared divergence
  phenomena and high-energy processes}},
  \href{https://doi.org/10.1016/0003-4916(61)90151-8}{\emph{Annals Phys.}
  {\bfseries 13} (1961) 379}.

\bibitem{Kuraev:1985hb}
E.A.~Kuraev and V.S.~Fadin, \emph{{On Radiative Corrections to e+ e- Single
  Photon Annihilation at High-Energy}}, {\emph{Sov. J. Nucl. Phys.} {\bfseries
  41} (1985) 466}.

\bibitem{Bardin:1993bh}
D.Y.~Bardin, M.S.~Bilenky, A.~Olchevski and T.~Riemann, \emph{{Off-shell W pair
  production in e+ e- annihilation: Initial state radiation}},
  \href{https://doi.org/10.1016/0370-2693(93)91305-7}{\emph{Phys. Lett. B}
  {\bfseries 308} (1993) 403}
  [\href{https://arxiv.org/abs/hep-ph/9507277}{{\ttfamily hep-ph/9507277}}].

\bibitem{Beenakker:1994vn}
W.~Beenakker and A.~Denner, \emph{{Standard model predictions for $W$ pair
  production in electron - positron collisions}},
  \href{https://doi.org/10.1142/S0217751X94001965}{\emph{Int. J. Mod. Phys. A}
  {\bfseries 9} (1994) 4837}.

\bibitem{Montagna:1994qu}
G.~Montagna, O.~Nicrosini, G.~Passarino and F.~Piccinini, \emph{{Semianalytical
  and Monte Carlo results for the production of four fermions in e+ e-
  collisions}}, \href{https://doi.org/10.1016/0370-2693(95)00098-6}{\emph{Phys.
  Lett. B} {\bfseries 348} (1995) 178}
  [\href{https://arxiv.org/abs/hep-ph/9411332}{{\ttfamily hep-ph/9411332}}].

\bibitem{Berends:1994pv}
F.A.~Berends, R.~Pittau and R.~Kleiss, \emph{{All electroweak four-fermion
  processes in electron-positron collisions}}, {\emph{Nucl. Phys.} {\bfseries
  B424} (1994) 308} [\href{https://arxiv.org/abs/hep-ph/9404313}{{\ttfamily
  hep-ph/9404313}}].

\bibitem{Schumann:2007mg}
S.~Schumann and F.~Krauss, \emph{{A parton shower algorithm based on
  Catani-Seymour dipole factorisation}}, {\emph{JHEP} {\bfseries 03} (2008)
  038} [\href{https://arxiv.org/abs/0709.1027}{{\ttfamily 0709.1027}}].

\bibitem{Nagy:2006kb}
Z.~Nagy and D.E.~Soper, \emph{{A new parton shower algorithm: Shower evolution,
  matching at leading and next-to-leading order level}},
  \href{https://arxiv.org/abs/hep-ph/0601021}{{\ttfamily hep-ph/0601021}}.

\bibitem{Dinsdale:2007mf}
M.~Dinsdale, M.~Ternick and S.~Weinzierl, \emph{{Parton showers from the dipole
  formalism}}, {\emph{Phys. Rev.} {\bfseries D76} (2007) 094003}
  [\href{https://arxiv.org/abs/0709.1026}{{\ttfamily 0709.1026}}].

\bibitem{Platzer:2011dq}
S.~Pl{\"a}tzer and M.~Sj{\"o}dahl, \emph{{The Sudakov Veto Algorithm
  Reloaded}}, \href{https://doi.org/10.1140/epjp/i2012-12026-x}{\emph{Eur.
  Phys. J. Plus} {\bfseries 127} (2012) 26}
  [\href{https://arxiv.org/abs/1108.6180}{{\ttfamily 1108.6180}}].

\bibitem{Ohl:1996fi}
T.~Ohl, \emph{{CIRCE version 1.0: Beam spectra for simulating linear collider
  physics}}, \href{https://doi.org/10.1016/S0010-4655(96)00167-1}{\emph{Comput.
  Phys. Commun.} {\bfseries 101} (1997) 269}
  [\href{https://arxiv.org/abs/hep-ph/9607454}{{\ttfamily hep-ph/9607454}}].

\bibitem{Badelek:2001xb}
{\scshape ECFA/DESY Photon Collider Working Group} collaboration, \emph{{TESLA
  Technical Design Report, Part VI, Chapter 1: Photon collider at TESLA}},
  {\emph{Int. J. Mod. Phys.} {\bfseries A19} (2004) 5097}
  [\href{https://arxiv.org/abs/1204.5201}{{\ttfamily 1204.5201}}].

\bibitem{Zarnecki:2002qr}
A.F.~{\.Z}arnecki, \emph{{CompAZ: Parametrization of the luminosity spectra for
  the photon collider}}, {\emph{Acta Phys. Polon.} {\bfseries B34} (2003) 2741}
  [\href{https://arxiv.org/abs/hep-ex/0207021}{{\ttfamily hep-ex/0207021}}].

\bibitem{Archibald:2008zzb}
J.~Archibald, S.~H{\"o}che, F.~Krauss, F.~Siegert, T.~Gleisberg,
  M.~Sch{\"o}nherr et~al., \emph{{Simulation of photon-photon interactions in
  hadron collisions with SHERPA}},
  \href{https://doi.org/10.1016/j.nuclphysbps.2008.07.027}{\emph{Nucl. Phys.
  Proc. Suppl.} {\bfseries 179-180} (2008) 218}.

\bibitem{vonWeizsacker:1934sx}
C.~von Weizs{\"a}cker, \emph{{Radiation emitted in collisions of very fast
  electrons}}, \href{https://doi.org/10.1007/BF01333110}{\emph{Z.Phys.}
  {\bfseries 88} (1934) 612}.

\bibitem{Budnev:1974de}
V.M.~Budnev, I.F.~Ginzburg, G.V.~Meledin and V.G.~Serbo, \emph{{The two photon
  particle production mechanism. Physical problems. Applications. Equivalent
  photon approximation}}, {\emph{Phys. Rept.} {\bfseries 15} (1974) 181}.

\bibitem{Gluck:1991ee}
M.~Gl{\"u}ck, E.~Reya and A.~Vogt, \emph{{Parton structure of the photon beyond
  the leading order}}, {\emph{Phys. Rev.} {\bfseries D45} (1992) 3986}.

\bibitem{Gluck:1991jc}
M.~Gl{\"u}ck, E.~Reya and A.~Vogt, \emph{{Photonic parton distributions}},
  {\emph{Phys. Rev.} {\bfseries D46} (1992) 1973}.

\bibitem{Krauss:2022ajk}
F.~Krauss, A.~Price and M.~Sch\"onherr, \emph{{YFS Resummation for Future
  Lepton-Lepton Colliders in SHERPA}},
  \href{https://arxiv.org/abs/2203.10948}{{\ttfamily 2203.10948}}.

\bibitem{Jadach:2000ir}
S.~Jadach, B.F.L.~Ward and Z.~Was, \emph{{Coherent exclusive exponentiation for
  precision Monte Carlo calculations}},
  \href{https://doi.org/10.1103/PhysRevD.63.113009}{\emph{Phys. Rev. D}
  {\bfseries 63} (2001) 113009}
  [\href{https://arxiv.org/abs/hep-ph/0006359}{{\ttfamily hep-ph/0006359}}].

\bibitem{Krauss:2018djz}
F.~Krauss, J.M.~Lindert, R.~Linten and M.~Sch{\"o}nherr, \emph{{Accurate
  simulation of W, Z and Higgs boson decays in Sherpa}},
  \href{https://doi.org/10.1140/epjc/s10052-019-6614-x}{\emph{Eur. Phys. J.}
  {\bfseries C79} (2019) 143}
  [\href{https://arxiv.org/abs/1809.10650}{{\ttfamily 1809.10650}}].

\bibitem{Schonherr:2008av}
M.~Sch{\"o}nherr and F.~Krauss, \emph{{Soft Photon Radiation in Particle Decays
  in SHERPA}}, \href{https://doi.org/10.1088/1126-6708/2008/12/018}{\emph{JHEP}
  {\bfseries 12} (2008) 018} [\href{https://arxiv.org/abs/0810.5071}{{\ttfamily
  0810.5071}}].

\bibitem{price2021yfs}
F.~Krauss, A.~Price and M.~Sch{\"o}nherr, \emph{To appear}, .

\bibitem{Price:2021ds}
A.~Price, \emph{{Precision Simulations for Future Colliders}}, .

\bibitem{Jadach:2019huc}
S.~Jadach and M.~Skrzypek, \emph{{Theory challenges at future lepton
  colliders}}, \href{https://doi.org/10.5506/APhysPolB.50.1705}{\emph{Acta
  Phys. Polon. B} {\bfseries 50} (2019) 1705}
  [\href{https://arxiv.org/abs/1911.09202}{{\ttfamily 1911.09202}}].

\bibitem{Jadach:1988gb}
S.~Jadach and B.~Ward, \emph{{Yfs2: The Second Order Monte Carlo for Fermion
  Pair Production at {LEP} / {SLC} With the Initial State Radiation of Two Hard
  and Multiple Soft Photons}},
  \href{https://doi.org/10.1016/0010-4655(90)90020-2}{\emph{Comput. Phys.
  Commun.} {\bfseries 56} (1990) 351}.

\bibitem{Jadach:1995sp}
S.~Jadach, W.~Placzek, M.~Skrzypek and B.F.L.~Ward, \emph{{Gauge invariant YFS
  exponentiation of (un)stable W+ W- production at and beyond LEP-2 energies}},
  \href{https://doi.org/10.1103/PhysRevD.54.5434}{\emph{Phys. Rev. D}
  {\bfseries 54} (1996) 5434}
  [\href{https://arxiv.org/abs/hep-ph/9606429}{{\ttfamily hep-ph/9606429}}].

\bibitem{Bardin:1993mc}
D.Y.~Bardin, W.~Beenakker and A.~Denner, \emph{{The Coulomb singularity in
  off-shell W pair production}},
  \href{https://doi.org/10.1016/0370-2693(93)91595-E}{\emph{Phys. Lett. B}
  {\bfseries 317} (1993) 213}.

\bibitem{Fadin:1993kg}
V.S.~Fadin, V.A.~Khoze and A.D.~Martin, \emph{{On $W^+ W^-$ production near
  threshold}}, \href{https://doi.org/10.1016/0370-2693(93)90575-3}{\emph{Phys.
  Lett. B} {\bfseries 311} (1993) 311}.

\bibitem{Fadin:1995fp}
V.S.~Fadin, V.A.~Khoze, A.D.~Martin and W.J.~Stirling, \emph{{Higher order
  Coulomb corrections to the threshold $e^{+} e^{-} \to W^{+} W^{-}$
  cross-section}},
  \href{https://doi.org/10.1016/0370-2693(95)01194-U}{\emph{Phys. Lett. B}
  {\bfseries 363} (1995) 112}
  [\href{https://arxiv.org/abs/hep-ph/9507422}{{\ttfamily hep-ph/9507422}}].

\bibitem{Chapovsky:1999kv}
A.P.~Chapovsky and V.A.~Khoze, \emph{{Screened Coulomb ansatz for the
  nonfactorizable radiative corrections to the off-shell W+ W- production}},
  \href{https://doi.org/10.1007/s100529900070}{\emph{Eur. Phys. J. C}
  {\bfseries 9} (1999) 449}
  [\href{https://arxiv.org/abs/hep-ph/9902343}{{\ttfamily hep-ph/9902343}}].

\bibitem{Krauss:2001iv}
F.~Krauss, R.~Kuhn and G.~Soff, \emph{{AMEGIC++ 1.0: A Matrix Element Generator
  In C++}}, {\emph{JHEP} {\bfseries 02} (2002) 044}
  [\href{https://arxiv.org/abs/hep-ph/0109036}{{\ttfamily hep-ph/0109036}}].

\bibitem{Gleisberg:2008fv}
T.~Gleisberg and S.~H{\"o}che, \emph{{Comix, a new matrix element generator}},
  \href{https://doi.org/10.1088/1126-6708/2008/12/039}{\emph{JHEP} {\bfseries
  12} (2008) 039} [\href{https://arxiv.org/abs/0808.3674}{{\ttfamily
  0808.3674}}].

\bibitem{Buccioni:2019sur}
F.~Buccioni, J.-N.~Lang, J.M.~Lindert, P.~Maierh\"ofer, S.~Pozzorini, H.~Zhang
  et~al., \emph{{OpenLoops 2}},
  \href{https://doi.org/10.1140/epjc/s10052-019-7306-2}{\emph{Eur. Phys. J. C}
  {\bfseries 79} (2019) 866}
  [\href{https://arxiv.org/abs/1907.13071}{{\ttfamily 1907.13071}}].

\bibitem{Cascioli:2011va}
F.~Cascioli, P.~Maierhofer and S.~Pozzorini, \emph{{Scattering Amplitudes with
  Open Loops}},
  \href{https://doi.org/10.1103/PhysRevLett.108.111601}{\emph{Phys. Rev. Lett.}
  {\bfseries 108} (2012) 111601}
  [\href{https://arxiv.org/abs/1111.5206}{{\ttfamily 1111.5206}}].

\bibitem{Actis:2016mpe}
S.~Actis, A.~Denner, L.~Hofer, J.-N.~Lang, A.~Scharf and S.~Uccirati,
  \emph{{RECOLA: REcursive Computation of One-Loop Amplitudes}},
  \href{https://doi.org/10.1016/j.cpc.2017.01.004}{\emph{Comput. Phys. Commun.}
  {\bfseries 214} (2017) 140}
  [\href{https://arxiv.org/abs/1605.01090}{{\ttfamily 1605.01090}}].

\bibitem{Kilian:2007gr}
W.~Kilian, T.~Ohl and J.~Reuter, \emph{{WHIZARD: Simulating Multi-Particle
  Processes at LHC and ILC}},
  \href{https://doi.org/10.1140/epjc/s10052-011-1742-y}{\emph{Eur. Phys. J. C}
  {\bfseries 71} (2011) 1742}
  [\href{https://arxiv.org/abs/0708.4233}{{\ttfamily 0708.4233}}].

\bibitem{Ohl:2002jp}
T.~Ohl and J.~Reuter, \emph{{Clockwork SUSY: Supersymmetric Ward and
  Slavnov-Taylor identities at work in Green's functions and scattering
  amplitudes}}, \href{https://doi.org/10.1140/epjc/s2003-01301-7}{\emph{Eur.
  Phys. J. C} {\bfseries 30} (2003) 525}
  [\href{https://arxiv.org/abs/hep-th/0212224}{{\ttfamily hep-th/0212224}}].

\bibitem{Moretti:2001zz}
M.~Moretti, T.~Ohl and J.~Reuter, \emph{{O'Mega: An Optimizing matrix element
  generator}}, {\emph{{}} (2001) }
  [\href{https://arxiv.org/abs/hep-ph/0102195}{{\ttfamily hep-ph/0102195}}].

\bibitem{Kilian:2012pz}
W.~Kilian, T.~Ohl, J.~Reuter and C.~Speckner, \emph{{QCD in the Color-Flow
  Representation}}, \href{https://doi.org/10.1007/JHEP10(2012)022}{\emph{JHEP}
  {\bfseries 10} (2012) 022} [\href{https://arxiv.org/abs/1206.3700}{{\ttfamily
  1206.3700}}].

\bibitem{Christensen:2010wz}
N.D.~Christensen, C.~Duhr, B.~Fuks, J.~Reuter and C.~Speckner,
  \emph{{Introducing an interface between WHIZARD and FeynRules}},
  \href{https://doi.org/10.1140/epjc/s10052-012-1990-5}{\emph{Eur. Phys. J. C}
  {\bfseries 72} (2012) 1990}
  [\href{https://arxiv.org/abs/1010.3251}{{\ttfamily 1010.3251}}].

\bibitem{Degrande:2011ua}
C.~Degrande, C.~Duhr, B.~Fuks, D.~Grellscheid, O.~Mattelaer and T.~Reiter,
  \emph{{UFO - The Universal FeynRules Output}},
  \href{https://doi.org/10.1016/j.cpc.2012.01.022}{\emph{Comput. Phys. Commun.}
  {\bfseries 183} (2012) 1201}
  [\href{https://arxiv.org/abs/1108.2040}{{\ttfamily 1108.2040}}].

\bibitem{Brass:2018hfw}
S.~Brass, C.~Fleper, W.~Kilian, J.~Reuter and M.~Sekulla, \emph{{Transversal
  Modes and Higgs Bosons in Electroweak Vector-Boson Scattering at the LHC}},
  \href{https://doi.org/10.1140/epjc/s10052-018-6398-4}{\emph{Eur. Phys. J. C}
  {\bfseries 78} (2018) 931}
  [\href{https://arxiv.org/abs/1807.02512}{{\ttfamily 1807.02512}}].

\bibitem{Alboteanu:2008my}
A.~Alboteanu, W.~Kilian and J.~Reuter, \emph{{Resonances and Unitarity in Weak
  Boson Scattering at the LHC}},
  \href{https://doi.org/10.1088/1126-6708/2008/11/010}{\emph{JHEP} {\bfseries
  11} (2008) 010} [\href{https://arxiv.org/abs/0806.4145}{{\ttfamily
  0806.4145}}].

\bibitem{Kilian:2014zja}
W.~Kilian, T.~Ohl, J.~Reuter and M.~Sekulla, \emph{{High-Energy Vector Boson
  Scattering after the Higgs Discovery}},
  \href{https://doi.org/10.1103/PhysRevD.91.096007}{\emph{Phys. Rev. D}
  {\bfseries 91} (2015) 096007}
  [\href{https://arxiv.org/abs/1408.6207}{{\ttfamily 1408.6207}}].

\bibitem{Fleper:2016frz}
C.~Fleper, W.~Kilian, J.~Reuter and M.~Sekulla, \emph{{Scattering of W and Z
  Bosons at High-Energy Lepton Colliders}},
  \href{https://doi.org/10.1140/epjc/s10052-017-4656-5}{\emph{Eur. Phys. J. C}
  {\bfseries 77} (2017) 120}
  [\href{https://arxiv.org/abs/1607.03030}{{\ttfamily 1607.03030}}].

\bibitem{Whalley:2005nh}
M.R.~Whalley, D.~Bourilkov and R.C.~Group, \emph{{The Les Houches accord PDFs
  (LHAPDF) and LHAGLUE}},  in \emph{{HERA and the LHC: A Workshop on the
  Implications of HERA and LHC Physics (Startup Meeting, CERN, 26-27 March
  2004; Midterm Meeting, CERN, 11-13 October 2004)}}, pp.~575--581, 8, 2005
  [\href{https://arxiv.org/abs/hep-ph/0508110}{{\ttfamily hep-ph/0508110}}].

\bibitem{Buckley:2014ana}
A.~Buckley, J.~Ferrando, S.~Lloyd, K.~Nordstr\"om, B.~Page, M.~R\"ufenacht
  et~al., \emph{{LHAPDF6: parton density access in the LHC precision era}},
  \href{https://doi.org/10.1140/epjc/s10052-015-3318-8}{\emph{Eur. Phys. J. C}
  {\bfseries 75} (2015) 132} [\href{https://arxiv.org/abs/1412.7420}{{\ttfamily
  1412.7420}}].

\bibitem{Gribov:1972rt}
V.N.~Gribov and L.N.~Lipatov, \emph{{e+ e- pair annihilation and deep inelastic
  e p scattering in perturbation theory}}, {\emph{Sov. J. Nucl. Phys.}
  {\bfseries 15} (1972) 675}.

\bibitem{Ohl:1998jn}
T.~Ohl, \emph{{Vegas revisited: Adaptive Monte Carlo integration beyond
  factorization}},
  \href{https://doi.org/10.1016/S0010-4655(99)00209-X}{\emph{Comput. Phys.
  Commun.} {\bfseries 120} (1999) 13}
  [\href{https://arxiv.org/abs/hep-ph/9806432}{{\ttfamily hep-ph/9806432}}].

\bibitem{Lepage:1980dq}
G.P.~Lepage, \emph{{VEGAS: AN ADAPTIVE MULTIDIMENSIONAL INTEGRATION PROGRAM}},
  .

\bibitem{Brass:2018xbv}
S.~Brass, W.~Kilian and J.~Reuter, \emph{{Parallel Adaptive Monte Carlo
  Integration with the Event Generator WHIZARD}},
  \href{https://doi.org/10.1140/epjc/s10052-019-6840-2}{\emph{Eur. Phys. J. C}
  {\bfseries 79} (2019) 344}
  [\href{https://arxiv.org/abs/1811.09711}{{\ttfamily 1811.09711}}].

\bibitem{ChokoufeNejad:2014skp}
B.~Chokoufe~Nejad, T.~Ohl and J.~Reuter, \emph{{Simple, parallel virtual
  machines for extreme computations}},
  \href{https://doi.org/10.1016/j.cpc.2015.05.015}{\emph{Comput. Phys. Commun.}
  {\bfseries 196} (2015) 58} [\href{https://arxiv.org/abs/1411.3834}{{\ttfamily
  1411.3834}}].

\bibitem{Frixione:1995ms}
S.~Frixione, Z.~Kunszt and A.~Signer, \emph{{Three-jet cross sections to
  next-to-leading order}},
  \href{https://doi.org/10.1016/0550-3213(96)00110-1}{\emph{Nucl. Phys. B}
  {\bfseries 467} (1996) 399}
  [\href{https://arxiv.org/abs/hep-ph/9512328}{{\ttfamily hep-ph/9512328}}].

\bibitem{Frixione:1997np}
S.~Frixione, \emph{{A General approach to jet cross-sections in QCD}},
  \href{https://doi.org/10.1016/S0550-3213(97)00574-9}{\emph{Nucl. Phys. B}
  {\bfseries 507} (1997) 295}
  [\href{https://arxiv.org/abs/hep-ph/9706545}{{\ttfamily hep-ph/9706545}}].

\bibitem{Frederix:2009yq}
R.~Frederix, S.~Frixione, F.~Maltoni and T.~Stelzer, \emph{{Automation of
  next-to-leading order computations in QCD: The FKS subtraction}},
  \href{https://doi.org/10.1088/1126-6708/2009/10/003}{\emph{JHEP} {\bfseries
  10} (2009) 003} [\href{https://arxiv.org/abs/0908.4272}{{\ttfamily
  0908.4272}}].

\bibitem{Cullen:2011ac}
G.~Cullen, N.~Greiner, G.~Heinrich, G.~Luisoni, P.~Mastrolia, G.~Ossola et~al.,
  \emph{{Automated One-Loop Calculations with GoSam}},
  \href{https://doi.org/10.1140/epjc/s10052-012-1889-1}{\emph{Eur. Phys. J. C}
  {\bfseries 72} (2012) 1889}
  [\href{https://arxiv.org/abs/1111.2034}{{\ttfamily 1111.2034}}].

\bibitem{Cullen:2014yla}
G.~Cullen et~al., \emph{{G$\scriptsize{O}$S$\scriptsize{AM}$-2.0: a tool for
  automated one-loop calculations within the Standard Model and beyond}},
  \href{https://doi.org/10.1140/epjc/s10052-014-3001-5}{\emph{Eur. Phys. J. C}
  {\bfseries 74} (2014) 3001}
  [\href{https://arxiv.org/abs/1404.7096}{{\ttfamily 1404.7096}}].

\bibitem{Denner:2017wsf}
A.~Denner, J.-N.~Lang and S.~Uccirati, \emph{{Recola2: REcursive Computation of
  One-Loop Amplitudes 2}},
  \href{https://doi.org/10.1016/j.cpc.2017.11.013}{\emph{Comput. Phys. Commun.}
  {\bfseries 224} (2018) 346}
  [\href{https://arxiv.org/abs/1711.07388}{{\ttfamily 1711.07388}}].

\bibitem{Rothe:2021sml}
V.~Rothe, \emph{{Automation of NLO QCD Corrections and the Application to
  $N$-Jet Processes at Lepton Colliders}}, Ph.D. thesis, Hamburg U., 2021.

\bibitem{Brass:xxx}
S.~Brass, P.~Bredt, B.~Chokoufe~Nejad, W.~Kilian, J.~Reuter, V.~Rothe et~al.,
  \emph{{Automation of NLO SM corrections for hadron and lepton colliders (in
  prep.)}}, .

\bibitem{Buckley:2010ar}
A.~Buckley, J.~Butterworth, D.~Grellscheid, H.~Hoeth, L.~Lonnblad, J.~Monk
  et~al., \emph{{Rivet user manual}},
  \href{https://doi.org/10.1016/j.cpc.2013.05.021}{\emph{Comput. Phys. Commun.}
  {\bfseries 184} (2013) 2803}
  [\href{https://arxiv.org/abs/1003.0694}{{\ttfamily 1003.0694}}].

\bibitem{Bierlich:2019rhm}
C.~Bierlich et~al., \emph{{Robust Independent Validation of Experiment and
  Theory: Rivet version 3}},
  \href{https://doi.org/10.21468/SciPostPhys.8.2.026}{\emph{SciPost Phys.}
  {\bfseries 8} (2020) 026} [\href{https://arxiv.org/abs/1912.05451}{{\ttfamily
  1912.05451}}].

\bibitem{Cacciari:2011ma}
M.~Cacciari, G.P.~Salam and G.~Soyez, \emph{{FastJet User Manual}},
  \href{https://doi.org/10.1140/epjc/s10052-012-1896-2}{\emph{Eur. Phys. J. C}
  {\bfseries 72} (2012) 1896}
  [\href{https://arxiv.org/abs/1111.6097}{{\ttfamily 1111.6097}}].

\bibitem{Frixione:1998jh}
S.~Frixione, \emph{{Isolated photons in perturbative QCD}},
  \href{https://doi.org/10.1016/S0370-2693(98)00454-7}{\emph{Phys. Lett. B}
  {\bfseries 429} (1998) 369}
  [\href{https://arxiv.org/abs/hep-ph/9801442}{{\ttfamily hep-ph/9801442}}].

\bibitem{Mangano:2001xp}
M.L.~Mangano, M.~Moretti and R.~Pittau, \emph{{Multijet matrix elements and
  shower evolution in hadronic collisions: $W b \bar{b}$ + $n$ jets as a case
  study}}, \href{https://doi.org/10.1016/S0550-3213(02)00249-3}{\emph{Nucl.
  Phys. B} {\bfseries 632} (2002) 343}
  [\href{https://arxiv.org/abs/hep-ph/0108069}{{\ttfamily hep-ph/0108069}}].

\bibitem{Frixione:2007vw}
S.~Frixione, P.~Nason and C.~Oleari, \emph{{Matching NLO QCD computations with
  Parton Shower simulations: the POWHEG method}},
  \href{https://doi.org/10.1088/1126-6708/2007/11/070}{\emph{JHEP} {\bfseries
  11} (2007) 070} [\href{https://arxiv.org/abs/0709.2092}{{\ttfamily
  0709.2092}}].

\bibitem{ChokoufeNejad:2015kpc}
B.~Chokoufe~Nejad, W.~Kilian, J.~Reuter and C.~Weiss, \emph{{Matching NLO QCD
  Corrections in WHIZARD with the POWHEG scheme}},
  \href{https://doi.org/10.22323/1.234.0317}{\emph{PoS} {\bfseries EPS-HEP2015}
  (2015) 317} [\href{https://arxiv.org/abs/1510.02739}{{\ttfamily
  1510.02739}}].

\bibitem{Kalinowski:2020lhp}
J.~Kalinowski, W.~Kotlarski, P.~Sopicki and A.F.~Zarnecki, \emph{{Simulating
  hard photon production with WHIZARD}},
  \href{https://doi.org/10.1140/epjc/s10052-020-8149-6}{\emph{Eur. Phys. J. C}
  {\bfseries 80} (2020) 634}
  [\href{https://arxiv.org/abs/2004.14486}{{\ttfamily 2004.14486}}].

\bibitem{1859727}
P.~Sopicki, J.~Kalinowski, W.~Kotlarski, K.~Meka\l{}a and A.F.~Zarnecki,
  \emph{{Simulating hard photon production with WHIZARD}},
  \href{https://doi.org/10.22323/1.390.0285}{\emph{PoS} {\bfseries ICHEP2020}
  (2021) 285}.

\bibitem{Bach:2017ggt}
F.~Bach, B.C.~Nejad, A.~Hoang, W.~Kilian, J.~Reuter, M.~Stahlhofen et~al.,
  \emph{{Fully-differential Top-Pair Production at a Lepton Collider: From
  Threshold to Continuum}},
  \href{https://doi.org/10.1007/JHEP03(2018)184}{\emph{JHEP} {\bfseries 03}
  (2018) 184} [\href{https://arxiv.org/abs/1712.02220}{{\ttfamily
  1712.02220}}].

\bibitem{Beneke:2007zg}
M.~Beneke, P.~Falgari, C.~Schwinn, A.~Signer and G.~Zanderighi,
  \emph{{Four-fermion production near the W pair production threshold}},
  \href{https://doi.org/10.1016/j.nuclphysb.2007.09.030}{\emph{Nucl. Phys. B}
  {\bfseries 792} (2008) 89} [\href{https://arxiv.org/abs/0707.0773}{{\ttfamily
  0707.0773}}].

\bibitem{Actis:2008rb}
S.~Actis, M.~Beneke, P.~Falgari and C.~Schwinn, \emph{{Dominant NNLO
  corrections to four-fermion production near the W-pair production
  threshold}},
  \href{https://doi.org/10.1016/j.nuclphysb.2008.08.006}{\emph{Nucl. Phys. B}
  {\bfseries 807} (2009) 1} [\href{https://arxiv.org/abs/0807.0102}{{\ttfamily
  0807.0102}}].

\bibitem{Alwall:2014hca}
J.~Alwall, R.~Frederix, S.~Frixione, V.~Hirschi, F.~Maltoni, O.~Mattelaer
  et~al., \emph{{The automated computation of tree-level and next-to-leading
  order differential cross sections, and their matching to parton shower
  simulations}}, \href{https://doi.org/10.1007/JHEP07(2014)079}{\emph{JHEP}
  {\bfseries 07} (2014) 079} [\href{https://arxiv.org/abs/1405.0301}{{\ttfamily
  1405.0301}}].

\bibitem{Frederix:2018nkq}
R.~Frederix, S.~Frixione, V.~Hirschi, D.~Pagani, H.S.~Shao and M.~Zaro,
  \emph{{The automation of next-to-leading order electroweak calculations}},
  \href{https://doi.org/10.1007/JHEP07(2018)185}{\emph{JHEP} {\bfseries 07}
  (2018) 185} [\href{https://arxiv.org/abs/1804.10017}{{\ttfamily
  1804.10017}}].

\bibitem{Frixione:2021zdp}
S.~Frixione, O.~Mattelaer, M.~Zaro and X.~Zhao, \emph{{Lepton collisions in
  MadGraph5\_aMC@NLO}},  \href{https://arxiv.org/abs/2108.10261}{{\ttfamily
  2108.10261}}.

\bibitem{epdf-pheno}
V.~Bertone, M.~Cacciari, S.~Frixione, G.~Stagnitto, M.~Zaro and X.~Zhao,
  \emph{{Studies of e+e- cross sections at the next-to-leading logarithmic
  accuracy}}, {\emph{in preparation} }.

\bibitem{deFlorian:2016gvk}
D.~de~Florian, G.F.R.~Sborlini and G.~Rodrigo, \emph{{Two-loop QED corrections
  to the Altarelli-Parisi splitting functions}},
  \href{https://doi.org/10.1007/JHEP10(2016)056}{\emph{JHEP} {\bfseries 10}
  (2016) 056} [\href{https://arxiv.org/abs/1606.02887}{{\ttfamily
  1606.02887}}].

\bibitem{Mele:1990cw}
B.~Mele and P.~Nason, \emph{{The Fragmentation function for heavy quarks in
  QCD}}, \href{https://doi.org/10.1016/0550-3213(91)90597-Q}{\emph{Nucl. Phys.
  B} {\bfseries 361} (1991) 626}.

\bibitem{Anlauf:1991wr}
H.~Anlauf, H.D.~Dahmen, P.~Manakos, T.~Mannel and T.~Ohl, \emph{{KRONOS: A
  Monte Carlo event generator for higher order electromagnetic radiative
  corrections to deep inelastic scattering at HERA}},
  \href{https://doi.org/10.1016/0010-4655(92)90095-G}{\emph{Comput. Phys.
  Commun.} {\bfseries 70} (1992) 97}.

\bibitem{Munehisa:1995si}
T.~Munehisa, J.~Fujimoto, Y.~Kurihara and Y.~Shimizu, \emph{{Improved QEDPS for
  radiative corrections in e+ e- annihilation}},
  \href{https://doi.org/10.1143/PTP.95.375}{\emph{Prog. Theor. Phys.}
  {\bfseries 95} (1996) 375}
  [\href{https://arxiv.org/abs/hep-ph/9603322}{{\ttfamily hep-ph/9603322}}].

\bibitem{CarloniCalame:2000pz}
C.M.~Carloni~Calame, C.~Lunardini, G.~Montagna, O.~Nicrosini and F.~Piccinini,
  \emph{Large angle bhabha scattering and luminosity at flavor factories},
  \href{https://doi.org/10.1016/S0550-3213(00)00356-4}{\emph{Nucl. Phys. B}
  {\bfseries 584} (2000) 459 }
  [\href{https://arxiv.org/abs/hep-ph/0003268}{{\ttfamily hep-ph/0003268}}].

\bibitem{Skrzypek:1992vk}
M.~Skrzypek, \emph{{Leading logarithmic calculations of QED corrections at
  LEP}}, {\emph{Acta Phys. Polon. B} {\bfseries 23} (1992) 135}.

\bibitem{Cacciari:1992pz}
M.~Cacciari, A.~Deandrea, G.~Montagna and O.~Nicrosini, \emph{{QED structure
  functions: A Systematic approach}},
  \href{https://doi.org/10.1209/0295-5075/17/2/007}{\emph{Europhys. Lett.}
  {\bfseries 17} (1992) 123}.

\bibitem{Denner:2000bj}
A.~Denner, S.~Dittmaier, M.~Roth and D.~Wackeroth, \emph{{Electroweak radiative
  corrections to $e^+ e^- \to W W \to 4$ fermions in double pole approximation:
  The RACOONWW approach}},
  \href{https://doi.org/10.1016/S0550-3213(00)00511-3}{\emph{Nucl. Phys. B}
  {\bfseries 587} (2000) 67}
  [\href{https://arxiv.org/abs/hep-ph/0006307}{{\ttfamily hep-ph/0006307}}].

\bibitem{Frixione:2021wzh}
S.~Frixione, \emph{{On factorisation schemes for the electron parton
  distribution functions in QED}},
  \href{https://doi.org/10.1007/JHEP07(2021)180}{\emph{JHEP} {\bfseries 07}
  (2021) 180} [\href{https://arxiv.org/abs/2105.06688}{{\ttfamily
  2105.06688}}].

\bibitem{CarloniCalame:2001ny}
C.M.~Carloni~Calame, \emph{An improved parton shower algorithm in qed},
  \href{https://doi.org/10.1016/S0370-2693(01)01108-X}{\emph{Phys. Lett. B}
  {\bfseries 520} (2001) 16 }
  [\href{https://arxiv.org/abs/hep-ph/0103117}{{\ttfamily hep-ph/0103117}}].

\bibitem{Balossini:2008xr}
G.~Balossini, C.~Bignamini, C.M.~Carloni~Calame, G.~Montagna, O.~Nicrosini and
  F.~Piccinini, \emph{{Photon pair production at flavour factories with per
  mille accuracy}},
  \href{https://doi.org/10.1016/j.physletb.2008.04.007}{\emph{Phys. Lett. B}
  {\bfseries 663} (2008) 209}
  [\href{https://arxiv.org/abs/0801.3360}{{\ttfamily 0801.3360}}].

\bibitem{Sudakov:1954sw}
V.V.~Sudakov, \emph{{Vertex parts at very high-energies in quantum
  electrodynamics}}, {\emph{Sov. Phys. JETP} {\bfseries 3} (1956) 65}.

\bibitem{Balossini:2006wc}
G.~Balossini, C.M.~Carloni~Calame, G.~Montagna, O.~Nicrosini and F.~Piccinini,
  \emph{{Matching perturbative and parton shower corrections to Bhabha process
  at flavour factories}},
  \href{https://doi.org/10.1016/j.nuclphysb.2006.09.022}{\emph{Nucl. Phys. B}
  {\bfseries 758} (2006) 227}
  [\href{https://arxiv.org/abs/hep-ph/0607181}{{\ttfamily hep-ph/0607181}}].

\bibitem{Frixione:2002ik}
S.~Frixione and B.R.~Webber, \emph{{Matching NLO QCD computations and parton
  shower simulations}},
  \href{https://doi.org/10.1088/1126-6708/2002/06/029}{\emph{JHEP} {\bfseries
  06} (2002) 029} [\href{https://arxiv.org/abs/hep-ph/0204244}{{\ttfamily
  hep-ph/0204244}}].

\bibitem{Nason:2004rx}
P.~Nason, \emph{{A New method for combining NLO QCD with shower Monte Carlo
  algorithms}},
  \href{https://doi.org/10.1088/1126-6708/2004/11/040}{\emph{JHEP} {\bfseries
  11} (2004) 040} [\href{https://arxiv.org/abs/hep-ph/0409146}{{\ttfamily
  hep-ph/0409146}}].

\bibitem{Altarelli:1989hv}
G.~Altarelli, R.~Kleiss and C.~Verzegnassi, eds., \emph{{Z PHYSICS AT LEP-1.
  PROCEEDINGS, WORKSHOP, GENEVA, SWITZERLAND, SEPTEMBER 4-5, 1989. VOL. 1:
  STANDARD PHYSICS}}, CERN Yellow Reports: Conference Proceedings, 9, 1989.
\newblock 10.5170/CERN-1989-008-V-1.

\bibitem{Greco:1987uw}
M.~Greco, \emph{{RADIATIVE CORRECTIONS TO e+ e- REACTIONS AT LEP / SLC
  ENERGIES}}, \href{https://doi.org/10.1007/BF02724480}{\emph{Riv. Nuovo Cim.}
  {\bfseries 11N5} (1988) 1}.

\bibitem{CarloniCalame:2019dom}
C.M.~Carloni~Calame, M.~Chiesa, G.~Montagna, O.~Nicrosini and F.~Piccinini,
  \emph{{Electroweak corrections to $e^+e^-\to\gamma\gamma$ as a luminosity
  process at FCC-ee}},
  \href{https://doi.org/10.1016/j.physletb.2019.134976}{\emph{Phys. Lett. B}
  {\bfseries 798} (2019) 134976}
  [\href{https://arxiv.org/abs/1906.08056}{{\ttfamily 1906.08056}}].

\bibitem{CarloniCalame:2006zq}
C.M.~Carloni~Calame, G.~Montagna, O.~Nicrosini and A.~Vicini, \emph{{Precision
  electroweak calculation of the charged current Drell-Yan process}},
  \href{https://doi.org/10.1088/1126-6708/2006/12/016}{\emph{JHEP} {\bfseries
  12} (2006) 016} [\href{https://arxiv.org/abs/hep-ph/0609170}{{\ttfamily
  hep-ph/0609170}}].

\bibitem{CarloniCalame:2007cd}
C.M.~Carloni~Calame, G.~Montagna, O.~Nicrosini and A.~Vicini, \emph{{Precision
  electroweak calculation of the production of a high transverse-momentum
  lepton pair at hadron colliders}},
  \href{https://doi.org/10.1088/1126-6708/2007/10/109}{\emph{JHEP} {\bfseries
  10} (2007) 109} [\href{https://arxiv.org/abs/0710.1722}{{\ttfamily
  0710.1722}}].

\bibitem{Boselli:2015aha}
S.~Boselli, C.M.~Carloni~Calame, G.~Montagna, O.~Nicrosini and F.~Piccinini,
  \emph{{Higgs boson decay into four leptons at NLOPS electroweak accuracy}},
  \href{https://doi.org/10.1007/JHEP06(2015)023}{\emph{JHEP} {\bfseries 06}
  (2015) 023} [\href{https://arxiv.org/abs/1503.07394}{{\ttfamily
  1503.07394}}].

\bibitem{TLEPDesignStudyWorkingGroup:2013myl}
{\scshape TLEP Design Study Working Group} collaboration, \emph{{First Look at
  the Physics Case of TLEP}},
  \href{https://doi.org/10.1007/JHEP01(2014)164}{\emph{JHEP} {\bfseries 01}
  (2014) 164} [\href{https://arxiv.org/abs/1308.6176}{{\ttfamily 1308.6176}}].

\bibitem{Proceedings:2019vxr}
A.~Blondel, J.~Gluza, S.~Jadach, P.~Janot and T.~Riemann, eds., \emph{{Theory
  for the FCC-ee}: {Report on the 11th FCC-ee Workshop Theory and
  Experiments}}, vol.~3/2020 of \emph{CERN Yellow Reports: Monographs},
  (Geneva), CERN, 5, 2019.
\newblock 10.23731/CYRM-2020-003.

\bibitem{Blondel:2018mad}
A.~Blondel et~al., \emph{{Standard model theory for the FCC-ee Tera-Z stage}},
  in \emph{{Mini Workshop on Precision EW and QCD Calculations for the FCC
  Studies : Methods and Techniques}}, vol.~3/2019 of \emph{CERN Yellow Reports:
  Monographs}, (Geneva), CERN, 9, 2018,
  \href{https://doi.org/10.23731/CYRM-2019-003}{DOI}
  [\href{https://arxiv.org/abs/1809.01830}{{\ttfamily 1809.01830}}].

\bibitem{Alacevich:2018vez}
M.~Alacevich, C.M.~Carloni~Calame, M.~Chiesa, G.~Montagna, O.~Nicrosini and
  F.~Piccinini, \emph{{Muon-electron scattering at NLO}},
  \href{https://doi.org/10.1007/JHEP02(2019)155}{\emph{JHEP} {\bfseries 02}
  (2019) 155} [\href{https://arxiv.org/abs/1811.06743}{{\ttfamily
  1811.06743}}].

\bibitem{Banerjee:2020tdt}
P.~Banerjee et~al., \emph{{Theory for muon-electron scattering \myat 10 ppm: A
  report of the MUonE theory initiative}},
  \href{https://doi.org/10.1140/epjc/s10052-020-8138-9}{\emph{Eur. Phys. J. C}
  {\bfseries 80} (2020) 591}
  [\href{https://arxiv.org/abs/2004.13663}{{\ttfamily 2004.13663}}].

\bibitem{CarloniCalame:2020yoz}
C.M.~Carloni~Calame, M.~Chiesa, S.M.~Hasan, G.~Montagna, O.~Nicrosini and
  F.~Piccinini, \emph{{Towards muon-electron scattering at NNLO}},
  \href{https://doi.org/10.1007/JHEP11(2020)028}{\emph{JHEP} {\bfseries 11}
  (2020) 028} [\href{https://arxiv.org/abs/2007.01586}{{\ttfamily
  2007.01586}}].

\bibitem{Budassi:2021twh}
E.~Budassi, C.M.~Carloni~Calame, M.~Chiesa, C.L.~Del~Pio, S.M.~Hasan,
  G.~Montagna et~al., \emph{{NNLO virtual and real leptonic corrections to
  muon-electron scattering}},
  \href{https://doi.org/10.1007/JHEP11(2021)098}{\emph{JHEP} {\bfseries 11}
  (2021) 098} [\href{https://arxiv.org/abs/2109.14606}{{\ttfamily
  2109.14606}}].

\bibitem{EuropeanStrategyforParticlePhysicsPreparatoryGroup:2019qin}
R.K.~Ellis et~al., \emph{{Physics Briefing Book}: {Input for the European
  Strategy for Particle Physics Update 2020}},
  \href{https://arxiv.org/abs/1910.11775}{{\ttfamily 1910.11775}}.

\bibitem{LCCPhysicsWorkingGroup:2019fvj}
{\scshape LCC Physics Working Group} collaboration, \emph{{Tests of the
  Standard Model at the International Linear Collider}},
  \href{https://arxiv.org/abs/1908.11299}{{\ttfamily 1908.11299}}.

\bibitem{dEnterria:2016sca}
D.~d'Enterria, \emph{{Physics at the FCC-ee}},  in \emph{{17th Lomonosov
  Conference on Elementary Particle Physics}}, pp.~182--191, 2017,
  \href{https://doi.org/10.1142/9789813224568_0028}{DOI}
  [\href{https://arxiv.org/abs/1602.05043}{{\ttfamily 1602.05043}}].

\bibitem{Freitas:2019bre}
A.~Freitas et~al., \emph{{Theoretical uncertainties for electroweak and
  Higgs-boson precision measurements at FCC-ee}},
  \href{https://arxiv.org/abs/1906.05379}{{\ttfamily 1906.05379}}.

\bibitem{Denner:1997ia}
A.~Denner, S.~Dittmaier and M.~Roth, \emph{{Non-factorizable photonic
  corrections to $e^+ e^-\to W W \to\,$four fermions}},
  \href{https://doi.org/10.1016/S0550-3213(98)00046-7}{\emph{Nucl. Phys.}
  {\bfseries B519} (1998) 39}
  [\href{https://arxiv.org/abs/hep-ph/9710521}{{\ttfamily hep-ph/9710521}}].

\bibitem{Beenakker:1997bp}
W.~Beenakker, A.P.~Chapovsky and F.A.~Berends, \emph{{Non-factorizable
  corrections to W~pair production}},
  \href{https://doi.org/10.1016/S0370-2693(97)01010-1}{\emph{Phys. Lett.}
  {\bfseries B411} (1997) 203}
  [\href{https://arxiv.org/abs/hep-ph/9706339}{{\ttfamily hep-ph/9706339}}].

\bibitem{Beenakker:1997ir}
W.~Beenakker, A.P.~Chapovsky and F.A.~Berends, \emph{{Non-factorizable
  corrections to W~pair production: Methods and analytic results}},
  \href{https://doi.org/10.1016/S0550-3213(97)80003-X,
  10.1016/S0550-3213(97)00628-7}{\emph{Nucl. Phys.} {\bfseries B508} (1997) 17}
  [\href{https://arxiv.org/abs/hep-ph/9707326}{{\ttfamily hep-ph/9707326}}].

\bibitem{Accomando:2004de}
E.~Accomando, A.~Denner and A.~Kaiser, \emph{{Logarithmic electroweak
  corrections to gauge-boson pair production at the LHC}},
  \href{https://doi.org/10.1016/j.nuclphysb.2004.11.019}{\emph{Nucl. Phys.}
  {\bfseries B706} (2005) 325}
  [\href{https://arxiv.org/abs/hep-ph/0409247}{{\ttfamily hep-ph/0409247}}].

\bibitem{Dittmaier:2015bfe}
S.~Dittmaier and C.~Schwan, \emph{{Non-factorizable photonic corrections to
  resonant production and decay of many unstable particles}},
  \href{https://doi.org/10.1140/epjc/s10052-016-3968-1}{\emph{Eur. Phys.~J.}
  {\bfseries C76} (2016) 144}
  [\href{https://arxiv.org/abs/1511.01698}{{\ttfamily 1511.01698}}].

\bibitem{Bohm:1987ck}
M.~{B\"ohm}, A.~Denner, T.~Sack, W.~Beenakker, F.A.~Berends and H.~Kuijf,
  \emph{{Electroweak Radiative Corrections to $e^+ e^- \to W^+ W^-$}},
  \href{https://doi.org/10.1016/0550-3213(88)90638-4}{\emph{Nucl. Phys. B}
  {\bfseries 304} (1988) 463}.

\bibitem{Fleischer:1988kj}
J.~Fleischer, F.~Jegerlehner and M.~Zralek, \emph{{Radiative Corrections to
  Helicity Amplitudes for W Pair Production in $e^+ e^-$ Annihilation}},
  \href{https://doi.org/10.1007/BF01548446}{\emph{Z. Phys. C} {\bfseries 42}
  (1989) 409}.

\bibitem{Denner:1990tx}
A.~Denner and T.~Sack, \emph{{The W boson width}},
  \href{https://doi.org/10.1007/BF01560267}{\emph{Z. Phys. C} {\bfseries 46}
  (1990) 653}.

\bibitem{Beenakker:1998gr}
W.~Beenakker, F.A.~Berends and A.P.~Chapovsky, \emph{{Radiative corrections to
  pair production of unstable particles: results for $e^+ e^-\to4\,$fermions}},
  \href{https://doi.org/10.1016/S0550-3213(99)00110-8}{\emph{Nucl. Phys.}
  {\bfseries B548} (1999) 3}
  [\href{https://arxiv.org/abs/hep-ph/9811481}{{\ttfamily hep-ph/9811481}}].

\bibitem{Jadach:1996hi}
S.~Jadach, W.~Placzek, M.~Skrzypek, B.~Ward and Z.~Was, \emph{{Exact O (alpha)
  gauge invariant YFS exponentiated Monte Carlo for (un)stable W+ W- production
  at and beyond LEP-2 energies}},
  \href{https://doi.org/10.1016/S0370-2693(97)01253-7}{\emph{Phys.Lett.}
  {\bfseries B417} (1998) 326}
  [\href{https://arxiv.org/abs/hep-ph/9705429}{{\ttfamily hep-ph/9705429}}].

\bibitem{Denner:2002cg}
A.~Denner, S.~Dittmaier, M.~Roth and D.~Wackeroth, \emph{{RACOONWW1.3: A Monte
  Carlo program for four fermion production at $e^+ e^-$ colliders}},
  \href{https://doi.org/10.1016/S0010-4655(03)00205-4}{\emph{Comput. Phys.
  Commun.} {\bfseries 153} (2003) 462}
  [\href{https://arxiv.org/abs/hep-ph/0209330}{{\ttfamily hep-ph/0209330}}].

\bibitem{Kurihara:2001um}
Y.~Kurihara, M.~Kuroda and D.~Schildknecht, \emph{{$e^+ e^-\to W^+ W^-\to 4f
  (+\gamma)$ at LEP2}},
  \href{https://doi.org/10.1016/S0370-2693(01)00531-7}{\emph{Phys. Lett.}
  {\bfseries B509} (2001) 87}
  [\href{https://arxiv.org/abs/hep-ph/0104201}{{\ttfamily hep-ph/0104201}}].

\bibitem{Grunewald:2000ju}
M.W.~Gr{\"u}newald et~al., \emph{{Reports of the Working Groups on Precision
  Calculations for LEP2 Physics: Proceedings. Four fermion production in
  electron positron collisions}},
  \href{https://arxiv.org/abs/hep-ph/0005309}{{\ttfamily hep-ph/0005309}}.

\bibitem{Denner:1999gp}
A.~Denner, S.~Dittmaier, M.~Roth and D.~Wackeroth, \emph{{Predictions for all
  processes $e^+e^-$ $\to$ fermions + $\gamma$}},
  \href{https://doi.org/10.1016/S0550-3213(99)00437-X}{\emph{Nucl. Phys.}
  {\bfseries B560} (1999) 33}
  [\href{https://arxiv.org/abs/hep-ph/9904472}{{\ttfamily hep-ph/9904472}}].

\bibitem{Schael:2013ita}
{\scshape ALEPH, DELPHI, L3, OPAL, LEP Electroweak} collaboration,
  \emph{{Electroweak Measurements in Electron-Positron Collisions at
  W-Boson-Pair Energies at LEP}},
  \href{https://doi.org/10.1016/j.physrep.2013.07.004}{\emph{Phys. Rept.}
  {\bfseries 532} (2013) 119}
  [\href{https://arxiv.org/abs/1302.3415}{{\ttfamily 1302.3415}}].

\bibitem{Berends:1987ab}
F.A.~Berends, W.~van Neerven and G.~Burgers, \emph{{Higher Order Radiative
  Corrections at LEP Energies}},
  \href{https://doi.org/10.1016/0550-3213(88)90313-6}{\emph{Nucl. Phys. B}
  {\bfseries 297} (1988) 429}.

\bibitem{Beenakker:1996kt}
W.~Beenakker et~al., \emph{{$W W$ cross-sections and distributions}},  in
  \emph{{CERN Workshop on LEP2 Physics (followed by 2nd meeting, 15-16 Jun 1995
  and 3rd meeting 2-3 Nov 1995) Geneva, Switzerland, February 2-3, 1995}},
  pp.~79--139, 1996 [\href{https://arxiv.org/abs/hep-ph/9602351}{{\ttfamily
  hep-ph/9602351}}].

\bibitem{Jadach:1998tz}
S.~Jadach, W.~{P\l{}aczek}, M.~Skrzypek, B.F.L.~Ward and Z.~{W\c as},
  \emph{{Final-state radiative effects for the exact $\mathcal{O}(\alpha)$
  Yennie-Frautschi-Suura exponentiated (un)stable $W^+ W^-$ production at and
  beyond LEP2 energies}},
  \href{https://doi.org/10.1103/PhysRevD.61.113010}{\emph{Phys. Rev.}
  {\bfseries D61} (2000) 113010}
  [\href{https://arxiv.org/abs/hep-ph/9907436}{{\ttfamily hep-ph/9907436}}].

\bibitem{Dittmaier:1991np}
S.~Dittmaier, M.~{B\"ohm} and A.~Denner, \emph{{Improved Born approximation for
  $e^+e^-\rightarrow W^+W^-$ in the LEP200 energy region}},
  \href{https://doi.org/10.1016/0550-3213(92)90066-K,
  10.1016/0550-3213(93)90156-J}{\emph{Nucl. Phys.} {\bfseries B376} (1992) 29}.

\bibitem{Denner:2001zp}
A.~Denner, S.~Dittmaier, M.~Roth and D.~Wackeroth, \emph{{Off-shell W pair
  production: Universal versus nonuniversal corrections}},  in \emph{{5th
  International Symposium on Radiative Corrections: Applications of Quantum
  Field Theory to Phenomenology}}, 1, 2001
  [\href{https://arxiv.org/abs/hep-ph/0101257}{{\ttfamily hep-ph/0101257}}].

\bibitem{Denner:2005es}
A.~Denner, S.~Dittmaier, M.~Roth and L.H.~Wieders, \emph{{Complete electroweak
  ${\cal O}(\alpha)$ corrections to charged-current $e^+e^- \to 4$ fermion
  processes}},
  \href{https://doi.org/10.1016/j.physletb.2005.03.007}{\emph{Phys. Lett. B}
  {\bfseries 612} (2005) 223}
  [\href{https://arxiv.org/abs/hep-ph/0502063}{{\ttfamily hep-ph/0502063}}].

\bibitem{Denner:2005fg}
A.~Denner, S.~Dittmaier, M.~Roth and L.H.~Wieders, \emph{{Electroweak
  corrections to charged-current $e^+ e^- \to 4$ fermion processes: Technical
  details and further results}},
  \href{https://doi.org/10.1016/j.nuclphysb.2011.09.001}{\emph{Nucl. Phys. B}
  {\bfseries 724} (2005) 247}
  [\href{https://arxiv.org/abs/hep-ph/0505042}{{\ttfamily hep-ph/0505042}}].

\bibitem{Denner:2006ic}
A.~Denner and S.~Dittmaier, \emph{{The Complex-mass scheme for perturbative
  calculations with unstable particles}},
  \href{https://doi.org/10.1016/j.nuclphysbps.2006.09.025}{\emph{Nucl. Phys.
  Proc. Suppl.} {\bfseries 160} (2006) 22}
  [\href{https://arxiv.org/abs/hep-ph/0605312}{{\ttfamily hep-ph/0605312}}].

\bibitem{Denner:2019vbn}
A.~Denner and S.~Dittmaier, \emph{{Electroweak Radiative Corrections for
  Collider Physics}},
  \href{https://doi.org/10.1016/j.physrep.2020.04.001}{\emph{Phys. Rept.}
  {\bfseries 864} (2020) 1} [\href{https://arxiv.org/abs/1912.06823}{{\ttfamily
  1912.06823}}].

\bibitem{Denner:2005nn}
A.~Denner and S.~Dittmaier, \emph{{Reduction schemes for one-loop tensor
  integrals}},
  \href{https://doi.org/10.1016/j.nuclphysb.2005.11.007}{\emph{Nucl. Phys.}
  {\bfseries B734} (2006) 62}
  [\href{https://arxiv.org/abs/hep-ph/0509141}{{\ttfamily hep-ph/0509141}}].

\bibitem{Passarino:1978jh}
G.~Passarino and M.~Veltman, \emph{{One Loop Corrections for e+ e- Annihilation
  Into mu+ mu- in the Weinberg Model}},
  \href{https://doi.org/10.1016/0550-3213(79)90234-7}{\emph{Nucl. Phys. B}
  {\bfseries 160} (1979) 151}.

\bibitem{Denner:2016kdg}
A.~Denner, S.~Dittmaier and L.~Hofer, \emph{{Collier: a fortran-based Complex
  One-Loop LIbrary in Extended Regularizations}},
  \href{https://doi.org/10.1016/j.cpc.2016.10.013}{\emph{Comput. Phys. Commun.}
  {\bfseries 212} (2017) 220}
  [\href{https://arxiv.org/abs/1604.06792}{{\ttfamily 1604.06792}}].

\bibitem{Pineda:1997bj}
A.~Pineda and J.~Soto, \emph{{Effective field theory for ultrasoft momenta in
  NRQCD and NRQED}},
  \href{https://doi.org/10.1016/S0920-5632(97)01102-X}{\emph{Nucl. Phys. B
  Proc. Suppl.} {\bfseries 64} (1998) 428}
  [\href{https://arxiv.org/abs/hep-ph/9707481}{{\ttfamily hep-ph/9707481}}].

\bibitem{Bauer:2000yr}
C.W.~Bauer, S.~Fleming, D.~Pirjol and I.W.~Stewart, \emph{An effective field
  theory for collinear and soft gluons: Heavy to light decays}, {\emph{Phys.
  Rev.} {\bfseries D63} (2001) 114020}
  [\href{https://arxiv.org/abs/hep-ph/0011336}{{\ttfamily hep-ph/0011336}}].

\bibitem{Beneke:2003xh}
M.~Beneke, A.P.~Chapovsky, A.~Signer and G.~Zanderighi, \emph{{Effective theory
  approach to unstable particle production}},
  \href{https://doi.org/10.1103/PhysRevLett.93.011602}{\emph{Phys. Rev. Lett.}
  {\bfseries 93} (2004) 011602}
  [\href{https://arxiv.org/abs/hep-ph/0312331}{{\ttfamily hep-ph/0312331}}].

\bibitem{Schwinn:2019qbp}
C.~Schwinn, \emph{{Prospects for higher-order corrections to W-pair production
  near threshold in the EFT approach}},
  \href{https://doi.org/10.23731/CYRM-2020-003.77}{\emph{CERN Yellow Reports:
  Monographs} {\bfseries 3} (2020) 77}.

\bibitem{Jadach:1993yv}
S.~Jadach, B.F.L.~Ward and Z.~Was, \emph{{The Monte Carlo program KORALZ,
  version 4.0, for the lepton or quark pair production at LEP / SLC energies}},
  \href{https://doi.org/10.1016/0010-4655(94)90190-2}{\emph{Comput. Phys.
  Commun.} {\bfseries 79} (1994) 503}.

\bibitem{Jadach:1999vf}
S.~Jadach, B.F.L.~Ward and Z.~Was, \emph{{The Precision Monte Carlo event
  generator K K for two fermion final states in e+ e- collisions}},
  \href{https://doi.org/10.1016/S0010-4655(00)00048-5}{\emph{Comput. Phys.
  Commun.} {\bfseries 130} (2000) 260}
  [\href{https://arxiv.org/abs/hep-ph/9912214}{{\ttfamily hep-ph/9912214}}].

\bibitem{Jadach:2013aha}
S.~Jadach, B.F.L.~Ward and Z.~W\c{a}s, \emph{{KK MC 4.22: Coherent exclusive
  exponentiation of electroweak corrections for f f-bar -->f' f'-bar at the LHC
  and muon colliders}},
  \href{https://doi.org/10.1103/PhysRevD.88.114022}{\emph{Phys. Rev.}
  {\bfseries D88} (2013) 114022}
  [\href{https://arxiv.org/abs/1307.4037}{{\ttfamily 1307.4037}}].

\bibitem{yfs:1961}
D.R.~Yennie, S.~Frautschi and H.~Suura, \emph{{The infrared divergence
  phenomena and high-energy processes}},
  \href{https://doi.org/10.1016/0003-4916(61)90151-8}{\emph{Ann. Phys. (NY)}
  {\bfseries 13} (1961) 379 }.

\bibitem{bhlumi2:1992}
S.~Jadach, E.~Richter-W\c{a}s, B.F.L.~Ward and Z.~W\c{a}s, \emph{Monte carlo
  program bhlumi-2.01 for bhabha scattering at low angles with
  yennie-frautschi-suura exponentiation}, {\emph{Comput. Phys. Commun.}
  {\bfseries 70} (1992) 305}.

\bibitem{bhlumi4:1996}
S.~Jadach, W.~Placzek, E.~Richter-W\c{a}s, B.F.L.~Ward and Z.~W\c{a}s,
  \emph{{Upgrade of the Monte Carlo program BHLUMI for Bhabha scattering at low
  angles to version 4.04}}, {\emph{Comput. Phys. Commun.} {\bfseries 102}
  (1997) 229}.

\bibitem{bhwide:1997}
S.~Jadach, W.~P\l{}aczek and B.F.L.~Ward, \emph{{BHWIDE 1.00: ${\cal
  O}(\alpha)$ YFS exponentiated Monte Carlo for Bhabha scattering at wide
  angles for LEP1/SLC and LEP2}},
  \href{https://doi.org/10.1016/S0370-2693(96)01382-2}{\emph{Phys. Lett.}
  {\bfseries B390} (1997) 298 }
  [\href{https://arxiv.org/abs/hep-ph/9608412}{{\ttfamily hep-ph/9608412}}].

\bibitem{yfsww3:2001}
S.~Jadach, W.~P{\l}aczek, M.~Skrzypek, B.F.L.~Ward and Z.~W\c{a}s, \emph{{The
  Monte Carlo Event Generator YFSWW3 Version 1.16 for W Pair Production and
  Decay at LEP-2/LC Energies}}, {\emph{Comput. Phys. Commun.} {\bfseries 140}
  (2001) 432}.

\bibitem{koralw:1998}
S.~Jadach, W.~Placzek, M.~Skrzypek, B.F.L.~Ward and Z.~W\c{a}s, \emph{Monte
  carlo program koralw 1.42 for all four-fermion final states in e+ e-
  collisions}, {\emph{Comput. Phys. Commun.} {\bfseries 119} (1999) 272}
  [\href{https://arxiv.org/abs/hep-ph/9906277}{{\ttfamily hep-ph/9906277}}].

\bibitem{kandy-2001}
S.~Jadach, W.~P{\l}aczek, M.~Skrzypek, B.F.L.~Ward and Z.~W\c{a}s, \emph{{The
  Monte Carlo program KoralW version 1.51 and the concurrent Monte Carlo KoralW
  and YFSWW3 with all background graphs and first order corrections to W pair
  production}}, {\emph{Comput. Phys. Commun.} {\bfseries 140} (2001) 475 }.

\bibitem{yfszz:1997}
S.~Jadach, W.~Placzek and B.F.L.~Ward, \emph{{Gauge invariant YFS
  exponentiation of (un)stable Z pair production at and beyond LEP-2
  energies}}, {\emph{Phys. Rev. D} {\bfseries 56} (1997) 6939 }
  [\href{https://arxiv.org/abs/hep-ph/9705430}{{\ttfamily hep-ph/9705430}}].

\bibitem{MC2000}
S.J.~(ed.), G.P.~(ed.) and R.P.~(ed.), \emph{Lep2 monte carlo workshop : Report
  of the working groups on precision calculations for lep2 physics},  2000.

\bibitem{SM1}
S.~Weinberg, \emph{{A Model of Leptons }}, {\emph{Phys. Rev. Lett.} {\bfseries
  19} (1967) 1264}.

\bibitem{SM2}
S.L.~Glashow, J.~Iliopoulos and L.~Maiani, \emph{{Weak Interactions with Lepton
  - Hadron Symmetry}}, {\emph{Phys. Rev.} {\bfseries D2} (1970) 1285}.

\bibitem{SM3}
S.L.~Glashow, \emph{{Partial Symmetries of Weak Interactions}}, {\emph{Nucl.
  Phys.} {\bfseries 22} (1961) 579}.

\bibitem{SM4}
A.~Salam, \emph{Elementary Particle Theory}, N. Svartholm (Almqvist and
  Wiksell), Stockholm (1968).

\bibitem{zfitter1}
D.~Bardin et~al., \emph{Dizet 6.21-electroweak one-loop corrections for $e^+e^-
  \rightarrow f^+f^-$ around the $z^0$ peak},  1990.

\bibitem{zfitter6:1999}
D.~Bardin et~al., \emph{Zfitter v.6.21: A semianalytical program for fermion
  pair production in e+ e- annihilation},  2001.

\bibitem{zfitter:2006}
A.~Arbusov et~al., \emph{Zfitter: A semi-analytical program for fermion pair
  production in e+ e- annihilation, from version 6.21 to version 6.42},
  {\emph{Comput. Phys. Commun.} {\bfseries 174} (2006) 728–758}
  [\href{https://arxiv.org/abs/hep-ph/0507146}{{\ttfamily hep-ph/0507146}}].

\bibitem{dizet642}
A.~Arbusov et~al., \emph{The monte carlo program kkmc, for the lepton or quark
  pair production at lep/slc energies – updates of electroweak calculations},
  {\emph{Comput. Phys. Commun.} {\bfseries 260} (2020) 107734}
  [\href{https://arxiv.org/abs/2007.07964}{{\ttfamily 2007.07964}}].

\bibitem{Arbuzov:2020coe}
A.~Arbuzov, S.~Jadach, Z.~W\c{a}s, B.F.L.~Ward and S.A.~Yost, \emph{{The Monte
  Carlo Program KKMC , for the Lepton or Quark Pair Production at LEP/SLC
  Energies\textemdash{}Updates of electroweak calculations}},
  \href{https://doi.org/10.1016/j.cpc.2020.107734}{\emph{Comput. Phys. Commun.}
  {\bfseries 260} (2021) 107734}
  [\href{https://arxiv.org/abs/2007.07964}{{\ttfamily 2007.07964}}].

\bibitem{mstw-mass}
A.D.~Martin et~al., \emph{Parton distributions incorporating qed
  contributions}, \href{https://doi.org/10.1140/epjc/s2004-02088-7}{\emph{Eur.
  Phys. J. C} {\bfseries 39} (2005) 155}
  [\href{https://arxiv.org/abs/hep-ph/0411040}{{\ttfamily hep-ph/0411040}}].

\bibitem{PDG:2016}
{\scshape Particle Data Group} collaboration, \emph{Review of particle
  physics}, {\emph{Chin. Phys. C} {\bfseries 40} (2016) 100001}.

\bibitem{kkmchh1}
S.~Jadach, B.F.L.~Ward, Z.~W\c{a}s and S.~Yost, \emph{{Systematic Studies of
  Exact ${\cal O}(\alpha^2L)$ CEEX EW Corrections in a Hadronic MC for
  Precision Z/$\gamma^*$ Physics at LHC Energies}}, {\emph{Phys. Rev. D}
  {\bfseries 99} (2019) 076016}
  [\href{https://arxiv.org/abs/hep-ph/1707.06502}{{\ttfamily
  hep-ph/1707.06502}}].

\bibitem{fjeger-fccwksp2019}
F.~Jegerlehner\href{https://doi.org/10.23731/CYRM-2020-003}{\emph{{CERN Yellow
  Reports: Monographs, eds. A. Blondel {\it et al.}, CERN-2020-003}} (2020) 9}
  [\href{https://arxiv.org/abs/1905.05078}{{\ttfamily 1905.05078}}].

\bibitem{Kobel:2000aw}
{\scshape Two Fermion Working Group} collaboration, \emph{{Two-Fermion
  Production in Electron-Positron Collisions}},  in \emph{{Proceedings, Monte
  Carlo Workshop: Report of the working groups on precision calculation for
  LEP-2 physics: CERN, Geneva, Switzerland, March 12-13, June 25-26, October
  12-13 Oct 1999}}, 2000, \href{https://doi.org/10.5170/CERN-2000-009.269}{DOI}
  [\href{https://arxiv.org/abs/hep-ph/0007180}{{\ttfamily hep-ph/0007180}}].

\bibitem{Jadach:2019bye}
S.~Jadach and M.~Skrzypek, \emph{{QED challenges at FCC-ee precision
  measurements}},
  \href{https://doi.org/10.1140/epjc/s10052-019-7255-9}{\emph{Eur. Phys. J. C}
  {\bfseries 79} (2019) 756}
  [\href{https://arxiv.org/abs/1903.09895}{{\ttfamily 1903.09895}}].

\bibitem{Jadach:2018jjo}
S.~Jadach, W.~P\l{}aczek, M.~Skrzypek, B.F.L.~Ward and S.A.~Yost, \emph{{The
  path to 0.01\% theoretical luminosity precision for the FCC-ee}},
  \href{https://doi.org/10.1016/j.physletb.2019.01.012}{\emph{Phys. Lett. B}
  {\bfseries 790} (2019) 314}
  [\href{https://arxiv.org/abs/1812.01004}{{\ttfamily 1812.01004}}].

\bibitem{Jadach:2021ayv}
S.~Jadach, W.~P\l{}aczek, M.~Skrzypek and B.F.L.~Ward, \emph{{Study of
  theoretical luminosity precision for electron colliders at higher energies}},
  \href{https://doi.org/10.1140/epjc/s10052-021-09860-9}{\emph{Eur. Phys. J. C}
  {\bfseries 81} (2021) 1047}.

\bibitem{Jadach:2019wol}
S.~Jadach, W.~P\l{}aczek and M.~Skrzypek, \emph{{QED exponentiation for
  quasi-stable charged particles: the $e^-e^+\rightarrow W^-W^+$ process}},
  \href{https://doi.org/10.1140/epjc/s10052-020-8034-3}{\emph{Eur. Phys. J. C}
  {\bfseries 80} (2020) 499}
  [\href{https://arxiv.org/abs/1906.09071}{{\ttfamily 1906.09071}}].

\bibitem{radcor2021-bw}
{B.F.L. Ward} et~al., \emph{{IR-Improved Amplitude-Based Resummation in Quantum
  Field Theory: New Results and New Issues}}, {\emph{SciPost} {\bfseries
  RADCOR2021} (2021) in press}
  [\href{https://arxiv.org/abs/2111.01277}{{\ttfamily 2111.01277}}].

\bibitem{fad-kur:1985}
E.A.~Kuraev and V.S.~Fadin{\emph{Sov. J. Nucl. Phys.} {\bfseries 41} (1985)
  466}.

\bibitem{altarelli-mart:1986}
G.~Altarelli and G.~Martinelli{\emph{Yellow Report CERN-86-02} (1986) 47}.

\bibitem{nicro-trent:1987}
O.~Nicrosini and L.~Trentadue{\emph{Phys. Lett. B} {\bfseries 196} (1987) 551}.

\bibitem{fad-khz:1987}
V.S.~Fadin and V.S.~Khoze, 1987.

\bibitem{berends-neerver-burgers:1988}
F.A.~Berends, W.L.~van Neerven and G.J.H.~Burgers, \emph{{Higher Order
  Radiative Corrections at LEP Energies}},
  \href{https://doi.org/10.1016/0550-3213(88)90313-6}{\emph{Nucl. Phys. B}
  {\bfseries 297} (1988) 429}.

\bibitem{bluemlein-freitas-vnNeervn:2011}
J.~Blumlein, A.~De~Freitas and W.~van Neerven, \emph{{Two-loop QED Operator
  Matrix Elements with Massive External Fermion Lines}},
  \href{https://doi.org/10.1016/j.nuclphysb.2011.10.009}{\emph{Nucl. Phys. B}
  {\bfseries 855} (2012) 508}
  [\href{https://arxiv.org/abs/1107.4638}{{\ttfamily 1107.4638}}].

\bibitem{frixione-2019}
{S. Frixione} et~al.{\emph{J. High Energy Phys.} {\bfseries 1911} (2019) 158}
  [\href{https://arxiv.org/abs/1909.03886}{{\ttfamily 1909.03886}}].

\bibitem{bertone-2019}
{V. Bertone} et~al.{\emph{J. High Energy Phys.} {\bfseries 2003} (2020) 135}
  [\href{https://arxiv.org/abs/1911.12040}{{\ttfamily 1911.12040}}].

\bibitem{frixione-2021}
{S. Frixione}{\emph{J. High Energy Phys.} {\bfseries 2021} (2021) 180}
  [\href{https://arxiv.org/abs/2105.06688}{{\ttfamily 2105.06688}}].

\bibitem{bardin-zplep1:1989}
D.~Bardin et~al., \emph{{Z Line Shape}},  in \emph{{Z Physics at LEP 1,
  CERN-89-08, v. 1, eds. G. Altarelli, R. Kleiss, and C. Verzegnassi, CERN,
  Geneva, 1989}}, p.~89, 1989.

\bibitem{bhabhacern-lep2:1996}
S.~Jadach, O.~Nicrosini et~al., \emph{{Event generators for Bhabha
  scattering}},  in \emph{{CERN Workshop on LEP2 Physics, CERN-1996-01, v. 2,
  CERN, Geneva, 1996}}, pp.~229 -- 298, 1996.

\bibitem{sabspv:1995}
M.~Cacciari et~al., \emph{Sabspv: A monte carlo integrator for small angle
  bhabha scattering}, {\emph{Comput.Phys.Commun.} {\bfseries 90} (1995) 301}.

\bibitem{babayaga-2019}
C.M.~Carloni~Calame et~al., \emph{{Status of the BabaYaga event generator}},
  {\emph{EPJ Web Conf. PHIPSI17} {\bfseries 218} (2019) 07004}.

\bibitem{Actis:2010gg}
S.~Actis et~al., \emph{{Quest for precision in hadronic cross sections at low
  energy: Monte Carlo tools vs. experimental data}},
  \href{https://doi.org/10.1140/epjc/s10052-010-1251-4}{\emph{Eur. Phys. J.}
  {\bfseries C66} (2010) 585}
  [\href{https://arxiv.org/abs/0912.0749}{{\ttfamily 0912.0749}}].

\bibitem{mcmule:website}
\emph{{The {\sc McMule} framework}},
  \href{https://mule-tools.gitlab.io}{https://mule-tools.gitlab.io}.

\bibitem{MUonE:LoI}
C.~Matteuzzi, G.~Venanzoni, D.~Abbaneo, G.~Abbiendi, G.~Bagliesi, D.~Banerjee
  et~al., \emph{{Letter of Intent: the MUonE project}},  Tech. Rep.
  \href{http://cds.cern.ch/record/2677471}{CERN-SPSC-2019-026. SPSC-I-252},
  CERN, Geneva (Jun, 2019).

\bibitem{Banerjee:2020rww}
P.~Banerjee, T.~Engel, A.~Signer and Y.~Ulrich, \emph{{QED at NNLO with
  McMule}}, \href{https://doi.org/10.21468/SciPostPhys.9.2.027}{\emph{SciPost
  Phys.} {\bfseries 9} (2020) 027}
  [\href{https://arxiv.org/abs/2007.01654}{{\ttfamily 2007.01654}}].

\bibitem{Fael:2018dmz}
M.~Fael, \emph{{Hadronic corrections to $\mu$-$e$ scattering at NNLO with
  space-like data}}, \href{https://doi.org/10.1007/JHEP02(2019)027}{\emph{JHEP}
  {\bfseries 02} (2019) 027}
  [\href{https://arxiv.org/abs/1808.08233}{{\ttfamily 1808.08233}}].

\bibitem{Fael:2019nsf}
M.~Fael and M.~Passera, \emph{{Muon-Electron Scattering at
  Next-To-Next-To-Leading Order: The Hadronic Corrections}},
  \href{https://doi.org/10.1103/PhysRevLett.122.192001}{\emph{Phys. Rev. Lett.}
  {\bfseries 122} (2019) 192001}
  [\href{https://arxiv.org/abs/1901.03106}{{\ttfamily 1901.03106}}].

\bibitem{Bonciani:2021okt}
R.~Bonciani et~al., \emph{{Two-Loop Four-Fermion Scattering Amplitude in QED}},
  \href{https://doi.org/10.1103/PhysRevLett.128.022002}{\emph{Phys. Rev. Lett.}
  {\bfseries 128} (2022) 022002}
  [\href{https://arxiv.org/abs/2106.13179}{{\ttfamily 2106.13179}}].

\bibitem{Engel:2019nfw}
T.~Engel, A.~Signer and Y.~Ulrich, \emph{{A subtraction scheme for massive
  QED}}, \href{https://doi.org/10.1007/JHEP01(2020)085}{\emph{JHEP} {\bfseries
  01} (2020) 085} [\href{https://arxiv.org/abs/1909.10244}{{\ttfamily
  1909.10244}}].

\bibitem{Buccioni:2017yxi}
F.~Buccioni, S.~Pozzorini and M.~Zoller, \emph{{On-the-fly reduction of open
  loops}}, \href{https://doi.org/10.1140/epjc/s10052-018-5562-1}{\emph{Eur.
  Phys. J. C} {\bfseries 78} (2018) 70}
  [\href{https://arxiv.org/abs/1710.11452}{{\ttfamily 1710.11452}}].

\bibitem{Penin:2005eh}
A.A.~Penin, \emph{{Two-loop photonic corrections to massive Bhabha
  scattering}},
  \href{https://doi.org/10.1016/j.nuclphysb.2005.11.016}{\emph{Nucl. Phys. B}
  {\bfseries 734} (2006) 185}
  [\href{https://arxiv.org/abs/hep-ph/0508127}{{\ttfamily hep-ph/0508127}}].

\bibitem{Becher:2007cu}
T.~Becher and K.~Melnikov, \emph{{Two-loop QED corrections to Bhabha
  scattering}},
  \href{https://doi.org/10.1088/1126-6708/2007/06/084}{\emph{JHEP} {\bfseries
  06} (2007) 084} [\href{https://arxiv.org/abs/0704.3582}{{\ttfamily
  0704.3582}}].

\bibitem{Engel:2018fsb}
T.~Engel, C.~Gnendiger, A.~Signer and Y.~Ulrich, \emph{{Small-mass effects in
  heavy-to-light form factors}},
  \href{https://doi.org/10.1007/JHEP02(2019)118}{\emph{JHEP} {\bfseries 02}
  (2019) 118} [\href{https://arxiv.org/abs/1811.06461}{{\ttfamily
  1811.06461}}].

\bibitem{Banerjee:2021mty}
P.~Banerjee, T.~Engel, N.~Schalch, A.~Signer and Y.~Ulrich, \emph{{Bhabha
  scattering at NNLO with next-to-soft stabilisation}},
  \href{https://doi.org/10.1016/j.physletb.2021.136547}{\emph{Phys. Lett. B}
  {\bfseries 820} (2021) 136547}
  [\href{https://arxiv.org/abs/2106.07469}{{\ttfamily 2106.07469}}].

\bibitem{Banerjee:2021qvi}
P.~Banerjee, T.~Engel, N.~Schalch, A.~Signer and Y.~Ulrich, \emph{{M\o{}ller
  scattering at NNLO}},
  \href{https://doi.org/10.1103/PhysRevD.105.L031904}{\emph{Phys. Rev. D}
  {\bfseries 105} (2022) L031904}
  [\href{https://arxiv.org/abs/2107.12311}{{\ttfamily 2107.12311}}].

\bibitem{Low:1958sn}
F.~Low, \emph{{Bremsstrahlung of very low-energy quanta in elementary particle
  collisions}}, \href{https://doi.org/10.1103/PhysRev.110.974}{\emph{Phys.
  Rev.} {\bfseries 110} (1958) 974}.

\bibitem{Burnett:1967km}
T.~Burnett and N.M.~Kroll, \emph{{Extension of the low soft photon theorem}},
  \href{https://doi.org/10.1103/PhysRevLett.20.86}{\emph{Phys. Rev. Lett.}
  {\bfseries 20} (1968) 86}.

\bibitem{Adler:1966gc}
S.L.~Adler and Y.~Dothan, \emph{{Low-energy theorem for the weak axial-vector
  vertex}}, \href{https://doi.org/10.1103/PhysRev.151.1267}{\emph{Phys. Rev.}
  {\bfseries 151} (1966) 1267}.

\bibitem{Engel:2021ccn}
T.~Engel, A.~Signer and Y.~Ulrich, \emph{{Universal structure of radiative QED
  amplitudes at one loop}},  \href{https://arxiv.org/abs/2112.07570}{{\ttfamily
  2112.07570}}.

\bibitem{Beneke:1997zp}
M.~Beneke and V.A.~Smirnov, \emph{{Asymptotic expansion of Feynman integrals
  near threshold}},
  \href{https://doi.org/10.1016/S0550-3213(98)00138-2}{\emph{Nucl. Phys.}
  {\bfseries B522} (1998) 321}
  [\href{https://arxiv.org/abs/hep-ph/9711391}{{\ttfamily hep-ph/9711391}}].

\bibitem{Bonocore:2014wua}
D.~Bonocore, E.~Laenen, L.~Magnea, L.~Vernazza and C.D.~White, \emph{{The
  method of regions and next-to-soft corrections in Drell-Yan production}},
  \href{https://doi.org/10.1016/j.physletb.2015.02.008}{\emph{Phys. Lett.}
  {\bfseries B742} (2015) 375}
  [\href{https://arxiv.org/abs/1410.6406}{{\ttfamily 1410.6406}}].

\bibitem{DelDuca:2017twk}
V.~Del~Duca, E.~Laenen, L.~Magnea, L.~Vernazza and C.D.~White,
  \emph{{Universality of next-to-leading power threshold effects for colourless
  final states in hadronic collisions}},
  \href{https://doi.org/10.1007/JHEP11(2017)057}{\emph{JHEP} {\bfseries 11}
  (2017) 057} [\href{https://arxiv.org/abs/1706.04018}{{\ttfamily
  1706.04018}}].

\bibitem{Bonocore:2021cbv}
D.~Bonocore and A.~Kulesza, \emph{{Soft photon bremsstrahlung at
  next-to-leading power}},  \href{https://arxiv.org/abs/2112.08329}{{\ttfamily
  2112.08329}}.

\bibitem{Kollatzsch:2022xxx}
S.~Kollatzsch and Y.~Ulrich{\emph{in preparation} (2022) }.

\bibitem{FERMILAB-PUB-22-116-SCD-T}
\emph{Monte carlo white paper}, .

\bibitem{TwoFermionWorkingGroup:2000nks}
{\scshape Two Fermion Working Group} collaboration, \emph{{Two-Fermion
  Production in Electron-Positron Collisions}: {Two-Fermion Working Group
  Report}},  in \emph{{LEP2 Monte Carlo Workshop}}, 9, 2000,
  \href{https://doi.org/10.5170/CERN-2000-009.269}{DOI}
  [\href{https://arxiv.org/abs/hep-ph/0007180}{{\ttfamily hep-ph/0007180}}].

\bibitem{CEPCStudyGroup:2018ghi}
{\scshape CEPC Study Group} collaboration, \emph{{CEPC Conceptual Design
  Report: Volume 2 - Physics \& Detector}},
  \href{https://arxiv.org/abs/1811.10545}{{\ttfamily 1811.10545}}.

\bibitem{Blondel:2019qlh}
A.~Blondel, A.~Freitas, J.~Gluza, T.~Riemann, S.~Heinemeyer, S.~Jadach et~al.,
  \emph{{Theory Requirements and Possibilities for the FCC-ee and other Future
  High Energy and Precision Frontier Lepton Colliders}},
  \href{https://arxiv.org/abs/1901.02648}{{\ttfamily 1901.02648}}.

\bibitem{Montagna:1993ai}
G.~Montagna, F.~Piccinini, O.~Nicrosini, G.~Passarino and R.~Pittau,
  \emph{{TOPAZ0: A Program for computing observables and for fitting
  cross-sections and forward - backward asymmetries around the Z0 peak}},
  \href{https://doi.org/10.1016/0010-4655(93)90060-P}{\emph{Comput. Phys.
  Commun.} {\bfseries 76} (1993) 328}.

\bibitem{Bardin:1999yd}
D.Y.~Bardin, P.~Christova, M.~Jack, L.~Kalinovskaya, A.~Olchevski, S.~Riemann
  et~al., \emph{{ZFITTER v.6.21: A Semianalytical program for fermion pair
  production in $e^+ e^-$ annihilation}},
  \href{https://doi.org/10.1016/S0010-4655(00)00152-1}{\emph{Comput. Phys.
  Commun.} {\bfseries 133} (2001) 229}
  [\href{https://arxiv.org/abs/hep-ph/9908433}{{\ttfamily hep-ph/9908433}}].

\bibitem{WorkingGrouponRadiativeCorrections:2010bjp}
{\scshape Working Group on Radiative Corrections, Monte Carlo Generators for
  Low Energies} collaboration, \emph{{Quest for precision in hadronic cross
  sections at low energy: Monte Carlo tools vs. experimental data}},
  \href{https://doi.org/10.1140/epjc/s10052-010-1251-4}{\emph{Eur. Phys. J. C}
  {\bfseries 66} (2010) 585} [\href{https://arxiv.org/abs/0912.0749}{{\ttfamily
  0912.0749}}].

\bibitem{Catani:1999ss}
S.~Catani and M.~Grazzini, \emph{{Infrared factorization of tree level QCD
  amplitudes at the next-to-next-to-leading order and beyond}},
  \href{https://doi.org/10.1016/S0550-3213(99)00778-6}{\emph{Nucl. Phys. B}
  {\bfseries 570} (2000) 287}
  [\href{https://arxiv.org/abs/hep-ph/9908523}{{\ttfamily hep-ph/9908523}}].

\bibitem{Catani:1998nv}
S.~Catani and M.~Grazzini, \emph{{Collinear factorization and splitting
  functions for next-to-next-to-leading order QCD calculations}},
  \href{https://doi.org/10.1016/S0370-2693(98)01513-5}{\emph{Phys. Lett. B}
  {\bfseries 446} (1999) 143}
  [\href{https://arxiv.org/abs/hep-ph/9810389}{{\ttfamily hep-ph/9810389}}].

\bibitem{Anastasiou:2015vya}
C.~Anastasiou, C.~Duhr, F.~Dulat, F.~Herzog and B.~Mistlberger, \emph{{Higgs
  Boson Gluon-Fusion Production in QCD at Three Loops}},
  \href{https://doi.org/10.1103/PhysRevLett.114.212001}{\emph{Phys. Rev. Lett.}
  {\bfseries 114} (2015) 212001}
  [\href{https://arxiv.org/abs/1503.06056}{{\ttfamily 1503.06056}}].

\bibitem{Duhr:2021vwj}
C.~Duhr and B.~Mistlberger, \emph{{Lepton-pair production at hadron colliders
  at N$^3$LO in QCD}},  \href{https://arxiv.org/abs/2111.10379}{{\ttfamily
  2111.10379}}.

\bibitem{Kinoshita:1958ru}
T.~Kinoshita and A.~Sirlin, \emph{{Radiative corrections to Fermi
  interactions}}, \href{https://doi.org/10.1103/PhysRev.113.1652}{\emph{Phys.
  Rev.} {\bfseries 113} (1959) 1652}.

\bibitem{Kinoshita:1962ur}
T.~Kinoshita, \emph{{Mass singularities of Feynman amplitudes}},
  \href{https://doi.org/10.1063/1.1724268}{\emph{J. Math. Phys.} {\bfseries 3}
  (1962) 650}.

\bibitem{Lee:1964is}
T.D.~Lee and M.~Nauenberg, \emph{{Degenerate Systems and Mass Singularities}},
  \href{https://doi.org/10.1103/PhysRev.133.B1549}{\emph{Phys. Rev.} {\bfseries
  133} (1964) B1549}.

\bibitem{Moch:2004pa}
S.~Moch, J.A.M.~Vermaseren and A.~Vogt, \emph{{The Three loop splitting
  functions in QCD: The Nonsinglet case}},
  \href{https://doi.org/10.1016/j.nuclphysb.2004.03.030}{\emph{Nucl. Phys. B}
  {\bfseries 688} (2004) 101}
  [\href{https://arxiv.org/abs/hep-ph/0403192}{{\ttfamily hep-ph/0403192}}].

\bibitem{Vogt:2004mw}
A.~Vogt, S.~Moch and J.A.M.~Vermaseren, \emph{{The Three-loop splitting
  functions in QCD: The Singlet case}},
  \href{https://doi.org/10.1016/j.nuclphysb.2004.04.024}{\emph{Nucl. Phys. B}
  {\bfseries 691} (2004) 129}
  [\href{https://arxiv.org/abs/hep-ph/0404111}{{\ttfamily hep-ph/0404111}}].

\bibitem{Ablinger:2014nga}
J.~Ablinger, A.~Behring, J.~Bl\"umlein, A.~De~Freitas, A.~von Manteuffel and
  C.~Schneider, \emph{{The 3-loop pure singlet heavy flavor contributions to
  the structure function $F_2(x,Q^2)$ and the anomalous dimension}},
  \href{https://doi.org/10.1016/j.nuclphysb.2014.10.008}{\emph{Nucl. Phys. B}
  {\bfseries 890} (2014) 48} [\href{https://arxiv.org/abs/1409.1135}{{\ttfamily
  1409.1135}}].

\bibitem{Ablinger:2017tan}
J.~Ablinger, A.~Behring, J.~Bl\"umlein, A.~De~Freitas, A.~von Manteuffel and
  C.~Schneider, \emph{{The three-loop splitting functions $P_{qg}^{(2)}$ and
  $P_{gg}^{(2, N_F)}$}},
  \href{https://doi.org/10.1016/j.nuclphysb.2017.06.004}{\emph{Nucl. Phys. B}
  {\bfseries 922} (2017) 1} [\href{https://arxiv.org/abs/1705.01508}{{\ttfamily
  1705.01508}}].

\bibitem{Butterworth:2015oua}
J.~Butterworth et~al., \emph{{PDF4LHC recommendations for LHC Run II}},
  \href{https://doi.org/10.1088/0954-3899/43/2/023001}{\emph{J. Phys. G}
  {\bfseries 43} (2016) 023001}
  [\href{https://arxiv.org/abs/1510.03865}{{\ttfamily 1510.03865}}].

\bibitem{Salam:2008qg}
G.P.~Salam and J.~Rojo, \emph{{A Higher Order Perturbative Parton Evolution
  Toolkit (HOPPET)}},
  \href{https://doi.org/10.1016/j.cpc.2008.08.010}{\emph{Comput. Phys. Commun.}
  {\bfseries 180} (2009) 120}
  [\href{https://arxiv.org/abs/0804.3755}{{\ttfamily 0804.3755}}].

\bibitem{Monni:2019whf}
P.F.~Monni, P.~Nason, E.~Re, M.~Wiesemann and G.~Zanderighi,
  \emph{{MiNNLO$_{PS}$: a new method to match NNLO QCD to parton showers}},
  \href{https://doi.org/10.1007/JHEP05(2020)143}{\emph{JHEP} {\bfseries 05}
  (2020) 143} [\href{https://arxiv.org/abs/1908.06987}{{\ttfamily
  1908.06987}}].

\bibitem{Grammer:1973db}
J.~Grammer, G. and D.~Yennie, \emph{{Improved treatment for the infrared
  divergence problem in quantum electrodynamics}},
  \href{https://doi.org/10.1103/PhysRevD.8.4332}{\emph{Phys. Rev. D} {\bfseries
  8} (1973) 4332}.

\bibitem{Bodwin:1984hc}
G.T.~Bodwin, \emph{{Factorization of the Drell-Yan Cross-Section in
  Perturbation Theory}},
  \href{https://doi.org/10.1103/PhysRevD.34.3932}{\emph{Phys. Rev. D}
  {\bfseries 31} (1985) 2616}.

\bibitem{Collins:1985ue}
J.C.~Collins, D.E.~Soper and G.F.~Sterman, \emph{{Factorization for Short
  Distance Hadron - Hadron Scattering}},
  \href{https://doi.org/10.1016/0550-3213(85)90565-6}{\emph{Nucl. Phys. B}
  {\bfseries 261} (1985) 104}.

\bibitem{Collins:1988ig}
J.C.~Collins, D.E.~Soper and G.F.~Sterman, \emph{{Soft Gluons and
  Factorization}},
  \href{https://doi.org/10.1016/0550-3213(88)90130-7}{\emph{Nucl. Phys. B}
  {\bfseries 308} (1988) 833}.

\bibitem{Laenen:2010uz}
E.~Laenen, L.~Magnea, G.~Stavenga and C.D.~White, \emph{{Next-to-eikonal
  corrections to soft gluon radiation: a diagrammatic approach}},
  \href{https://doi.org/10.1007/JHEP01(2011)141}{\emph{JHEP} {\bfseries 1101}
  (2011) 141} [\href{https://arxiv.org/abs/1010.1860}{{\ttfamily 1010.1860}}].

\bibitem{DelDuca:1990gz}
V.~Del~Duca, \emph{{High-energy Bremsstrahlung Theorems for Soft Photons}},
  \href{https://doi.org/10.1016/0550-3213(90)90392-Q}{\emph{Nucl. Phys. B}
  {\bfseries 345} (1990) 369}.

\bibitem{Bauer:2000ew}
C.W.~Bauer, S.~Fleming and M.E.~Luke, \emph{{Summing Sudakov logarithms in B
  ---> X(s gamma) in effective field theory}},
  \href{https://doi.org/10.1103/PhysRevD.63.014006}{\emph{Phys. Rev. D}
  {\bfseries 63} (2000) 014006}
  [\href{https://arxiv.org/abs/hep-ph/0005275}{{\ttfamily hep-ph/0005275}}].

\bibitem{Bauer:2001ct}
C.W.~Bauer and I.W.~Stewart, \emph{{Invariant operators in collinear effective
  theory}}, \href{https://doi.org/10.1016/S0370-2693(01)00902-9}{\emph{Phys.
  Lett. B} {\bfseries 516} (2001) 134}
  [\href{https://arxiv.org/abs/hep-ph/0107001}{{\ttfamily hep-ph/0107001}}].

\bibitem{Bauer:2001yt}
C.W.~Bauer, D.~Pirjol and I.W.~Stewart, \emph{{Soft collinear factorization in
  effective field theory}},
  \href{https://doi.org/10.1103/PhysRevD.65.054022}{\emph{Phys. Rev. D}
  {\bfseries 65} (2002) 054022}
  [\href{https://arxiv.org/abs/hep-ph/0109045}{{\ttfamily hep-ph/0109045}}].

\bibitem{Beneke:2002ph}
M.~Beneke, A.~Chapovsky, M.~Diehl and T.~Feldmann, \emph{{Soft collinear
  effective theory and heavy to light currents beyond leading power}},
  \href{https://doi.org/10.1016/S0550-3213(02)00687-9}{\emph{Nucl. Phys. B}
  {\bfseries 643} (2002) 431}
  [\href{https://arxiv.org/abs/hep-ph/0206152}{{\ttfamily hep-ph/0206152}}].

\bibitem{Larkoski:2014bxa}
A.J.~Larkoski, D.~Neill and I.W.~Stewart, \emph{{Soft Theorems from Effective
  Field Theory}}, \href{https://doi.org/10.1007/JHEP06(2015)077}{\emph{JHEP}
  {\bfseries 06} (2015) 077} [\href{https://arxiv.org/abs/1412.3108}{{\ttfamily
  1412.3108}}].

\bibitem{Moult:2019mog}
I.~Moult, I.W.~Stewart and G.~Vita, \emph{{Subleading Power Factorization with
  Radiative Functions}},
  \href{https://doi.org/10.1007/JHEP11(2019)153}{\emph{JHEP} {\bfseries 11}
  (2019) 153} [\href{https://arxiv.org/abs/1905.07411}{{\ttfamily
  1905.07411}}].

\bibitem{Beneke:2019oqx}
M.~Beneke, A.~Broggio, S.~Jaskiewicz and L.~Vernazza, \emph{{Threshold
  factorization of the Drell-Yan process at next-to-leading power}},
  \href{https://doi.org/10.1007/JHEP07(2020)078}{\emph{JHEP} {\bfseries 07}
  (2020) 078} [\href{https://arxiv.org/abs/1912.01585}{{\ttfamily
  1912.01585}}].

\bibitem{Broggio:2021fnr}
A.~Broggio, S.~Jaskiewicz and L.~Vernazza, \emph{{Next-to-leading power
  two-loop soft functions for the Drell-Yan process at threshold}},
  \href{https://doi.org/10.1007/JHEP10(2021)061}{\emph{JHEP} {\bfseries 10}
  (2021) 061} [\href{https://arxiv.org/abs/2107.07353}{{\ttfamily
  2107.07353}}].

\bibitem{Bonocore:2015esa}
D.~Bonocore, E.~Laenen, L.~Magnea, S.~Melville, L.~Vernazza and C.~White,
  \emph{{A factorization approach to next-to-leading-power threshold
  logarithms}}, \href{https://doi.org/10.1007/JHEP06(2015)008}{\emph{JHEP}
  {\bfseries 06} (2015) 008}
  [\href{https://arxiv.org/abs/1503.05156}{{\ttfamily 1503.05156}}].

\bibitem{Bonocore:2016awd}
D.~Bonocore, E.~Laenen, L.~Magnea, L.~Vernazza and C.D.~White,
  \emph{{Non-abelian factorisation for next-to-leading-power threshold
  logarithms}}, \href{https://doi.org/10.1007/JHEP12(2016)121}{\emph{JHEP}
  {\bfseries 12} (2016) 121}
  [\href{https://arxiv.org/abs/1610.06842}{{\ttfamily 1610.06842}}].

\bibitem{Liu:2019oav}
Z.L.~Liu and M.~Neubert, \emph{{Factorization at subleading power and
  endpoint-divergent convolutions in $h\to\gamma\gamma$ decay}},
  \href{https://doi.org/10.1007/JHEP04(2020)033}{\emph{JHEP} {\bfseries 04}
  (2020) 033} [\href{https://arxiv.org/abs/1912.08818}{{\ttfamily
  1912.08818}}].

\bibitem{Liu:2020tzd}
Z.L.~Liu, B.~Mecaj, M.~Neubert and X.~Wang, \emph{{Factorization at subleading
  power, Sudakov resummation, and endpoint divergences in soft-collinear
  effective theory}},
  \href{https://doi.org/10.1103/PhysRevD.104.014004}{\emph{Phys. Rev. D}
  {\bfseries 104} (2021) 014004}
  [\href{https://arxiv.org/abs/2009.04456}{{\ttfamily 2009.04456}}].

\bibitem{Moult:2019uhz}
I.~Moult, I.W.~Stewart, G.~Vita and H.X.~Zhu, \emph{{The Soft Quark Sudakov}},
  \href{https://doi.org/10.1007/JHEP05(2020)089}{\emph{JHEP} {\bfseries 05}
  (2020) 089} [\href{https://arxiv.org/abs/1910.14038}{{\ttfamily
  1910.14038}}].

\bibitem{Beneke:2020ibj}
M.~Beneke, M.~Garny, S.~Jaskiewicz, R.~Szafron, L.~Vernazza and J.~Wang,
  \emph{{Large-x resummation of off-diagonal deep-inelastic parton scattering
  from d-dimensional refactorization}},
  \href{https://arxiv.org/abs/2008.04943}{{\ttfamily 2008.04943}}.

\bibitem{Laenen:2008gt}
E.~Laenen, G.~Stavenga and C.D.~White, \emph{{Path integral approach to eikonal
  and next-to-eikonal exponentiation}},
  \href{https://doi.org/10.1088/1126-6708/2009/03/054}{\emph{JHEP} {\bfseries
  03} (2009) 054} [\href{https://arxiv.org/abs/0811.2067}{{\ttfamily
  0811.2067}}].

\bibitem{Gardi:2010rn}
E.~Gardi, E.~Laenen, G.~Stavenga and C.D.~White, \emph{{Webs in multiparton
  scattering using the replica trick}},
  \href{https://doi.org/10.1007/JHEP11(2010)155}{\emph{JHEP} {\bfseries 1011}
  (2010) 155} [\href{https://arxiv.org/abs/1008.0098}{{\ttfamily 1008.0098}}].

\bibitem{Bahjat-Abbas:2019fqa}
N.~Bahjat-Abbas, D.~Bonocore, J.~Sinninghe~Damst\'{e}, E.~Laenen, L.~Magnea,
  L.~Vernazza et~al., \emph{{Diagrammatic resummation of leading-logarithmic
  threshold effects at next-to-leading power}},
  \href{https://doi.org/10.1007/JHEP11(2019)002}{\emph{JHEP} {\bfseries 11}
  (2019) 002} [\href{https://arxiv.org/abs/1905.13710}{{\ttfamily
  1905.13710}}].

\bibitem{vanBeekveld:2021mxn}
M.~van Beekveld, L.~Vernazza and C.D.~White, \emph{{Threshold resummation of
  new partonic channels at next-to-leading power}},
  \href{https://doi.org/10.1007/JHEP12(2021)087}{\emph{JHEP} {\bfseries 12}
  (2021) 087} [\href{https://arxiv.org/abs/2109.09752}{{\ttfamily
  2109.09752}}].

\bibitem{Gervais:2017yxv}
H.~Gervais, \emph{{Soft Photon Theorem for High Energy Amplitudes in Yukawa and
  Scalar Theories}},
  \href{https://doi.org/10.1103/PhysRevD.95.125009}{\emph{Phys. Rev.}
  {\bfseries D95} (2017) 125009}
  [\href{https://arxiv.org/abs/1704.00806}{{\ttfamily 1704.00806}}].

\bibitem{Laenen:2020nrt}
E.~Laenen, J.~Sinninghe~Damst\'e, L.~Vernazza, W.~Waalewijn and L.~Zoppi,
  \emph{{Towards all-order factorization of QED amplitudes at next-to-leading
  power}}, \href{https://doi.org/10.1103/PhysRevD.103.034022}{\emph{Phys. Rev.
  D} {\bfseries 103} (2021) 034022}
  [\href{https://arxiv.org/abs/2008.01736}{{\ttfamily 2008.01736}}].

\bibitem{Collins:1989bt}
J.C.~Collins, \emph{{Sudakov form-factors}}, {\emph{Adv.Ser.Direct.High Energy
  Phys.} {\bfseries 5} (1989) 573}
  [\href{https://arxiv.org/abs/hep-ph/0312336}{{\ttfamily hep-ph/0312336}}].

\bibitem{Dixon:2008gr}
L.J.~Dixon, L.~Magnea and G.F.~Sterman, \emph{{Universal structure of
  subleading infrared poles in gauge theory amplitudes}},
  \href{https://doi.org/10.1088/1126-6708/2008/08/022}{\emph{JHEP} {\bfseries
  0808} (2008) 022} [\href{https://arxiv.org/abs/0805.3515}{{\ttfamily
  0805.3515}}].

\bibitem{Landau:1959fi}
L.~Landau, \emph{{On analytic properties of vertex parts in quantum field
  theory}},
  \href{https://doi.org/10.1016/B978-0-08-010586-4.50103-6}{\emph{Nucl. Phys.}
  {\bfseries 13} (1960) 181}.

\bibitem{Coleman:1965xm}
S.~Coleman and R.~Norton, \emph{{Singularities in the physical region}},
  \href{https://doi.org/10.1007/BF02750472}{\emph{Nuovo Cim.} {\bfseries 38}
  (1965) 438}.

\bibitem{Sterman:1978bi}
G.F.~Sterman, \emph{{Mass Divergences in Annihilation Processes. 1. Origin and
  Nature of Divergences in Cut Vacuum Polarization Diagrams}},
  \href{https://doi.org/10.1103/PhysRevD.17.2773}{\emph{Phys. Rev.} {\bfseries
  D17} (1978) 2773}.

\end{thebibliography}

\providecommand{\href}[2]{#2}\begingroup\raggedright\endgroup

\end{document}